\documentclass[preprint,aps,showpacs,amsmath,amssymb,prd,eqsecnum]{revtex4-1}

\usepackage{natbib}
\allowdisplaybreaks

\begin{document}

\title{Local momentum space and the vector field}

\author{David J. Toms}
\homepage{http://www.staff.ncl.ac.uk/d.j.toms}
\email{d.j.toms@newcastle.ac.uk}
\affiliation{
School of Mathematics and Statistics,
Newcastle University,
Newcastle upon Tyne, U.K. NE1 7RU}

\date{\today}

\begin{abstract}
The local momentum space expansion for the real vector field is considered. Using Riemann normal coordinates we obtain an expansion of the Feynman Green function up and including terms that are quadratic in the curvature. The results are valid for a non-minimal operator such as that arising from a general Feynman type gauge fixing condition. The result is used to derive the first three terms in the asymptotic expansion for the coincidence limit of the heat kernel without taking the trace, thus obtaining the untraced heat kernel coefficients. The spacetime dimension is kept general before specializing to four dimensions for comparison with previously known results. As a further application we re-examine the anomalous trace of the stress-energy-momentum tensor for the Maxwell field and comment on the gauge dependence.
\end{abstract}


\maketitle

\section{Introduction}\label{sec-intro}

Since the introduction of the local momentum space method into quantum field theory in curved spacetime by Bunch and Parker~\cite{BunchParker}, the technique has been used in a variety of different applications. The original application was to consider the renormalization of interacting quantum fields~\cite{bunch1981local,bunch1981bphz,tomsYM,leen1983renormalization,ParkerTomsbook} and to study the renormalization group behaviour of gauge theories in curved spacetime~\cite{PhysRevLett.52.1269,PhysRevD.29.1584}. It has also been used in part to calculate the dependence of the one loop effective action on the scalar curvature and demonstrate curvature induced asymptotic freedom~\cite{calzetta1985curvature,calzetta1986quantum}. It has been used to study the Wigner function in curved spacetime~\cite{PhysRevD.37.2901}, and to obtain an expansion of the effective action at zero~\cite{hu1984effective} and finite temperature~\cite{hu1987finite,moss1992effective}. The application to Kaluza-Klein theory was given in \cite{toms1983induced,awada1984induced,huggins1988vilkovisky}. More recently it has been used to investigate quantum gravitational effects on gauge coupling constants~\cite{toms2010quantum,PhysRevD.84.084016}, and directly related to the present paper was one of the methods used to calculate heat kernel coefficients for non-minimal operators~\cite{MossToms}.

The purpose of the present paper is to extend the local momentum space technique for the Green function to the case of real vector fields with a general gauge parameter. These results can be used to check the possible gauge dependence in calculations, and can in some cases be used to justify the standard choice of the Feynman gauge. We will also generalize some of the results of \cite{MossToms} to the case of the untraced heat kernel coefficients for the real vector field for a non-minimal operator. Most previous attention has focussed mainly on the case of traced heat kernel coefficients. (See for example, \cite{BarvinskyVilkovisky,gilkey1991heat,fulling1992kernel,fulling1992kernel2,branson1994heat,gusynin1997computation}.) An exception is \cite{endo1984gauge} for electromagnetism in four spacetime dimensions, and \cite{gusynin1999complete} where general, but extremely lengthy, expressions are given for the untraced coefficients. The results that we quote below agree with those of \cite{MossToms} when the trace is taken, and are valid for any spacetime dimension. The four dimensional special case reproduces the results of \cite{endo1984gauge} and \cite{BarvinskyVilkovisky}.

As a particular application of our results we will re-examine the anomalous trace of the stress-energy-momentum tensor, $T_{\mu\nu}$, for the quantized Maxwell field, and comment on the $\Box R$ controversy. The regularization and renormalization of $T_{\mu\nu}$ for the Maxwell field has a long and controversial history which is briefly reviewed in \cite{Duff20}. Of particular relevance to the present paper are \cite{brown1977stress,adler1977regularization,adler1978trace,endo1984gauge,nielsen1988gauge}. We will comment more on this in Secs.~\ref{sec5} and \ref{sec6}.

The outline of our paper is as follows. Sec.~\ref{sec2} sets out a review of the general formalism that we use to obtain the local momentum space expansion of a general Green function. Expressions are obtained for the first few terms, including all those necessary to compute the first three untraced heat kernel coefficients. In Sec.~\ref{sec3} we specialize to the real vector field and evaluate the first three heat kernel coefficients. A number of limiting cases of physical interest are presented and comparison is made with some previously known results as a check on the calculations. In Sec.~\ref{sec4} we consider the local momentum space expansion for the Maxwell field. Sec.~\ref{sec5} applies our results to the evaluation of the trace anomaly for electromagnetism and comments are made on the interpretation in Sec.~\ref{sec6} where our results are discussed briefly. The calculations are quite lengthy and some of the more cumbersome results are given in the appendices.

\section{General formalism}\label{sec2}

Consider a generic Bose field $\varphi^i(x)$. Here $i$ represents any type of indices. In the case we will look at the index will be a vector field index, but the formalism does not require this and we can be more general at this stage. Suppose that we use $\Delta^{i}{}_{j}$ to represent the relevant differential operator for the field $\varphi^i$. We choose a Riemannian spacetime metric, take the spacetime dimension to be $N$, and adopt the curvature conventions of \cite{MTW}. For the case of the real vector field that is the focus of this paper we have
\begin{equation}
\Delta^{\mu}{}_{\nu}=-\delta^{\mu}_{\nu}\,\Box+q\,\nabla^\mu\nabla_\nu+Q^{\mu}{}_{\nu}.\label{2.1.1}
\end{equation}
Here $q$ is some real parameter that comes from the gauge fixing in the quantum theory, and $Q^{\mu}{}_{\nu}$ is some function of $x$ with the indicated transformation properties under general coordinate transformations. In the special case of Maxwell theory, $Q_{\mu\nu}=R_{\mu\nu}$ which leads to considerable simplifications. We will keep $Q_{\mu\nu}$ general at this stage.

The heat kernel $K^{i}{}_{j}(x,x';\tau)$ for the differential operator $\Delta^{i}{}_{j}$ is a solution to 
\begin{equation}
\Delta^{i}{}_{j}K^{j}{}_{k}(x,x';\tau)=-\frac{\partial}{\partial\tau}K^{i}{}_{k}(x,x';\tau),\label{djt1.1}
\end{equation}
with the boundary condition
\begin{equation}
K^{i}{}_{j}(x,x';\tau=0)=\delta^{i}{}_{j}\delta(x,x').\label{djt1.2}
\end{equation}
Here $\delta(x,x')$ is the biscalar Dirac delta distribution. The importance of the heat kernel is that under fairly general assumptions it admits an asymptotic expansion as $\tau\rightarrow 0$ of the form
\begin{equation}
K^{i}{}_{j}(x,x;\tau)\sim(4\pi\tau)^{-N/2}\sum_{k=0}^{\infty}\tau^k(E_k)^{i}{}_{j}(x).\label{djt1.3}
\end{equation}
The coefficients $(E_k)^{i}{}_{j}(x)$ are the heat kernel coefficients that are local expressions determined solely by the form of the operator $\Delta^{i}{}_{j}$. (Note that we do not consider any contributions from a possible boundary here.)

The method that we will use here makes use of the Green function for the operator $\Delta^{i}{}_{j}$ rather than the heat kernel directly. Because the Green function is useful in calculations that are of interest in quantum field theory these results for the Green function will be useful later in Sec.~\ref{sec4} and Sec.~\ref{sec5}. There is a simple relationship between the two. Normally the Green function is defined as the solution to 
\begin{equation}\label{djt1.4}
\Delta^{i}{}_{j}G^{j}{}_{k}(x,x')=\delta^{i}_{j}\delta(x,x').
\end{equation}
This Green function is the analytic continuation of the normal Feynman Green function (or propagator) to imaginary time. We will consider it further in Secs.~\ref{sec4} and \ref{sec5}. It proves convenient to define an auxiliary Green function $G(x,x';s)$ as the solution to
\begin{equation}\label{djt1.5}
(\Delta^{i}{}_{k}-s\delta^{i}_{k})G^{k}{}_{j}(x,x';s)=\delta^{i}_{j}\delta(x,x').
\end{equation}
The usual Green function $G^{i}{}_{j}(x,x')$ in \eqref{djt1.4} is related clearly to the auxiliary Green function $G^{i}{}_{j}(x,x';s)$ by
\begin{equation}\label{djt1.6}
G^{i}{}_{j}(x,x')=G^{i}{}_{j}(x,x';s=0).
\end{equation}
The relation between the auxiliary Green function and the heat kernel is
\begin{equation}\label{djt1.7}
G^{i}{}_{j}(x,x';s)=\int\limits_{0}^{\infty}d\tau\;e^{s\tau}\,K^{i}{}_{j}(x,x';\tau),
\end{equation}
which can be recognized as a one-sided Laplace transform \cite{DuffNaylor}. The inverse of this, giving the heat kernel in terms of the auxiliary Green function, can be obtained as 
\begin{equation}\label{djt1.8}
K^{i}{}_{j}(x,x';\tau)=\int\limits_{c-i\infty}^{c+i\infty}\frac{ds}{2\pi i}\;e^{-s\tau}\;G^{i}{}_{j}(x,x';s).
\end{equation}
Here $c$ is chosen to be a real constant smaller than the lowest eigenvalue of the differential operator $\Delta^{i}{}_{j}$ and the contour is closed in the right hand side of the complex $s$-plane. It is easily verified, using (\ref{djt1.4}), that the heat kernel obeys (\ref{djt1.1}). The boundary condition \eqref{djt1.2} follows by using the expansion of the Green function in terms of eigenfunctions of the operator $\Delta^{i}{}_{j}$.

We will be interested in the case where 
\begin{equation}
(\Delta)^{i}{}_{j}=(A^{\alpha\beta})^{i}{}_{j}\,\partial_\alpha\partial_\beta +(B^{\alpha})^{i}{}_{j}\,\partial_\alpha +(C)^{i}{}_{j},\label{djt1.9}
\end{equation}
for some coefficients $(A^{\alpha\beta})^{i}{}_{j},\ (B^{\alpha})^{i}{}_{j}$ and $(C)^{i}{}_{j}$. For the operator for the real vector field in \eqref{2.1.1} this form follows simply by writing out the covariant derivatives in terms of ordinary ones. The results are (remembering that $i$ and $j$ are vector indices in this example)
\begin{eqnarray}
\big(A^{\mu\nu}\big)^{\lambda}{}_{\tau}&=&-g^{\mu\nu}\,\delta^{\lambda}_{\tau}+\frac{1}{2}q\big( g^{\lambda\mu}\,\delta^{\nu}_{\tau}+g^{\lambda\nu}\,\delta^{\mu}_{\tau}\big),\label{2.1.2}\\
\big(B^{\mu}\big)^{\lambda}{}_{\tau}&=&-2\,g^{\mu\nu}\,\Gamma^{\lambda}_{\nu\tau}+g^{\alpha\beta}\,\Gamma^{\mu}_{\alpha\beta}\,\delta^{\lambda}_{\tau}+q\,g^{\mu\lambda}\,\Gamma^{\nu}_{\nu\tau},\label{2.1.3}\\
\big(C\big)^{\lambda}{}_{\tau}&=&Q^{\lambda}{}_{\tau}+g^{\alpha\beta}\,\Gamma^{\gamma}_{\alpha\beta}\,\Gamma^{\lambda}_{\gamma\tau}-g^{\alpha\beta}\,\Gamma^{\lambda}_{\beta\gamma}\,\Gamma^{\gamma}_{\alpha\tau}-g^{\alpha\beta}\,\Gamma^{\lambda}_{\alpha\tau,\beta}\nonumber\\
&&\qquad+q\,g^{\lambda\sigma}\,\Gamma^{\nu}_{\nu\tau,\sigma}.\label{2.1.4}
\end{eqnarray}

When there is no ambiguity we will omit the component indices $i$ and $j$ and deal with the coefficients in \eqref{djt1.9} as matrices. Without loss of generality we can assume $A^{\beta\alpha}=A^{\alpha\beta}$. The method that we will adopt makes use of the local momentum space approach of Bunch and Parker~\cite{BunchParker} to calculate the auxiliary Green function in \eqref{djt1.5}. Introduce normal coordinates at the point $x'$ with $x^\mu=x^{\prime\mu}+y^\mu$.  The coefficients in \eqref{djt1.9} can all be expanded about $x^\mu=x^{\prime\mu}$, or equivalently $y^\mu=0$. This gives
\begin{eqnarray}
(A^{\alpha\beta})^{i}{}_{j}&=&(A_{0}^{\alpha\beta})^{i}{}_{j} +\sum_{n=2}^{\infty}(A^{\alpha\beta}{}_{\mu_1\cdots\mu_n})^{i}{}_{j}y^{\mu_1}\cdots y^{\mu_n},\label{djt1.10}\\
(B^{\alpha})^{i}{}_{j}&=&\sum_{n=1}^{\infty}(B^{\alpha}{}_{\mu_1\cdots\mu_n})^{i}{}_{j}y^{\mu_1}\cdots y^{\mu_n},\label{djt1.11}\\
(C)^{i}{}_{j}&=&(C_{0})^{i}{}_{j} +\sum_{n=1}^{\infty}(C_{\mu_1\cdots\mu_n})^{i}{}_{j}y^{\mu_1}\cdots y^{\mu_n}\;.\label{djt1.12}
\end{eqnarray}
These are not the most general possibilities for these expansions, but are sufficient to deal with the vector field case of the present paper. The absence of a linear term in $y^\alpha$ in \eqref{djt1.10} can be understood as a consequence of the fact that in examples of interest to quantum field theory $A^{\alpha\beta}$ depends only on the spacetime metric whose expansion in Riemann normal coordinates has the first non-trivial term quadratic in $y^\alpha$. (See the result in \eqref{2.1.2} for the vector field and the expansion for $g^{\mu\nu}$ in \eqref{3.5}.) The absence of a zeroth order term in \eqref{djt1.11} arises because $B^\alpha$ involves the connection whose Riemann normal coordinate expansion begins at order $y^\alpha$. (See the expression for the vector field in \eqref{2.1.3} and the expansion for $\Gamma^{\lambda}_{\mu\nu}$ in \eqref{3.6}.) 

The Green function, that obeys \eqref{djt1.5}, is Fourier expanded as usual,
\begin{equation}
G^{i}{}_{j}(x,x';s)=\int\frac{d^Np}{(2\pi)^N}\,e^{ip\cdot y}\,G^{i}{}_{j}(p;s),\label{djt1.13}
\end{equation}
except that the Fourier transform $G^{i}{}_{j}(p;s)$ can depend on the origin of the normal coordinates $x'$, but we will not indicate this dependence explicitly. The advantage of introducing the Fourier transform, as in flat spacetime, is that it turns the differential equation for the Green function into an algebraic equation for its Fourier expansion. Because of the similarity with the normal Fourier transform in flat spacetime quantum field theory, this is called the local momentum space expansion \cite{BunchParker}.

The aim now is to use \eqref{djt1.9} making use of the expansions \eqref{djt1.10}--\eqref{djt1.12} in \eqref{djt1.5} and use the local momentum space expansion \eqref{djt1.13} for the auxiliary Green function. The factors of $y^{\mu_1}\cdots y^{\mu_n}$ that occur in these expansions can be dealt with by using
\begin{equation}
y^{\mu_1}\cdots y^{\mu_n}\,e^{ip\cdot y}=(-i)^n\frac{\partial^n}{\partial p_{\mu_1}\cdots\partial p_{\mu_n}}\,e^{ip\cdot y},\label{djt1.14}
\end{equation}
followed by partial integrations with respect to $p$ to remove derivatives from the exponential factor. The following result is obtained (indices $i$ and $j$ suppressed)
\begin{eqnarray}
I&=&-A_0^{\mu\nu}p_\mu p_\nu G(p;s)-sG(p;s) +A^{\mu\nu}{}_{\alpha\beta}\,\frac{\partial^2}{\partial p_\alpha\partial p_\beta}\big\lbrack p_\mu p_\nu G(p;s)\big\rbrack\nonumber\\
&&+i\,A^{\mu\nu}{}_{\alpha\beta\gamma}\,\frac{\partial^3}{\partial p_\alpha\partial p_\beta \partial p_\gamma}\big\lbrack p_\mu p_\nu G(p;s)\big\rbrack\nonumber\\
&&-A^{\mu\nu}{}_{\alpha\beta\gamma\delta}\,\frac{\partial^4}{\partial p_\alpha\partial p_\beta \partial p_\gamma 
\partial p_\delta}\big\lbrack p_\mu p_\nu G(p;s)\big\rbrack\nonumber\\
&&-B^{\mu}{}_{\alpha}\,\frac{\partial}{\partial p_\alpha}\big\lbrack p_\mu G(p;s)\big\rbrack
-i\,B^{\mu}{}_{\alpha\beta}\,\frac{\partial^2}{\partial p_\alpha\partial p_\beta}\big\lbrack p_\mu G(p;s)\big\rbrack \nonumber\\
&&+B^{\mu}{}_{\alpha\beta\gamma}\,\frac{\partial^3}{\partial p_\alpha\partial p_\beta\partial p_\gamma}\big\lbrack p_\mu G(p;s)\big\rbrack +C_0G(p;s)\nonumber\\
&&+i\,C_\alpha\,\frac{\partial}{\partial p_\alpha}G(p;s)- C_{\alpha\beta}\,\frac{\partial^2}{\partial p_\alpha\partial p_\beta}G(p;s)+\cdots.\label{djt1.15}
\end{eqnarray}
The next step is to assume an expansion for $G(p;s)$ that is
\begin{equation}
G(p;s)=G_0(p;s)+G_2(p;s)+G_3(p;s)+G_4(p;s)+\cdots.\label{djt1.16}
\end{equation}
We will define $G_0(p;s)$ by
\begin{equation}
I=\big(-A_0^{\mu\nu}p_\mu p_\nu -sI\big)G_0(p;s).\label{djt1.17}
\end{equation}
This choice, although arbitrary, has a significant advantage over other choices that could be made as we will discuss later. The expansion \eqref{djt1.16} can be viewed as an asymptotic expansion in inverse powers of $p$ beginning with $G_0$ at order $p^{-2}$. From \eqref{djt1.15} it is clear that if the expansion $G=G_0+\cdots$ is used, simple power counting shows that we have terms of order $1,\ p^{-2},\ p^{-3},\ p^{-4}$ and so on. We can interpret a general term $G_n$ in \eqref{djt1.16} as the term in the asymptotic expansion of $G(p;s)$ that behaves like $p^{-2-n}$ for large $p$.

The terms of order $p^{-2}$ in \eqref{djt1.15} read
\begin{eqnarray}
0&=&\big(-A_{0}^{\mu\nu}p_\mu p_\nu-sI\big)G_2(p;s)+ A^{\mu\nu}{}_{\alpha\beta}\frac{\partial^2}{\partial p_\alpha\partial p_\beta}\big\lbrack p_\mu p_\nu G_0(p;s)\big\rbrack\nonumber\\
&&-B^{\mu}{}_{\alpha}\frac{\partial}{\partial p_\alpha}\big\lbrack p_\mu G_0(p;s)\big\rbrack +C_0 G_0(p;s).\label{djt1.18}
\end{eqnarray}

The terms of order $p^{-3}$ read
\begin{eqnarray}
0&=&\big(-A_{0}^{\mu\nu}p_\mu p_\nu-sI\big)G_3(p;s)+ iA^{\mu\nu}{}_{\alpha\beta\gamma}\frac{\partial^3}{\partial p_\alpha\partial p_\beta\partial p_\gamma}\big\lbrack p_\mu p_\nu G_0(p;s)\big\rbrack\nonumber\\
&&-iB^{\mu}{}_{\alpha\beta}\frac{\partial^2}{\partial p_\alpha \partial p_\beta}\big\lbrack p_\mu G_0(p;s)\big\rbrack +iC_\alpha\frac{\partial}{\partial p_\alpha} G_0(p;s).\label{djt1.18b}
\end{eqnarray}

The terms of order $p^{-4}$ read
\begin{eqnarray}
0&=&\big(-A_{0}^{\mu\nu}p_\mu p_\nu-sI\big)G_4(p;s)+ A^{\mu\nu}{}_{\alpha\beta}\frac{\partial^2}{\partial p_\alpha\partial p_\beta}\big\lbrack p_\mu p_\nu G_2(p;s)\big\rbrack\nonumber\\
&&-A^{\mu\nu}{}_{\alpha\beta\gamma\delta}\frac{\partial^4}{\partial p_\alpha\partial p_\beta\partial p_\gamma \partial p_\delta}\big\lbrack p_\mu p_\nu G_0(p;s)\big\rbrack -B^{\mu}{}_{\alpha}\frac{\partial}{\partial p_\alpha}\big\lbrack p_\mu G_2(p;s)\big\rbrack\nonumber\\
&&+B^{\mu}{}_{\alpha\beta\gamma}\frac{\partial^3}{\partial p_\alpha \partial p_\beta \partial p_\gamma}\big\lbrack p_\mu G_0(p;s)\big\rbrack +C_0G_2(p;s)\nonumber\\
&&\qquad-C_{\alpha\beta}\frac{\partial^2}{\partial p_\alpha \partial p_\beta} G_0(p;s).\label{djt1.19}
\end{eqnarray}

The equations that we have found for the different orders can be solved recursively. We first solve for $G_0$ using \eqref{djt1.17} for a given $A_0^{\mu\nu}$. This then determines $G_2$ from \eqref{djt1.18} and $G_3$ from \eqref{djt1.18b}. $G_4$ is found from \eqref{djt1.19} and the knowledge of $G_0$ and $G_2$. It is clear how higher order terms could be obtained, although with increasing algebraic complexity.

Because of \eqref{djt1.18} we can immediately write down
\begin{equation}
G_2(p;s)=G_{21}(p;s)+G_{22}(p;s)+G_{23}(p;s),\label{djt1.20}
\end{equation}
where 
\begin{eqnarray}
G_{21}(p;s)&=&-G_0(p;s)\,A^{\mu\nu}{}_{\alpha\beta}\,\frac{\partial^2}{\partial p_\alpha\partial p_\beta}\big\lbrack p_\mu p_\nu G_0(p;s)\big\rbrack,\label{djt1.21}\\
G_{22}(p;s)&=&G_0(p;s)\,B^{\mu}{}_{\alpha}\,\frac{\partial}{\partial p_\alpha}\big\lbrack p_\mu  G_0(p;s)\big\rbrack,\label{djt1.22}\\
G_{23}(p;s)&=&-G_0(p;s)\,C_0\, G_0(p;s).\label{djt1.23}
\end{eqnarray}
Obviously it is not strictly necessary to split the terms in $G_2$ up as we have done but it does prove helpful in doing so in order to demonstrate the source of various terms in the heat kernel coefficients, as well as to isolate any sources of calculational error.

Similarly from (\ref{djt1.18b}) we have 
\begin{equation}
G_3(p;s)=G_{31}(p;s)+G_{32}(p;s)+G_{33}(p;s),\label{3.1.24}
\end{equation}
where 
\begin{eqnarray}
G_{31}(p;s)&=&-iG_0(p;s)\,A^{\mu\nu}{}_{\alpha\beta\gamma}\,\frac{\partial^3}{\partial p_\alpha\partial p_\beta \partial p_\gamma} 
\big\lbrack p_\mu p_\nu G_0(p;s)\big\rbrack,\label{3.1.25}\\
G_{32}(p;s)&=&iG_0(p;s)\,B^{\mu}{}_{\alpha\beta}\,\frac{\partial^2}{\partial p_\alpha \partial p_\beta}
\big\lbrack p_\mu  G_0(p;s)\big\rbrack,\label{3.1.26}\\
G_{33}(p;s)&=&-iG_0(p;s)\,C_\alpha\,\frac{\partial}{\partial p_\alpha} G_0(p;s).\label{3.1.27}
\end{eqnarray}

Finally from \eqref{djt1.19} we have
\begin{equation}
G_4(p;s)=\sum_{n=1}^{6}G_{4n}(p;s),\label{djt1.24}
\end{equation}
where 
\begin{eqnarray}
G_{41}(p;s)&=&-G_0(p;s)\,A^{\mu\nu}{}_{\alpha\beta}\,\frac{\partial^2}{\partial p_\alpha\partial p_\beta}\big\lbrack p_\mu p_\nu G_2(p;s)\big\rbrack,\label{djt1.25}\\
G_{42}(p;s)&=&G_0(p;s)\,A^{\mu\nu}{}_{\alpha\beta\gamma\delta}\,\frac{\partial^4}{\partial p_\alpha 
\partial p_\beta \partial p_\gamma \partial p_\delta}\big\lbrack p_\mu p_\nu G_0(p;s)\big\rbrack,\label{djt1.26}\\
G_{43}(p;s)&=&G_0(p;s)\,B^{\mu}{}_{\alpha}\,\frac{\partial}{\partial p_\alpha}\big\lbrack p_\mu  G_2(p;s)\big\rbrack,\label{djt1.27}\\
G_{44}(p;s)&=&-G_0(p;s)\,B^{\mu}{}_{\alpha\beta\gamma}\,\frac{\partial^3}{\partial p_\alpha \partial p_\beta \partial p_\gamma}\big\lbrack p_\mu  G_0(p;s)\big\rbrack,\label{djt1.28}\\
G_{45}(p;s)&=&-G_0(p;s)\,C_0\, G_2(p;s),\label{djt1.29}\\
G_{46}(p;s)&=&G_0(p;s)\,C_{\alpha\beta}\,\frac{\partial^2}{\partial p_\alpha\partial p_\beta} G_0(p;s).\label{djt1.30}
\end{eqnarray}

We will consider a direct link between the expansions we have been considering and the heat kernel coefficients defined in \eqref{djt1.3} in the following section.

\section{Heat kernel coefficients}\label{sec3}

If we take $x^\mu\rightarrow x^{\prime\mu}$ in \eqref{djt1.8} and then substitute the local momentum expansion \eqref{djt1.13} we have
\begin{equation}
K(x',x';\tau)=\int\frac{d^Np}{(2\pi)^N} \int\limits_{c-i\infty}^{c+i\infty}\frac{ds}{2\pi i}\;e^{-s\tau}\,G(p;s).\label{djt1.31}
\end{equation}
It is now possible to see that if we use the expansion \eqref{djt1.16} then
\begin{equation}
E_k=(4\pi\tau)^{N/2}\tau^{-k}\int\frac{d^Np}{(2\pi)^N}\int\limits_{c-i\infty}^{c+i\infty}\frac{ds}{2\pi i}\,e^{-s\tau}\,G_{2k}(p;s).\label{djt1.32}
\end{equation}
The proof of this assertion makes use of a rescaling of $s\rightarrow \tau^{-1}s$ followed by $p_\mu\rightarrow\tau^{-1/2}p_\mu$. The crucial feature that makes \eqref{djt1.32} true is that $G_0$ as defined in \eqref{djt1.17} obeys the scaling relation
\begin{equation}
G_0(\tau^{-1/2}p;s/\tau)=\tau G_0(p;s).\label{djt1.33}
\end{equation}

When calculating the heat kernel coefficients it is advantageous to adopt a local orthonormal frame by introducing the $N$-bein $e^{a}{}_{\mu}(x)$. This is defined so that
\begin{equation}
g_{\mu\nu}(x)=\delta_{ab}e^{a}{}_{\mu}(x)e^{b}{}_{\nu}(x)\;.\label{3.1}
\end{equation}
The inverse $N$-bein $e_{a}{}^{\mu}(x)$ satisfies
\begin{equation}
e_{a}{}^{\mu}(x)e^{b}{}_{\mu}(x)=\delta^{b}_{a}\;,\label{3.2}
\end{equation}
so that 
\begin{equation}
\delta_{ab}=g_{\mu\nu}(x)e_{a}{}^{\mu}(x)e_{b}{}^{\nu}(x)\;.\label{3.3}
\end{equation}
Spacetime indices are raised and lowered with the spacetime metric $g_{\mu\nu}$ and its inverse as usual; orthonormal frame indices, that we use the indices $a,b,c,\ldots$ for, are raised and lowered with the Kronecker delta. Any tensor can be referred back to the orthonormal frame by converting its spacetime indices to orthonormal frame indices with suitable $N$-beins. The advantage of using the $N$-bein formalism is that it avoids the use of the bivector of geodetic parallel displacement that plays a role in DeWitt's~\cite{DeWittdynamical} calculation of the heat kernel coefficients. We will comment more on this point later.

\subsection{Riemann normal coordinate expansions}\label{secRNC}

In order to obtain the expansion of the Green function in Riemann normal coordinates we will need the expansions for the metric tensor, Christoffel symbol, and $N$-bein. These are, using $x^\mu=x^{\prime\mu}+y^\mu$ with $x^{\prime\mu}$ viewed as the origin of the Riemann normal coordinate system,
\begin{eqnarray}
g_{\mu\nu}(y)&=&\delta_{\mu\nu}-\frac{1}{3}y^\alpha y^\beta\,R_{\mu\alpha\nu\beta}\big|_{x'} -\frac{1}{6} y^\alpha y^\beta y^\gamma\,R_{\mu\alpha\nu\beta;\gamma}\big|_{x'}\nonumber\\
&&+y^\alpha y^\beta y^\gamma y^\delta\Big(-\frac{1}{20}R_{\mu\alpha\nu\beta;\gamma\delta} +\frac{2}{45}R_{\mu\alpha\beta\lambda}R_{\nu\gamma\delta}{}^{\lambda}\Big)\Big|_{x'}+\cdots,\label{3.4}\\
g^{\mu\nu}(y)&=&\delta^{\mu\nu}+\frac{1}{3}\,y^\alpha y^\beta\,R^{\mu}{}_{\alpha}{}^{\nu}{}_{\beta}\big|_{x'} +\frac{1}{6}\, y^\alpha y^\beta y^\gamma\, R^{\mu}{}_{\alpha}{}^{\nu}{}_{\beta;\gamma}\big|_{x'}\nonumber\\
&&+y^\alpha y^\beta y^\gamma y^\delta\Big(\frac{1}{20}\,R^{\mu}{}_{\alpha}{}^{\nu}{}_{\beta;\gamma\delta} +\frac{1}{15}\,R^{\mu}{}_{\alpha\beta\lambda}R^{\nu}{}_{\gamma\delta}{}^{\lambda} \Big)\Big|_{x'}+\cdots.\label{3.5}\\
\Gamma^{\lambda}_{\mu\nu}(y)=&=&-\frac{1}{3}y^\alpha\big( R^{\lambda}{}_{\mu\nu\alpha} +R^{\lambda}{}_{\nu\mu\alpha}\big)\big|_{x'}\nonumber\\
&&-\frac{1}{12} y^\alpha y^\beta \big( 2R^{\lambda}{}_{\mu\nu\alpha;\beta} +2R^{\lambda}{}_{\nu\mu\alpha;\beta} +R^{\lambda}{}_{\alpha\nu\beta;\mu} \nonumber\\
&&\qquad +R^{\lambda}{}_{\alpha\mu\beta;\nu} - R_{\mu\alpha\nu\beta}{}^{;\lambda}  \big)\big|_{x'}\nonumber\\
&&+ y^\alpha y^\beta y^\gamma\Big( -\frac{1}{20} R^{\lambda}{}_{\mu\nu\gamma;\alpha\beta} -\frac{1}{20} R^{\lambda}{}_{\nu\mu\gamma;\alpha\beta} -\frac{1}{40} R^{\lambda}{}_{\beta\nu\gamma;\mu\alpha}\nonumber\\
&&-\frac{1}{40} R^{\lambda}{}_{\beta\mu\gamma;\nu\alpha}- \frac{1}{40} R^{\lambda}{}_{\beta\nu\gamma;\alpha\mu} -\frac{1}{40} R^{\lambda}{}_{\beta\mu\gamma;\alpha\nu} +\frac{1}{40} R_{\mu\beta\nu\gamma}{}^{;\lambda}{}_{\alpha}\nonumber\\
&& +\frac{1}{40} R_{\mu\beta\nu\gamma}{}^{;}{}_{\alpha}{}^{\lambda} +\frac{1}{45}R^{\lambda}{}_{\alpha\mu\sigma}R_{\nu\beta\gamma}{}^{\sigma} +\frac{1}{45}R^{\lambda}{}_{\alpha\nu\sigma}R_{\mu\beta\gamma}{}^{\sigma}\nonumber\\
&&  -\frac{4}{45}R^{\lambda}{}_{\alpha\beta\sigma}R_{\nu\gamma\mu}{}^{\sigma} -\frac{4}{45}R^{\lambda}{}_{\alpha\beta\sigma}R_{\mu\gamma\nu}{}^{\sigma} -\frac{1}{45}R^{\lambda}{}_{\sigma\mu\beta}R_{\nu\alpha\gamma}{}^{\sigma}\nonumber\\
&& -\frac{1}{45}R^{\lambda}{}_{\sigma\nu\beta}R_{\mu\alpha\gamma}{}^{\sigma} +\frac{2}{45}R^{\lambda}{}_{\mu\beta\sigma}R_{\nu\alpha\gamma}{}^{\sigma} +\frac{2}{45}R^{\lambda}{}_{\nu\beta\sigma}R_{\mu\alpha\gamma}{}^{\sigma}\Big)\Big|_{x'}+\cdots,\label{3.6}\\
e^{a}{}_{\mu}(y)&=&e^{a}{}_{\nu}(x')\Big\lbrace \delta^{\nu}_{\mu} -\frac{1}{6}\,R^{\nu}{}_{\alpha\mu\beta}\,y^\alpha y^\beta -\frac{1}{12}\,R^{\nu}{}_{\alpha\mu\beta;\gamma}\,y^\alpha y^\beta y^\gamma\nonumber\\
&+&\Big(-\frac{1}{40}\,R^{\nu}{}_{\alpha\mu\beta;\gamma\delta} + \frac{1}{120}\,R^{\nu}{}_{\alpha\beta\sigma}R_{\mu\gamma\delta}{}^{\sigma}\Big)y^\alpha y^\beta y^\gamma y^\delta + \cdots\Big\rbrace,\label{3.7}\\
e_{a}{}^{\mu}(y)&=&e_{a}{}^{\nu}(x')\Big\lbrace \delta^{\mu}_{\nu} +\frac{1}{6}\,R_{\nu\alpha}{}^{\mu}{}_{\beta}\,y^\alpha y^\beta +\frac{1}{12}\,R_{\nu\alpha}{}^{\mu}{}_{\beta;\gamma}\,y^\alpha y^\beta y^\gamma\nonumber\\
&+&\Big(\frac{1}{40}\,R_{\nu\alpha}{}^{\mu}{}_{\beta;\gamma\delta} + \frac{7}{360}\,R_{\nu\alpha\beta\sigma}R^{\mu}{}_{\gamma\delta}{}^{\sigma}\Big)y^\alpha y^\beta y^\gamma y^\delta + \cdots\Big\rbrace.\label{3.8}
\end{eqnarray}
The $N$-bein $e_{a}{}^{\mu}(x)$ is defined in terms of $e_{a}{}^{\mu}(x')$ by parallel transport along the geodesic that connects the origin $x^{\prime\mu}$ to $x^\mu$. This is how the expansions in \eqref{3.7} and \eqref{3.8} were calculated (using \eqref{3.6}). It also explains the relationship between the $N$-bein formalism and the bivector of geodesic parallel displacement. The orthonormal frame components of a vector field $A^\mu(x)$ are given by 
\begin{equation}
A^a=e^{a}{}_{\mu}(x)A^\mu(x)\;.\label{3.9a}
\end{equation}
This must be the same as we find at the point $x'$ if the $N$-bein, or equivalently the tangent space basis vectors, are defined through parallel transport. This gives the relation
\begin{equation}
A^a=e^{a}{}_{\mu}(x')A^\mu(x')\;.\label{3.9b}
\end{equation}
Combining \eqref{3.9a} and \eqref{3.9b} shows that
\begin{equation}
A^\mu(x)=e_{a}{}^{\mu}(x)A^a=e_{a}{}^{\mu}(x)e^{a}{}_{\nu}(x')A^\nu(x')\;.\label{3.9c}
\end{equation}
The expression $e_{a}{}^{\mu}(x)e^{a}{}_{\nu}(x')$ can be interpreted as the bivector of geodesic parallel displacement. There is an obvious inverse relation to \eqref{3.9c}. There is no need to use the bivector of geodesic parallel displacement explicitly if the $N$-bein formalism is adopted. We will return to this in Sec.~\ref{sec4}.

We can write the Green function as
\begin{equation}
G_{\mu\nu}(x,x')=e^{a}{}_{\mu}(x)e^{b}{}_{\nu}(x')G_{ab}(x,x')\;.\label{3.10}
\end{equation}
The indices $i$ and $j$ in the general expression for the operator in Sec.~\ref{sec2} can now be interpreted as orthonormal frame indices $a$ and $b$. Because the orthonormal frame indices are raised and lowered with the Kronecker delta there is no real need to distinguish between upper and lower indices. By transforming the operator in \eqref{2.1.1} or the defining equation for the Green function \eqref{djt1.5} the expressions for the coefficients $A,B,C$ in the general expression \eqref{djt1.9} become (with all expressions evaluated at the general point $x$)
\begin{eqnarray}
(A^{\mu\nu})_{ab}&=&-g^{\mu\nu}\delta_{ab}+\frac{q}{2}\left(e_{a}{}^{\mu}e_{b}{}^{\nu} + e_{a}{}^{\nu}e_{b}{}^{\mu}\right),\label{3.11}\\
(B^{\mu})_{ab}&=&g^{\alpha\beta}\Gamma^{\mu}_{\alpha\beta}\delta_{ab}-2\,e_{a}{}^{\lambda} g^{\alpha\mu}e_{b\lambda\,;\alpha} -q\,e_{a}{}^{\sigma}e_{b}{}^{\lambda}\Gamma^{\mu}_{\lambda\sigma}\\
&&\quad+q\,e_{a}{}^{\mu}e_{b}{}^{\lambda}{}_{;\lambda}+q\,e_{a}{}^{\lambda}e_{b}{}^{\mu}{}_{;\lambda}\;,\label{3.12}\\
(C)_{ab}&=&Q_{ab}-e_{a}{}^{\mu}g^{\alpha\beta}e_{b\mu\,;\alpha\beta} +q\,e_{a}{}^{\mu}e_{b}{}^{\lambda}{}_{;\lambda\mu}.\label{3.13}
\end{eqnarray}
Here we have used
\begin{equation}
G_{ab\,;\alpha\beta}=G_{ab\,,\alpha\beta}-\Gamma^{\sigma}_{\alpha\beta}G_{ab\,,\sigma},\label{3.14}
\end{equation}
and defined
\begin{equation}
Q_{ab}=e_{a}{}^{\mu}e_{b}{}^{\nu}Q_{\mu\nu}.\label{3.15}
\end{equation}
Note that we have been careful to symmetrize the second term of \eqref{3.11} in $\mu$ and $\nu$. Also, if we assume that $Q_{\nu\mu}=Q_{\mu\nu}$ is symmetric, then $Q_{ab}=Q_{ba}$ will also be symmetric. The semicolon on the $N$-bein denotes the spacetime covariant derivative rather than the full covariant derivative which must vanish. So for example,
\begin{equation}
e^{a}{}_{\mu\,;\nu}=e^{a}{}_{\mu\,,\nu}-\Gamma^{\lambda}_{\mu\nu}\,e^{b}{}_{\lambda}\;.\label{3.15b}
\end{equation}
The full covariant derivative of $e^{a}{}_{\mu}$ involves the spin connection $(\omega_\mu)_{ab}$. The full covariant derivative of $e^{a}{}_{\mu}$ is given by
\begin{equation}
\nabla_\nu\, e^{a}{}_{\mu}=0=e^{a}{}_{\mu\,,\nu}-\Gamma^{\lambda}_{\mu\nu}\,e^{a}{}_{\lambda} + (\omega_{\nu})^{a}{}_{b}e^{b}{}_{\mu}. \label{3.16}
\end{equation}
Making use of \eqref{3.15} gives
\begin{equation}
e^{a}{}_{\mu\,;\nu}=-(\omega_\nu)^{a}{}_{b}\,e^{b}{}_{\mu}\;.\label{3.17}
\end{equation}
This last result can be used to eliminate the spacetime covariant derivatives in \eqref{3.12} and \eqref{3.13} in terms of the spin connection and its derivatives which allows some simplifications in obtaining the Riemann normal coordinate expansions of $B$ and $C$, although the results in \eqref{3.12} and \eqref{3.13} can be used equally well. The expansion of $(\omega_\mu)_{ab}$ in Riemann normal coordinates can be shown to be
\begin{eqnarray}
(\omega_{\mu})_{ab}(y)&=& e_{a}{}^{\lambda}(x')e_{b}{}^{\sigma}(x')\Big\lbrace
-\frac{1}{2}\,R_{\mu\alpha\lambda\sigma}y^\alpha -\frac{1}{3}R_{\mu\alpha\lambda\sigma\,;\beta}\,y^{\alpha}y^{\beta}\nonumber\\
&&-\frac{1}{8}\big\lbrack R_{\mu\alpha\lambda\sigma\,;\beta\gamma}-\frac{1}{3}R^{\tau}{}_{\alpha\mu\beta}R_{\tau\gamma\lambda\sigma}\big\rbrack\,y^{\alpha}y^{\beta}\,y^{\gamma}+\cdots\Big\rbrace.\label{3.19}
\end{eqnarray}
All curvature terms appearing on the right hand side of \eqref{3.19} are evaluated at the origin $x'$ of the Riemann normal coordinate system.

This gives sufficient results to evaluate the expansion of $A,B,C$ in Riemann normal coordinates and to read off the coefficients needed in the expansion of the Green function as described in Sec.~\ref{sec2}. The necessary results are summarized in Appendix~\ref{ABC}.

\subsection{Auxiliary Green function expansions}\label{secauxGexp}

From \eqref{djt1.17} if we use \eqref{4.1.21a} it can be seen that (with $e_{a}{}^{\mu}$ evaluated at $x'$)
\begin{equation}
(G_0)_{ab}=\delta_{a b}S+q e_{a}{}^{ \mu} e_{b}{}^{ \nu} p_{\mu} p_{\nu}\, S T\;,\label{4.1.1.1}
\end{equation}
where we have defined
\begin{eqnarray}
S&=&(p^2-s)^{-1}\;,\label{4.1.1.2}\\
T&=&\lbrack(1-q)p^2-s\rbrack^{-1}.\label{4.1.1.3}
\end{eqnarray}
We now use this result in \eqref{djt1.20}--\eqref{djt1.23} along with the relevant expressions for $A,B,C$ taken from \eqref{4.1.21a},\eqref{4.1.24} and \eqref{4.1.35a}. The result turns out to be given by
\begin{eqnarray}
e^{a}{}_{\alpha}(x')e^{b}{}_{\beta}(x'){(G_{2})}_{a b} &=&- {Q}_{\alpha \beta} S{}^{2} - q\,{Q}_{\alpha}{}^{ \mu}\, S{}^{2} T\,{p}_{\beta} {p}_{\mu}  - q\,{Q}^{\mu}{}_{ \beta}\, S{}^{2} T\, {p}_{\alpha} {p}_{\mu} \nonumber\\
&&-q{}^{2}\, {Q}^{\mu \nu}\, S{}^{2} T{}^{2}\, {p}_{\alpha} {p}_{\beta} {p}_{\mu} {p}_{\nu} \nonumber\\
&& +\frac{1}{3} R \Big\lbrace S{}^{2} {\delta}_{\alpha \beta} + q\,ST\lbrack S +( 1 -  q)  T \rbrack\,{p}_{\alpha} {p}_{\beta} \Big\rbrace\nonumber\\
&& + \frac{q}{6}\, {R}_{\alpha \beta}\, S T 
  + q\,{R}_{\alpha}{}^{ \mu}\, S{}^{2} T\, {p}_{\beta} {p}_{\mu} + q\,{R}_{\beta}{}^{ \mu}\, S{}^{2} T\, {p}_{\alpha} {p}_{\mu} \nonumber\\
&&+\frac{1}{3}\, {R}^{\mu \nu} \Big\lbrace - 2\,  S{}^{2} {\delta}_{\alpha \beta}\nonumber\\
 &&\quad-  q\, \lbrack  2\, S{}^{2}  +(  2 - 5q) S T 
  + 2(1-q)^2 T{}^{2}\rbrack\,T\,{p}_{\alpha} {p}_{\beta}\Big\rbrace S\, {p}_{\mu} {p}_{\nu}\nonumber\\
&&-\frac{q}{3}\, {R}_{\alpha}{}^{ \mu}{}_{ \beta}{}^{ \nu}  \lbrack S+(1-q)T\rbrack  S T\,{p}_{\mu} {p}_{\nu}\;.\label{4.1.1.4}
\end{eqnarray}
As usual, all terms on the right hand side are evaluated at the origin of RNC, $x'$. It is advantageous to do the tedious calculations here and in what follows with Cadabra~\cite{Peeters1,Peeters2}. The much lengthier expressions for $G_3$ and $G_4$ are given in Appendix~\ref{auxG4}.

\subsection{Laplace transforms and momentum space results}

The aim now is to use the results for the terms in the expansion of the Green function found in \eqref{4.1.1.1}, \eqref{4.1.1.4} and \eqref{G4ab} to find the first three heat kernel coefficients. The result for $G_3$ is not needed here, although it will be needed in Secs.~\ref{sec4} and \ref{sec5}. From our results it is clear that we first need to evaluate the inverse Laplace transform of powers of $S$ in \eqref{4.1.1.2} and $T$ in \eqref{4.1.1.3}. We will define
\begin{eqnarray}
L_{nm}&=&\int\limits_{c-i\infty}^{c+i\infty}\frac{ds}{2\pi i}\,e^{-\tau s}\,S^nT^m\;,\label{4.1.1.5}\\
&=&\int\limits_{c-i\infty}^{c+i\infty}\frac{ds}{2\pi i}\,e^{-\tau s}\,{\mathcal S}^n(p^2){\mathcal S}^m(k^2)\;,\label{4.1.1.6}
\end{eqnarray}
where we will take 
\begin{equation}
{\mathcal S}(p^2)=\lbrack p^2-s\rbrack^{-1}\;.\label{4.1.1.7a}
\end{equation}
and define
\begin{equation}
k^2=(1-q)p^2\;.\label{4.1.1.7}
\end{equation}
It is now possible to set up a recursion relation that allows us to calculate $L_{nm}$ in terms of $L_{11}$:
\begin{equation}
L_{nm}=\frac{(-1)^{n+m}}{(n-1)!(m-1)!}\left(\frac{\partial}{\partial\,p^2}\right)^{n-1} \left(\frac{\partial}{\partial\,k^2}\right)^{m-1}\,L_{11}.\label{4.1.1.8}
\end{equation}
For $L_{11}$ we find
\begin{equation}
L_{11}=\frac{e^{-\tau\,p^2}-e^{-\tau\,k^2}}{k^2-p^2}\,.\label{4.1.1.9}
\end{equation}
This is sufficient for determining the necessary inverse Laplace transforms when $n,m\ne0$. If $m=0$ we have
\begin{equation}
L_{n0}=I_n=\frac{\tau^n}{(n-1)!}\,e^{-\tau\,p^2}.\label{4.1.1.10}
\end{equation}
In a similar way, when $n=0$ we have
\begin{equation}
L_{0m}=\frac{\tau^m}{(m-1)!}\,e^{-\tau\,k^2}=\frac{\tau^m}{(m-1)!}\,e^{-(1-q)\tau\,p^2}.\label{4.1.1.11a}
\end{equation}
We have therefore reduced the evaluation of $L_{nm}$ into a simple recursive procedure. 

From the results found for $L_{nm}$ it should be clear that when the momentum integration is considered we encounter more complicated expressions than is found in the non-minimal case when $q=0$ and a single exponential factor is obtained. We now encounter expressions like
\begin{equation}
J_{\mu_1\cdots\mu_{2n}}(k,n;\tau)=\int\frac{d^Np}{(2\pi)^N}\,\frac{p_{\mu_1}\cdots p_{\mu_{2n}}}{(p^2)^k}\,e^{-\tau p^2}\;.\label{4.1.1.11}
\end{equation}
(terms with an odd number of momenta will integrate to zero.) It is now necessary to consider convergence of this expression. The exponential factor ensures convergence as $p\rightarrow\infty$. As $p\rightarrow0$ the integrand behaves like $p^{2n-2k}$ and there is a factor of $p^{N-1}$ from the volume element. The integral in \eqref{4.1.1.11} is only convergent at the lower limit if $2k<2n+N$. In cases where $2k\ge 2n+N$ the integral in \eqref{4.1.1.11} will diverge. However the overall expression for the heat kernel coefficient cannot diverge. This means that there will be integrals that separately diverge but that when combined are convergent. The situation is completely analogous to
\begin{equation}
F(a,b)=\int\limits_{0}^{\infty}\frac{d\tau}{\tau}\left( e^{-\tau a}-e^{-\tau b}\right),\label{4.1.1.12}
\end{equation}
where $a$ and $b$ are both assumed to be positive (or to have positive real parts if they are complex). The integral on the right hand side exists since the integrand vanishes exponentially fast as $\tau\rightarrow\infty$ and is integrable around $\tau=0$ because of the difference of the two exponentials.
If we split up the integral in \eqref{4.1.1.12} into two separate integrals then each integral will diverge at the lower limit; yet the combined expression in \eqref{4.1.1.12} is convergent and is easily seen to be given by
\begin{equation*}
F(a,b)=\ln b-\ln a\;.
\end{equation*} 
We will discuss this in more detail below.

By symmetry we must have
\begin{equation}
J_{\mu_1\cdots\mu_{2n}}=J(k,n;\tau)\delta_{\mu_1\cdots\mu_{2n}}\;,\label{4.1.1.13}
\end{equation}
for some coefficients $J(k,n;\tau)$. Here $\delta_{\mu_1\cdots\mu_{2n}}$ is expressed as the sum of products of $n$ Kronecker deltas with all possible pairings of indices. For example,
\begin{equation*}
\delta_{\mu_1\mu_2\mu_3\mu_{4}}=\delta_{\mu_1\mu_{2}}\delta_{\mu_3\mu_{4}}+ \delta_{\mu_1\mu_{3}}\delta_{\mu_2\mu_{4}} +\delta_{\mu_1\mu_{4}}\delta_{\mu_2\mu_{3}}.
\end{equation*}
We also have the recursive result that
\begin{eqnarray*}
\delta_{\mu_1\cdots\mu_{2n}}&=&\delta_{\mu_1\mu_{2}}\delta_{\mu_3\cdots\mu_{2n}}+ \delta_{\mu_1\mu_{3}}\delta_{\mu_2\mu_4\cdots\mu_{2n}}+\cdots+\delta_{\mu_1\mu_{2n}}\delta_{\mu_2\cdots\mu_{2n-1}},\\
&=&\sum_{k=2}^{2n}\delta_{\mu_1\mu_{k}}\delta_{\mu_2\cdots\hat{\mu}_{k}\cdots\mu_{2n}},
\end{eqnarray*}
where the $\hat{\mu}_k$ means that the index corresponding to $k$ is omitted from the sum. There are $(2n-1)$ terms in the sum. For later use we record that
\begin{equation}
\delta^{\mu_1\mu_2}\delta_{\mu_1\cdots\mu_{2n}}=(N+2n-2)\delta_{\mu_3\cdots\mu_{2n}}.\label{condelta}
\end{equation}
By contracting both sides of \eqref{4.1.1.13} with $\delta^{\mu_1\mu_2}$, and making use of \eqref{condelta} it can be seen that
\begin{equation}
J(k-1,n-1;\tau)=(N+2n-2)J(k,n;\tau)\;.\label{4.1.1.14}
\end{equation}
This gives us a reduction formula that we can use to reduce $J(k,n;\tau)$ to $J(0,n';\tau)$ for some $n'<n$ provided that the original expression is convergent. Noting that 
\begin{eqnarray}
J(0,n';\tau)=\frac{1}{(4\pi\tau)^{N/2}(2\tau)^{n'}}\;,\label{4.1.1.15}
\end{eqnarray}
completes the recursive definition for $J(k,n;\tau)$. 

If the original expression \eqref{4.1.1.13} is not convergent then we can only reduce $J(k,n;\tau)$ to $J(k',0;\tau)$ for some $1\le k'<k$ which is a divergent integral. However as already mentioned there must be another expression of a similar structure such that the combined expression is convergent, analogous to the simple result in \eqref{4.1.1.12}. The calculation of $E_2$ involves $J(1,0;\tau)-J(1,0;(1-q)\tau)$. It is possible to show that, assuming $N\ne2$,
\begin{eqnarray}
J(1,0;\tau)-J(1,0;(1-q)\tau)&=&\int\frac{d^Np}{(2\pi)^N}\,\frac{1}{(p^2)}\big\lbrack e^{-\tau p^2} -e^{-(1-q)\tau p^2}\big\rbrack \;,\nonumber\\
&=&\frac{2\tau}{(N-2)}(4\pi\tau)^{-N/2}\big\lbrack1-(1-q)^{1-N/2}\big\rbrack\;.\label{4.1.1.16}
\end{eqnarray}
In the special case of $N=2$ it is possible to evaluate the integral in the first line directly and to show that the result agrees with what is obtained by taking the limit of \eqref{4.1.1.16} as $N\rightarrow2$. The fact that $J(1,0;\tau)$ and $J(1,0;(1-q)\tau)$ only occur in the combination $J(1,0;\tau)-J(1,0;(1-q)\tau)$ which is given by the finite expression \eqref{4.1.1.16} is a useful check on any calculation.

The calculation of $E_2$ also involves $J(2,0;\tau)$ and $J(2,0;(1-q)\tau)$. This time the finite result does not simply involve the difference. Instead we find that 
\begin{equation}
J(2,0;\tau)-J(2,0;(1-q)\tau)+\tau q\,J(1,0;\tau)=F_2(\tau,q)\;,\label{4.1.1.17}
\end{equation}
where
\begin{equation}
F_2(\tau,q)=\frac{2\tau^2}{(N-2)}\,(4\pi\tau)^{-N/2}\Big\lbrace q+\frac{2}{(N-4)}\lbrack 1-(1-q)^{2-N/2}\big\rbrack\Big\rbrace\;,\label{4.1.1.18}
\end{equation}
assuming $N\ne2,4$. In the special cases of $N=2,4$ a direct evaluation of the integrals occurring on the left hand side of \eqref{4.1.1.17} agrees with taking the limit of \eqref{4.1.1.18}. (Both limits are of course finite as expected.)

\subsection{$E_{0\,\mu\nu}$ expression}

Because the expression for $G_0$ in \eqref{4.1.1.1} does not involve the curvature explicitly, we would expect that the result for the first heat kernel coefficient $E_{0\,\mu\nu}$ should agree with that found in flat spacetime. The relevant expression is the $k=0$ case of \eqref{djt1.32}.

From \eqref{djt1.32} with $k=0$, and using the expression \eqref{4.1.1.1} for $G_{0\,ab}$ we have
\begin{eqnarray}
E_{0\,ab}(x')&=&(4\pi\tau)^{N/2}\int\frac{d^Np}{(2\pi)^N}\int\limits_{c-i\infty}^{c+i\infty}\frac{ds}{2\pi i}\,e^{-s\tau}\big\lbrack \delta_{a b}S+q e_{a}{}^{ \mu}(x') e_{b}{}^{ \nu}(x') p_{\mu} p_{\nu}\, S T\big\rbrack\nonumber\\
&=&(4\pi\tau)^{N/2}\int\frac{d^Np}{(2\pi)^N}\big\lbrack \delta_{a b}L_{10}+q e_{a}{}^{ \mu}(x') e_{b}{}^{ \nu}(x') p_{\mu} p_{\nu}\, L_{11}\big\rbrack\;.\label{4.1.1.19}
\end{eqnarray}
The definition of $L_{nm}$ in \eqref{4.1.1.5} has been used here. Next we use \eqref{4.1.1.8} and \eqref{4.1.1.9}  to find
\begin{eqnarray}
E_{0\,ab}(x')&=&(4\pi\tau)^{N/2}\int\frac{d^Np}{(2\pi)^N}\Big\lbrace \delta_{a b}\,e^{-p^2 \tau }\nonumber\\
&&+ e_{a}{}^{ \mu}(x') e_{b}{}^{ \nu}(x') \frac{p_{\mu} p_{\nu}}{p^2 }\, \big\lbrack  e^{-(1-q)\,\tau\,p^2}-e^{-p^2 \tau} \big\rbrack\Big\rbrace\;.\label{4.1.1.20}
\end{eqnarray}
The momentum integration is performed using \eqref{4.1.1.11} and \eqref{4.1.1.13} to find
\begin{equation}
E_{0\,ab}(x')=(4\pi\tau)^{N/2}\,\delta_{a b}\,\big\lbrack J(0,0;\tau)+  J(1,1;(1-q)\tau)-J(1,1;\tau)\big\rbrack\;.\label{4.1.1.21}
\end{equation}
The Kronecker delta that arises in the integration has been used to reduce the $N$-bein terms to $\delta_{ab}$. Now use  the recursion relation in \eqref{4.1.1.14} with \eqref{4.1.1.15} to find
\begin{equation}
E_{0\,ab}(x')=\delta_{a b}\,\Big\lbrace 1+  \frac{1}{N}\lbrack(1-q)^{-N/2}-1\rbrack\Big\rbrace\;.\label{4.1.1.22}
\end{equation}
This is in complete agreement with a simple flat spacetime calculation as expected.

\subsection{$E_{1\,\mu\nu}$ expression}

With $k=1$ in \eqref{djt1.32} we have
\begin{equation}
E_{1\,ab}(x')=\tau^{-1}\,(4\pi\tau)^{N/2}\int\frac{d^Np}{(2\pi)^N}\int\limits_{c-i\infty}^{c+i\infty}\frac{ds}{2\pi i}\,e^{-s\tau}\,G_{2\,ab}\;,\label{4.1.1.29}
\end{equation}
where $G_{2\,ab}$ is given by \eqref{4.1.1.4}. This is still sufficiently simple to do by hand. The steps are identical to those that we have just described for $E_0$ above. We will leave out the intermediate steps and simply give the end result after changing from the local orthonormal basis back to the coordinate basis, which is
\begin{equation}
E_{1}{}^{\mu}{}_{\nu}=T_{11}\,Q^{\mu}{}_{\nu}+T_{12}\,R^{\mu}{}_{\nu}+T_{13}\,Q\,\delta^{\mu}{}_{\nu}+T_{14}\,R\,\delta^{\mu}{}_{\nu}\;,\label{4.2.30}
\end{equation}
where we have defined
\begin{equation}
Q=Q^{\lambda}{}_{\lambda}\;,\label{4.2.30a}
\end{equation}
analogously to the definition of $R=R^{\lambda}{}_{\lambda}$. The coefficients in this expression are given for a general spacetime of dimension $N$ by
\begin{eqnarray}
T_{11}&=&\frac{\lbrack 4N-q\,(N-2 )(N^2-2)\rbrack}{N \left(N^2-4\right) q}
\nonumber\\
&&+(1-q)^{-{N}/{2}}\,\frac{(2 N q-4 N+4 q) }{N \left(N^2-4\right) q}\;,\label{4.2.31a}\\
T_{12}&=&\frac{\lbrack -12 N-q(5 N^2-8N-12)\rbrack}{3N (N^2-4)  q}\nonumber\\
&&+(1-q)^{-{N}/{2}}\,\frac{ \lbrack N (N+2) q^2-(N+2) (N+6) q+12 N\rbrack}{3 N \left(N^2-4\right) q},\label{4.2.32b}\\
T_{13}&=&-\,\frac{\lbrack4+q(N-2)\rbrack}{N(N^2-4)q}+(1-q)^{-{N}/{2}}\,\frac{\lbrack4-q(N+2)\rbrack }{N(N^2-4)q},\label{4.2.32c}\\
T_{14}&=&\frac{\lbrack24+q(N^3-N^2-12)\rbrack}{6 N \left(N^2-4\right) q}\nonumber\\
&&-(1-q)^{-{N}/{2}}\,\frac{ \lbrack 24+(N+2)(Nq-N-6)q\rbrack}{6 N \left(N^2-4\right) q}\;.\label{4.2.32d}
\end{eqnarray}

The results for $N=2$ must be found by taking the limit of these expressions as $N\rightarrow2$. The results in this special case are
\begin{eqnarray}
T_{11}&=&-\frac{(2-q)}{4 (1-q)}+\frac{\log (1-q)}{2 q}\;,\label{4.2.33a}\\
T_{12}&=&\frac{(3 q-2)}{4 (1-q)}+\frac{(q-3) \log (1-q)}{6 q}\;,\label{4.2.33b}\\
T_{13}&=&\frac{(q-2)}{8 (1-q)}-\frac{\log (1-q)}{4 q}\;,\label{4.2.33c} \\
T_{14}&=&\frac{(10-7q)}{24(1-q)}+\frac{(3-q) \log (1-q)}{12 q}\;.\label{4.2.33d}
\end{eqnarray}

For the physically interesting case of $N=4$ we find
\begin{eqnarray}
T_{11}&=&\frac{\lbrack-7 q^2+18 q-12\rbrack}{12 (1-q)^2}\;,\label{4.2.34a}\\
&=&-\frac{\gamma ^2}{12}-\frac{\gamma }{2}-1\;,\label{4.2.34aa}\\
T_{12}&=&\frac{(4-3 q) q}{12 (1-q)^2}\;,\label{4.2.34b}\\
&=&\frac{1}{12} \gamma  (\gamma +4)\;,\label{4.2.34bb}\\
T_{13}&=&-\,\frac{q^2}{24 (1-q)^2}\;,\label{4.2.34c}\\
&=&-\frac{\gamma ^2}{24}\;,\label{4.2.34cc}\\
T_{14}&=&\frac{(3q^2-6q+4)}{24(1-q)^2}\;,\label{4.2.34d}\\
&=&\frac{\gamma ^2}{24}+\frac{\gamma }{12}+\frac{1}{6}\;.\label{4.2.34dd}
\end{eqnarray}
In the second equality for each term we have written the result in terms of the variable
\begin{equation}
\gamma=q/(1-q)\;,\label{4.2.34e}
\end{equation}
that was used in \cite{BarvinskyVilkovisky}.

Finally we give the first terms in the expansion about $q=0$:
\begin{eqnarray}
T_{11}&=&-1-\frac{q}{2}+\cdots\;,\label{4.2.35a}\\
T_{12}&=&\frac{q}{3}+\cdots\;,\label{4.2.35b}\\
T_{13}&=&-\frac{q^2}{24}+\cdots\;,\label{4.2.35c}\\
T_{14}&=&\frac{1}{6}+\frac{q}{12}+\cdots\;.\label{4.2.35d}
\end{eqnarray}
In the $q=0$ limit we recover the result for the non-minimal vector field.

\subsection{$E_{2\,\mu\nu}$ expression}\label{E2exp}

With $k=2$ in \eqref{djt1.32} we have
\begin{equation}
E_{2\,ab}(x')=\tau^{-2}\,(4\pi\tau)^{N/2}\int\frac{d^Np}{(2\pi)^N}\int\limits_{c-i\infty}^{c+i\infty}\frac{ds}{2\pi i}\,e^{-s\tau}\,G_{4\,ab}\;,\label{4.2.36}
\end{equation}
where $G_{4\,ab}$ is given by \eqref{G4ab}. This is clearly rather a lengthy expression to deal with by hand. However the basic method is exactly the same as that which we have described for the lower heat kernel coefficients. Cadabra~\citep{Peeters1,Peeters2} can be used efficiently to deal with the necessary algebraic complexity. The result turns out to be
\begin{eqnarray}
E_{2\,\mu\nu}&=&T_{21}\,RR_{\mu\nu} + T_{22}\,R_{\mu\lambda}R^{\lambda}{}_{\nu} + T_{23}\, R_{\mu\alpha\nu\beta}R^{\alpha\beta}\nonumber\\
&&+T_{24}\,R_{\mu\alpha\beta\gamma} R_{\nu}{}^{\alpha\beta\gamma}+T_{25}\,R^2\,g_{\mu\nu}+T_{26}\,R_{\alpha\beta}R^{\alpha\beta}\,g_{\mu\nu}\nonumber\\
&&+T_{27}\,R_{\alpha\beta\gamma\delta}R^{\alpha\beta\gamma\delta}\,g_{\mu\nu}+ T_{28}\,\nabla_{\mu}\nabla_{\nu}R+T_{29}\,\Box R_{\mu\nu}\nonumber\\
&&+T_{210}\,\Box R\,g_{\mu\nu}+T_{211}\,Q_{\alpha\beta}Q^{\alpha\beta}\,g_{\mu\nu}+ T_{212}\,Q_{\mu\lambda}Q^{\lambda}{}_{\nu}\nonumber\\
&&+T_{213}\,Q^2\,g_{\mu\nu}+T_{214}\,Q\,R_{\mu\nu} +T_{215}\,Q\,R\,g_{\mu\nu}+T_{216}\,R_{\mu\alpha\nu\beta}Q^{\alpha\beta}\nonumber\\
&&+T_{217}\,Q_{\alpha\beta}R^{\alpha\beta}\,g_{\mu\nu}+T_{218}\,(R_{\mu\lambda}Q^{\lambda}{}_{\nu} + R_{\nu\lambda}Q^{\lambda}{}_{\mu})+T_{219}\,R\,Q_{\mu\nu}\nonumber\\
&&+T_{220}\,Q\,Q_{\mu\nu}+T_{221}\,(\nabla_{\mu}\nabla_{\lambda}Q^{\lambda}{}_{\nu} + \nabla_{\nu}\nabla_{\lambda}Q^{\lambda}{}_{\mu})\nonumber\\
&&+T_{222}\,(\Box Q\,g_{\mu\nu}+2\,\nabla_{\mu}\nabla_{\nu} Q+2\,g_{\mu\nu}\,\nabla_{\alpha}\nabla_\beta Q^{\alpha\beta})\nonumber\\
&&+T_{223}\,\Box Q_{\mu\nu}\;.\label{4.2.37}
\end{eqnarray}
$Q$ is as defined in \eqref{4.2.30a}. The coefficients appearing in \eqref{4.2.37} are somewhat lengthy and are given for general spacetime dimensions in Appendix~\ref{Tresults}.

The $N=4$ results are also a special case, and coincide with taking the $N\rightarrow4$ limit of \eqref{4.2.38a}--\eqref{4.2.38w}.
\begin{eqnarray}
T_{21}&=&\frac{(12-34 q+36 q^2-13 q^3)}{96 (1-q)^2 q} +\frac{\left(2 q^2-12 q+9\right) \log (1-q)}{72 q^2}\;,\label{4.2.40a}\\
T_{22}&=&\frac{(15 q^4-38 q^3+50 q^2-48 q+24)}{144 (1-q)^2 q^2} -\frac{\left(4 q^3-30 q^2+15 q-30\right) \log (1-q)}{180 q^3}\;,\label{4.2.40b}\\
T_{23}&=&\frac{(3 q+2) \left(q^2+6 q-6\right)}{72 (1-q) q^2} +\frac{(q^3+15 q^2-15 q-15) \log (1-q)}{90 q^3}\;,\label{4.2.40c}\\
T_{24}&=&-\frac{1}{12}+\frac{1}{90} \log (1-q)\;,\label{4.2.40d}\\
T_{25}&=&\frac{(37 q^3-116 q^2+118 q-36)}{1152 (1-q)^2 q} -\frac{\left(2 q^2-12 q+9\right) \log (1-q)}{288 q^2}\;,\label{4.2.40e}\\
T_{26}&=&\frac{(180-526q+452q^2-91 q^3)}{2880 (1-q)^2 q} + \frac{\left(2 q^2-60 q+45\right) \log (1-q)}{720 q^2}\;,\label{4.2.40f}\\
T_{27}&=&\frac{1}{180}-\frac{1}{360} \log (1-q)\;,\label{4.2.40g}\\
T_{28}&=&\frac{(12+30q-20q^2-19 q^3)}{144 (1-q) q^2} +\frac{\left(5+15q-2 q^3\right) \log (1-q)}{60 q^3}\;,\label{4.2.40h}\\
T_{29}&=&\frac{(36-78q+36q^2+13 q^3)}{144 (1-q) q^2} +\frac{(q^3+5 q^2-25 q+15) \log (1-q)}{60 q^3}\;,\label{4.2.40i}\\
T_{210}&=&-\,\frac{(59 q^3-64 q^2-90 q+60)}{720 (1-q) q^2}-\,\frac{(q^3-5 q^2-5 q+5) \log (1-q)}{60 q^3}\;,\label{4.2.40j}\\
T_{211}&=&\frac{(3 q^3+4 q^2-18 q+12)}{192 (1-q)^2 q}+ \frac{\log (1-q)}{16 q^2}\;,\label{4.2.40k}\\
T_{212}&=&\frac{11 q^3-40 q^2+42 q-12}{48 (1-q)^2 q}-\frac{\log (1-q)}{4 q^2}\;,\label{4.2.40l}\\ 
T_{213}&=&-\frac{(q^3+4 q^2-18 q+12)}{384 (1-q)^2 q} -\frac{\log (1-q)}{32 q^2}\;,\label{4.2.40m}\\
T_{214}&=&\frac{(q^3-16 q^2+26 q-12)}{96 (1-q)^2 q} + \frac{(2 q-3) \log (1-q)}{24 q^2}\;,\label{4.2.40n}\\
T_{215}&=&\frac{(2-q) \left(3 q^2-10 q+6\right)}{192 (1-q)^2 q}+ \frac{(3-2 q) \log (1-q)}{48 q^2}\;,\label{4.2.40o}\\
T_{216}&=&\frac{(3 q+2) (6-6q-q^2)}{72 (1-q) q^2} + \frac{(1+q-q^2) \log (1-q)}{6 q^3}\;,\label{4.2.40p}\\
T_{217}&=&\frac{(6-13 q+6 q^2)}{24 q(1-q)} + \frac{(3-5 q+q^2) \log (1-q)}{12 q^2}\;,\label{4.2.40q}\\
T_{218}&=&\frac{(12 q^4+7 q^3-52 q^2+42 q-12)}{144 (1-q)^2 q^2} -\frac{(1-q)^2 \log (1-q)}{12 q^3}\;,\label{4.2.40r}\\
T_{219}&=&-\,\frac{(12-10q-8q^2+7 q^3)}{96 (1-q)^2 q} + \frac{(2 q-3) \log (1-q)}{24 q^2}\;,\label{4.2.40s}\\
T_{220}&=&\frac{(12-18q+4q^2+3 q^3) }{96 (1-q)^2 q} + \frac{\log (1-q)}{8 q^2}\;,\label{4.2.40t}\\
T_{221}&=&-\,\frac{(6+3q-7q^2-3 q^3)}{36 (1-q) q^2} - \frac{(2+2q-q^2) \log (1-q)}{12 q^3}\;,\label{4.2.40u}\\
T_{222}&=&\frac{(q-2) \left(7 q^2+6 q-6\right)}{288 (1-q) q^2}+\frac{\left(1-q-q^2\right) \log (1-q)}{24 q^3}\;,\label{4.2.40v}\\
T_{223}&=&-\,\frac{(36-78q+60q^2-11 q^3)}{144 (1-q) q^2}  -\frac{\left(3-5q+q^2\right) \log (1-q)}{12 q^3}\;.\label{4.2.40w}
\end{eqnarray}

For evaluating the divergent part of the one-loop effective action only the trace of the heat kernel coefficient is involved. It is straightforward to see from \eqref{4.2.37} that there are 10 independent terms that can be given as
\begin{eqnarray}
{\rm tr}(E_2)&=&t_{21}\,R^2+t_{22}\,R_{\mu\nu}R^{\mu\nu} + t_{23}\,R_{\mu\nu\lambda\sigma}R^{\mu\nu\lambda\sigma} + t_{24}\,\Box R + t_{25}\, RQ\nonumber\\
&&\hspace{-36pt}+ t_{26}\,R_{\mu\nu}Q^{\mu\nu} + t_{27}\, Q^2 + t_{28}\,Q_{\mu\nu}Q^{\mu\nu} + t_{29} \Box Q + t_{210}\,Q^{\mu\nu}{}_{;\mu\nu}\;.\label{4.2.41}
\end{eqnarray}
The coefficients $t_{21},\ldots,t_{210}$ are given in terms of the coefficients $T_{21},\ldots,T_{223}$ defined in \eqref{4.2.37} and given explicitly in \eqref{4.2.38a}--\eqref{4.2.38w}. They are
\begin{eqnarray}
t_{21}&=&T_{21}+N\,T_{25}\nonumber\\
&=&\frac{\left(N^4-N^3-16 N^2+16 N-72\right) q+144}{72 N \left(N^2-4\right) q}\nonumber\\
&&+\frac{(1-q)^{-{N}/{2}} }{72 N ( N^2-4) q}\Big\lbrack N(N-2)(N+2) q^3-2 N (N+2) (N+4) q^2\nonumber\\
&&\qquad+(N+2) \left(N^2+10 N+36\right) q-144\Big\rbrack \;,\label{4.2.42a}\\
t_{22}&=&T_{22}+T_{23}+N\,T_{26}\nonumber\\
&=&\frac{\left(-N^4+N^3-116 N^2+296 N+360\right) q-360 N}{180 N \left(N^2-4\right) q}\nonumber\\
&&+\frac{(1-q)^{-{N}/{2}}}{180 N \left(N^2-4\right) q} \Big\lbrack -(N-2) N (N+2) q^3+2 N (N+2) (N+28) q^2\nonumber\\
&&\qquad-(N+2) \left(N^2+58 N+180\right) q+360 N\Big\rbrack\;,\label{4.2.42b}\\
t_{23}&=&T_{24}+N\,T_{27}\nonumber\\
&=&\frac{1}{180} \Big\lbrack(1-q)^{2-{N}/{2}}+N-16\Big\rbrack\;,\label{4.2.42c}\\
t_{24}&=&T_{28}+T_{29}+N\,T_{210}\nonumber\\
&=&\frac{120 \left(N^2-3 N+6\right) q+\left(N^5-5 N^4+15 N^3-70 N^2+104 N-240\right) q^2+240 (N-2)}{30 (N-4) N \left(N^2-4\right) q^2}\nonumber\\
&&+\frac{(1-q)^{1-{N}/{2}}}{30 (N-4) N \left(N^2-4\right) q^2} \Big\lbrack -N (N+1) \left(N^2-4\right) q^3\nonumber\\
&&\qquad+N (N+1) (N+2) (N+8) q^2-120 (N+2) q-240 (N-2)\Big\rbrack\;,\label{4.2.42d}\\
t_{25}&=&T_{214}+T_{219}+N\,T_{215}\nonumber\\
&=&\frac{(-N^3 q+N^2 q+12 q-24)}{6 N \left(N^2-4\right) q}\nonumber\\
&&+\frac{(1-q)^{-{N}/{2}}}{6 N \left(N^2-4\right) q}\Big\lbrack N^2 q^2-N^2 q+2 N q^2-8 N q-12 q+24\Big\rbrack \;,\label{4.2.42e}\\
t_{26}&=&T_{216}+2\,T_{218}+N\,T_{217}\nonumber\\
&=&\frac{\lbrack \left(5 N^2-8 N-12\right) q+12 N\rbrack}{3 N (N^2-4)  q}\nonumber\\
&&\quad+\frac{(1-q)^{-{N}/{2}}}{3 N \left(N^2-4\right) q} \Big\lbrack -N(N+2) q^2+(N+2) (N+6) q-12 N\Big\rbrack\;,\label{4.2.42f}\\
t_{27}&=&T_{220}+N\,T_{213}\nonumber\\
&=&\frac{(5 N^2 q-8 N q+12 N-12 q)}{3 N \left(N^2-4\right) q}\nonumber\\
&&+\frac{(1-q)^{-{N}/{2}}}{3 N \left(N^2-4\right) q}\Big\lbrack -N^2 q^2+N^2 q-2 N q^2+8 N q-12 N+12 q\Big\rbrack \;,\label{4.2.42g}\\
t_{28}&=&T_{212}+N\,T_{211}\nonumber\\
&=&\frac{\lbrack-2 N q+4 N-4 q\rbrack (1-q)^{-\frac{N}{2}}}{2 N \left(N^2-4\right) q}+\frac{(N^3 q-2 N^2 q-2 N q-4 N+4 q)}{2 N \left(N^2-4\right) q}\;,\label{4.2.42h}\\
t_{29}&=&T_{223}+(N+2)\,T_{222}\nonumber\\
&=&-\frac{\lbrack\left(N^4-5 N^3+2 N^2+32 N-96\right) q^2+192 q-96\rbrack}{6 (N-4) N \left(N^2-4\right) q^2}\nonumber\\
&&-\frac{(1-q)^{1-{N}/{2}}}{6 (N-4) N \left(N^2-4\right) q^2} \Big\lbrack N \left(N^2+6 N+8\right) q^2-48 (N+2) q+96\Big\rbrack\;,\label{4.2.42i}\\
t_{210}&=&2\,T_{221}+2N\,T_{222}\nonumber\\
&=&-\frac{\lbrack5 N^3 q^2-24 N^2 (q-1) q+4 N \left(q^2-18 q+12\right)+48 (q-1) q\rbrack}{3 (N-4) N \left(N^2-4\right) q^2}\nonumber\\
&&-\frac{(1-q)^{1-\frac{N}{2}}}{3 (N-4) N \left(N^2-4\right) q^2} \Big\lbrack N^3 q^2-4 N \left(q^2-6 q+12\right)+48 q\Big\rbrack\;.\label{4.2.42j}
\end{eqnarray}
These results were given in \cite{MossToms} with slightly different notation.

As a check on our results we can take the $q\rightarrow0$ limit of these coefficients which should give the result for the trace of the minimal vector field. (See for example \cite{Gilkey75}.) It can be shown that
\begin{eqnarray}
T_{21}&=&\frac{N}{72}+\frac{1}{144} (N+8) q+\cdots \;,\label{4.2.43a}\\
T_{22}&=&-\frac{N}{180}+\left(\frac{17}{180}-\frac{N}{360}\right) q +\cdots\;,\label{4.2.43b}\\
T_{23}&=&\frac{N-15}{180}+\frac{1}{360} (N-4) q+\cdots \;,\label{4.2.43c}\\
T_{24}&=&\frac{N}{30}+\frac{1}{60} (N+1) q+\cdots \;,\label{4.2.43d}\\
T_{25}&=&-\frac{1}{6}-\frac{q}{12}+\cdots \;,\label{4.2.43e}\\
T_{26}&=&-\frac{q}{3}+\left(-\frac{N}{24}-\frac{1}{4}\right) q^2+\cdots \;,\label{4.2.43f}\\
T_{27}&=&\frac{q^2}{48}+\frac{1}{192} (N+4) q^3+\cdots \;,\label{4.2.43g}\\
T_{28}&=&\frac{1}{2}+\frac{q}{4}+\cdots\;,\label{4.2.43h}\\
T_{29}&=&-\frac{1}{6}-\frac{q}{12}+\cdots \;,\label{4.2.43i}\\
T_{210}&=&\frac{q}{6}+\frac{q^2}{12}+\cdots \;.\label{4.2.43j}
\end{eqnarray}
where we keep the first two terms in the expansion of \eqref{4.2.42a}--\eqref{4.2.42j} about $q=0$. The $q=0$ results agree with that for the minimal operator as they should.

In the physically interesting case of $N=4$ we find, using the definition \eqref{4.2.34e} for comparison with \cite{BarvinskyVilkovisky}, 
\begin{eqnarray}
t_{21}&=&\frac{1}{144} \left(3 \gamma ^2+12 \gamma +8\right)\;\label{4.2.44a}\\
t_{22}&=&\frac{1}{360} \left(15 \gamma ^2+30 \gamma -8\right)\;,\label{4.2.44b}\\
t_{23}&=&-\frac{11}{180}\;,\label{4.2.44c}\\
t_{24}&=&\frac{95 \gamma ^2+288 \gamma +60}{360 \gamma }-\frac{\left(6 \gamma ^2+9 \gamma +2\right) \log \left(\gamma +1\right)}{12 \gamma ^2}\;,\label{4.2.44d}\\
t_{25}&=&\frac{1}{24} \left(-\gamma ^2-2 \gamma -4\right)\;,\label{4.2.44e}\\
t_{26}&=&-\frac{1}{12} \gamma  (\gamma +4)\;,\label{4.2.44f}\\
t_{27}&=&\frac{\gamma ^2}{48}\;,\label{4.2.44g}\\
t_{28}&=&\frac{1}{24} \left(\gamma ^2+6 \gamma +12\right)\;,\label{4.2.44h}\\
t_{29}&=&\frac{-7 \gamma ^2-9 \gamma +6}{36 \gamma }+\frac{\left(\gamma ^2-1\right) \log \left(\gamma +1\right)}{6 \gamma ^2}\;,\label{4.2.44i}\\
t_{210}&=&-\frac{5 \gamma ^2+42 \gamma +24}{36 \gamma }+\frac{\left(5 \gamma ^2+9 \gamma +4\right) \log \left(\gamma +1\right)}{6 \gamma ^2}\;.\label{4.2.44j}\\
\end{eqnarray}
The results for the coefficients that do not involve total derivatives agree with \cite{BarvinskyVilkovisky}. The terms with total derivatives were not needed by \cite{BarvinskyVilkovisky} and the result here agrees with the evaluation of \cite{MossToms}. Only those terms that involve total derivatives involve the logarithm; thus, the logarithm does not enter into the divergent part of the effective action.

\section{Feynman Green function expansions}\label{sec4}

In quantum field theory the Feynman Green function, rather than the auxiliary Green function of Sec.~\ref{secauxGexp}, is the relevant Green's function for developing perturbation theory. This is obtained as in \eqref{djt1.6} by taking $s=0$ in the auxiliary Green function. This relates $T$ to $S$ with a factor of $1/(1-q)$ according to our definitions in \eqref{4.1.1.2} and \eqref{4.1.1.3}. The expressions for the components of the Green function expansions given in Sec.~\ref{secauxGexp} simplify considerably. In addition, we need to be careful if we want to evaluate $G_{\mu\nu}$ from $G_{ab}$ because of the presence of the $N$-bein factors in \eqref{3.10}. It is necessary to use the expansion \eqref{3.7} of the $N$-bein factor of $e^{a}{}_{\mu}(x)$ about the origin of the RNC system at $x'$ to obtain the correct terms in the Green function expansion. In addition, we will specialize to the case of Maxwell electromagnetism by taking 
\begin{equation}
Q_{\mu\nu}=R_{\mu\nu}\;.\label{4.1}
\end{equation}
This too leads to a simplification of terms from the more general results that we have been considering.

We will define $\gamma$ as in \eqref{4.2.34e} as in \cite{BarvinskyVilkovisky}. The result for $G_2$ is
\begin{eqnarray}
G_{2\;\mu\nu}&=&R \Big\lbrack\frac{1}{3}  \delta_{\mu \nu}+\frac{2}{3}\,\gamma\,S p_{\mu} p_{\nu}\Big\rbrack S^2\nonumber\\
&&- R^{\alpha \beta} 
\Big\lbrack \frac{2}{3}   \delta_{\mu \nu}+2\gamma\,  S p_{\mu} p_{\nu} \Big\rbrack S^3 p_{\alpha} p_{\beta} \nonumber\\
&&-\frac{4}{3}R_{\mu \nu} S^2+\frac{4}{3}R_{\mu}{}^{ \alpha}{}_{ \nu}{}^{ \beta} S^3 p_{\alpha} p_{\beta}.\label{4.2}
\end{eqnarray}

For $G_3$ we have
\begin{eqnarray}
G_{3\;\mu\nu}&=&-iR^{;\alpha}\Big\lbrack   \delta_{\mu \nu}+3\gamma\,S p_{\mu} p_{\nu} \Big\rbrack S^{3} p_{\alpha}\nonumber\\
&&+\frac{2i}{3}\gamma\,R_{;\nu} S^3 p_{\mu} + i R^{\alpha \beta\,; \lambda}\Big\lbrack 2  \delta_{\mu \nu}+8\gamma\,S p_{\mu} p_{\nu}  \Big\rbrack S^4 p_{\alpha} p_{\beta} p_{\lambda}\nonumber\\
&&+4i\,R_{\mu \nu}{}^{;\alpha} S^3 p_{\alpha} -2i\gamma\,R^{\alpha \beta}{}_{;\nu} S^4 p_{\mu} p_{\alpha} p_{\beta} \nonumber\\
&&-\frac{4i}{3}R_{\mu}{}^{ \alpha}{}_{;\nu} S^3 p_{\alpha}-4i R_{\mu}{}^{ \alpha}{}_{ \nu}{}^{ \beta\,;\lambda} S^4 p_{\alpha} p_{\beta} p_{\lambda}\;.\label{4.3}
\end{eqnarray}

The expression for $G_4$ is still lengthy and is
\begin{eqnarray}
G_{4\;\mu\nu}&=&\Box R \Big\lbrack\frac{2}{5}  \delta_{\mu \nu} + \frac{6 \gamma }{5}S p_{\mu} p_{\nu} \Big\rbrack S^3 + R^{2} \Big\lbrack\frac{1}{9}  \delta_{\mu \nu} + \frac{\gamma}{3}S p_{\mu} p_{\nu}\Big\rbrack S^3\nonumber\\
&&-R R^{\alpha \beta} \Big\lbrack\frac{2}{3}   \delta_{\mu \nu}  + \frac{8 \gamma }{3}Sp_{\mu} p_{\nu} \Big\rbrack S^4p_{\alpha} p_{\beta}-\frac{8}{9}RR_{\mu \nu} S^3\nonumber\\
&&+ R^{\alpha \beta} R_{\alpha \beta} \Big\lbrack\frac{2}{45}  \delta_{\mu \nu} + \frac{2 \gamma }{15} S p_{\mu} p_{\nu}\Big\rbrack S^3\nonumber\\
&&- R^{\alpha \beta} R_{\alpha}{}^{ \lambda} \Big\lbrack\frac{8}{5}  \delta_{\mu \nu}  +\frac{32 \gamma}{5}S  p_{\mu} p_{\nu} \Big\rbrack S^4 p_{\beta} p_{\lambda}\nonumber\\
&&+ R^{\alpha \beta} R^{\lambda \sigma} \Big\lbrack\frac{4}{3} \delta_{\mu \nu}  + \frac{20 \gamma }{3} S p_{\mu} p_{\nu}\Big\rbrack S^5 p_{\alpha} p_{\beta} p_{\lambda} p_{\sigma}\nonumber\\
&&+\frac{8}{3}R^{\alpha \beta} R_{\mu \nu} S^4 p_{\alpha} p_{\beta}+\frac{4 \gamma }{3}R^{\alpha \beta} R_{\nu \alpha} S^4 p_{\beta} p_{\mu}+\frac{32}{15}R_{\mu}{}^{ \alpha} R_{\nu \alpha} S^3\nonumber\\
&&+ R^{\alpha \lambda \beta \sigma} R_{\alpha \beta} \Big\lbrack\frac{4}{5}  \delta_{\mu \nu}  +\frac{16 \gamma }{5}S p_{\mu} p_{\nu} \Big\rbrack S^4 p_{\lambda} p_{\sigma}\nonumber\\
&&+\frac{4}{3} R R_{\mu}{}^{ \alpha}{}_{ \nu}{}^{ \beta}  S^4 p_{\alpha} p_{\beta}-\frac{28}{45}R_{\mu}{}^{ \alpha}{}_{ \nu}{}^{ \beta}R_{\alpha \beta} S^3\nonumber\\
&&+\frac{2 \left(17 \gamma ^2+69 \gamma +57\right)}{45 (\gamma +1)}R_{\mu \alpha \nu}{}^{ \lambda} R^{\alpha \beta} S^4 p_{\beta} p_{\lambda}\nonumber\\
&&-\frac{22}{5}R_{\mu}{}^{ \beta \alpha \lambda} R_{\nu \alpha} S^4 p_{\beta} p_{\lambda}-\frac{18 \gamma}{5} R_{\mu}{}^{ \lambda \alpha \sigma} R_{\alpha}{}^{ \beta} S^5 p_{\beta} p_{\lambda} p_{\nu} p_{\sigma}\nonumber\\
&&-\frac{2 \left(8 \gamma ^2-54 \gamma -57\right)}{45 (\gamma +1)}R_{\mu}{}^{ \lambda}{}_{ \nu \alpha} R^{\alpha \beta} S^4 p_{\beta} p_{\lambda}\nonumber\\
&&-\frac{16}{3}R_{\mu}{}^{ \lambda}{}_{ \nu}{}^{ \sigma} R^{\alpha \beta} S^5 p_{\alpha} p_{\beta} p_{\lambda} p_{\sigma} + \frac{2 \gamma  (2 \gamma +3)}{9 (\gamma +1)}R_{\mu \nu \alpha}{}^{ \lambda} R^{\alpha \beta} S^4 p_{\beta} p_{\lambda}\nonumber\\
&&+\frac{4 \gamma }{3}R_{\nu}{}^{ \alpha \beta \lambda} R_{\alpha \beta} S^4 p_{\lambda} p_{\mu} - \frac{2}{5}   R_{\mu \alpha} R_{\nu }{}^{\beta \alpha \lambda} S^4 p_{\beta} p_{\lambda} \nonumber\\
&&+\frac{26 \gamma }{15}R_{\nu}{}^{ \lambda \alpha \sigma} R_{\alpha}{}^{ \beta} S^5 p_{\beta} p_{\lambda} p_{\mu} p_{\sigma} +  R^{\alpha \beta \lambda \sigma} R_{\alpha \beta \lambda \sigma} 
\Big\lbrack\frac{1}{15} \delta_{\mu \nu}+\frac{\gamma }{5} S p_{\mu} p_{\nu}\Big\rbrack S^3\nonumber\\
&&- R^{\alpha \beta \lambda \sigma} R_{\alpha \beta \lambda}{}^{ \tau} \Big\lbrack\frac{8}{15}   \delta_{\mu \nu} + \frac{32 \gamma }{15} S p_{\mu} p_{\nu} \Big\rbrack S^4  p_{\sigma} p_{\tau}\nonumber\\
&&+ R^{\alpha \beta \lambda \sigma} R_{\alpha}{}^{ \tau}{}_{ \lambda}{}^{ \delta} \Big\lbrack\frac{16}{15}  \delta_{\mu \nu}  +\frac{16 \gamma }{3}S p_{\mu} p_{\nu} \Big\rbrack S^5 p_{\beta} p_{\delta} p_{\sigma} p_{\tau}\nonumber\\
&&-\frac{2}{5} (\gamma +6)R^{\alpha \beta \lambda \sigma} R_{\mu \alpha \nu \lambda} S^4 p_{\beta} p_{\sigma} -\frac{2}{5} R_{\mu}{}^{ \alpha \beta \lambda} R_{\nu \alpha \beta \lambda} S^3\nonumber\\
&&+\frac{4 \left(8 \gamma ^2+31 \gamma +18\right)}{45 (\gamma +1)} R_{\mu}{}^{ \alpha \beta \lambda} R_{\nu \alpha \beta}{}^{ \sigma} S^4 p_{\lambda} p_{\sigma}\nonumber\\
&&-\frac{(14 \gamma ^2-74 \gamma -108)}{45(1+ \gamma)}R_{\mu}{}^{ \alpha \beta \lambda} R_{\nu \beta \alpha}{}^{ \sigma} S^4 p_{\lambda} p_{\sigma}\nonumber\nonumber\\
&&-\frac{(5 \gamma ^2+25 \gamma +18)}{9(1+ \gamma)} R_{\mu}{}^{ \alpha \beta \lambda} R_{\nu}{}^{ \sigma}{}_{ \beta \lambda} S^4 p_{\alpha} p_{\sigma}+\frac{16}{5} R_{\mu}{}^{ \alpha \beta \lambda} R_{\nu}{}^{ \sigma}{}_{ \beta}{}^{ \tau} S^5 p_{\alpha} p_{\lambda} p_{\sigma} p_{\tau}\nonumber\\
&&+\frac{(6-\gamma) }{5}\Box R_{\mu}{}^{ \alpha}{}_{ \nu}{}^{ \beta} S^4 p_{\alpha} p_{\beta} - \frac{48}{5} R_{\mu}{}^{ \lambda}{}_{ \nu}{}^{ \sigma\,;\alpha\beta} S^5 p_{\alpha} p_{\beta} p_{\lambda} p_{\sigma}\nonumber\\
&&-R^{;\alpha\beta} \Big\lbrack\frac{12}{5}  \delta_{\mu \nu} + \frac{48 \gamma }{5} S p_{\mu} p_{\nu} \Big\rbrack  S^4 p_{\alpha} p_{\beta} + \frac{1}{5} R_{;\mu\nu} S^3+3\gamma R^{;\alpha}{}_{\nu} S^4 p_{\alpha} p_{\mu} \nonumber\\
&&- \Box R^{\alpha \beta }\Big\lbrack\frac{4}{5}  \delta_{\mu \nu} +\frac{16 \gamma }{5}S  p_{\mu} p_{\nu}\Big\rbrack S^4 p_{\alpha} p_{\beta} -\frac{8}{5} \Box R_{\mu \nu} S^3\nonumber\\
&&+ R^{\lambda \sigma\,;\alpha\beta}\Big\lbrack\frac{24}{5}   \delta_{\mu \nu}  + 24\gamma S p_{\mu} p_{\nu} \Big\rbrack S^5 p_{\alpha} p_{\beta} p_{\lambda} p_{\sigma}\nonumber\\
&&+\frac{(\gamma +50)}{5} R_{\mu \nu}{}^{;\alpha\beta} S^4 p_{\alpha} p_{\beta}+ \frac{9\gamma}{5}( R^{\beta \lambda}{}_{;\mu}{}^{\alpha} - R^{\beta \lambda\,;\alpha}{}_{\mu}) S^5 p_{\alpha} p_{\beta} p_{\lambda} p_{\nu} \nonumber\\
&&-\frac{2}{5}R_{\nu}{}^{ \beta}{}_{;\mu}{}^{\alpha} S^4 p_{\alpha} p_{\beta} -\frac{31 \gamma}{5} R^{\beta \lambda}{}_{;\nu}{}^{\alpha} S^5 p_{\alpha} p_{\beta} p_{\lambda} p_{\mu}-\frac{22}{5} R_{\mu}{}^{ \beta}{}_{;\nu}{}^{\alpha} S^4 p_{\alpha} p_{\beta}\nonumber\\
&&- \frac{\gamma}{5} R_{\nu}{}^{ \beta\,;\alpha}{}_{\mu} S^4 p_{\alpha} p_{\beta}+\frac{3\gamma}{5} R^{\alpha \beta}{}_{;\nu\mu} S^4 p_{\alpha} p_{\beta} - \frac{\gamma}{5} R_{\mu}{}^{ \beta\,;\alpha}{}_{\nu} S^4 p_{\alpha} p_{\beta}\nonumber\\
&&- \frac{2\gamma}{5}R^{\alpha \beta}{}_{;\mu\nu} S^4 p_{\alpha} p_{\beta}-\frac{9 \gamma}{5} R^{\beta \lambda\,;\alpha}{}_{\nu} S^5 p_{\alpha} p_{\beta} p_{\lambda} p_{\mu}\;.\label{4.4}
\end{eqnarray}
As a check on either expressions for $G_{4\mu\nu}$ we can calculate the $E_{2\mu\nu}$ coefficient for  $N=6$ from the pole that arises in dimensional regularization~\cite{TomsPRDscalar}. This comes from taking the normal coordinate momentum space expansions, letting $y\rightarrow0$ and replacing
\begin{eqnarray*}
p_{\alpha} p_{\beta} p_{\gamma} p_{\mu} p_{\nu} p_{\lambda} S^6 &\rightarrow& \frac{1}{480} \delta_{\alpha\beta\gamma\mu\nu\lambda},\\
p_{\alpha} p_{\beta} p_{\gamma} p_{\lambda} S^5 &\rightarrow& \frac{1}{48} \delta_{\alpha\beta\gamma\lambda},\\
p_{\alpha} p_{\beta} S^4  &\rightarrow& \frac{1}{6} \delta_{\alpha\beta},\\
S^3 &\rightarrow& 1.
\end{eqnarray*} 
Here $\delta_{\alpha\beta\gamma\mu\nu\lambda}$ and $\delta_{\alpha\beta\gamma\lambda}$ are totally symmetrized sums of ordinary Kronecker deltas taken over all distinct index pairs as defined above. The result for $E_{2\mu\nu}$ is one half of the result of this calculation. It takes the form
\begin{eqnarray*}
E_{2\mu\nu}&=&\left(1+\frac{\gamma}{2}\right)\,\delta_{\mu\nu}\left\lbrack \frac{1}{72}R^2-\frac{1}{180}R_{\alpha\beta}R^{\alpha\beta} + \frac{1}{180}R_{\alpha \beta \gamma \lambda}R^{\alpha \beta \gamma \lambda}+\frac{1}{30}\Box R\right\rbrack\\
&&-\left(\frac{1}{6}+\frac{\gamma}{36}\right) \,RR_{\mu\nu} + \left(\frac{1}{2}+\frac{\gamma}{45}\right) \,R_{\mu}{}^{\alpha}R_{\nu\alpha}-\frac{\gamma}{90}R_{\mu\alpha\nu\beta}R^{\alpha\beta}\\
&&-\left(\frac{1}{12}+\frac{\gamma}{90}\right) \,R_{\mu}{}^{\alpha\beta\lambda}R_{\nu\alpha\beta\lambda} + \frac{\gamma}{30}\,R_{;\mu\nu} -\left(\frac{1}{6}+\frac{\gamma}{60}\right) \,\Box R_{\mu\nu}.
\end{eqnarray*}
The coefficients here can be shown to agree with those found in Sec.~\ref{E2exp} when we let $N=6$. In the case $q\rightarrow0$, which implies that $\gamma\rightarrow0$, we recover the result for the minimal vector field in agreement with the general results given by \cite{Gilkey75} for example.

\section{Trace anomaly for Maxwell field}\label{sec5}

As an application of the results found above we will discuss the trace anomaly for the Maxwell field at one loop order. The classical action for the electromagnetic field is
\begin{equation}
S_M=-\frac{1}{4}\int\,d^Nx\,g^{1/2}F_{\mu\nu}F^{\mu\nu}\,,\label{5.1}
\end{equation}
with $F_{\mu\nu}$ the usual field strength. To this must be added the gauge fixing condition, that we take to be
\begin{equation}
S_{GF}=-\frac{1}{2\xi}\int\,d^Nx\,g^{1/2}(\nabla_\mu A^\mu)^2\,,\label{5.2}
\end{equation}
and the associated ghost action,
\begin{equation}
S_{GH}=\xi^{-1/2}\int\,d^Nx\,g^{1/2}\,\bar{\eta}(-\Box)\eta\,.\label{5.3}
\end{equation}
With these choices the one-loop effective action, if computed using the normal assumptions~\cite{tomsYM}, reads
\begin{equation}
\Gamma^{(1)}=\frac{1}{2}\ln\det\,\Delta^{\mu}{}_{\nu}-\ln\det\,(-\xi^{-1/2}\Box)\;,\label{5.4}
\end{equation}
where $\Delta^{\mu}{}_{\nu}$ takes the form of \eqref{2.1.1} with
\begin{eqnarray}
q&=&1-1/\xi\,\label{5.5}\\
Q^{\mu}{}_{\nu}&=&R^{\mu}{}_{\nu}\,.\label{5.6}
\end{eqnarray}
The special form for $Q^{\mu}{}_{\nu}$ given here allows great simplification, the root cause being that the Green function for the electromagnetic and ghost fields are related by a Ward identity~\cite{dewitt1960radiation,adler1977regularization,BarvinskyVilkovisky}. The divergent part of $\Gamma^{(1)}$ proves to be important in the evaluation of the renormalized stress-energy-momentum tensor. Using the heat kernel expansion and dimensional regularization, with the spacetime dimension $N=4+\epsilon$, it follows that the pole part of $\Gamma^{(1)}$ is given by
\begin{eqnarray}
{\rm PP}(\Gamma^{(1)})&=&\frac{1}{8\pi^2\epsilon}\int d^4x\,g^{1/2}\,\Big\lbrace -\frac{5}{72}R^2+\frac{11}{45}R_{\mu\nu}R^{\mu\nu}-\frac{13}{360}R_{\mu\nu\lambda\sigma}R^{\mu\nu\lambda\sigma} \nonumber\\
&&+\big\lbrack \frac{1}{24}\ln\xi-\frac{1}{20}\big\rbrack\Box R\Big\rbrace\;.\label{5.7}
\end{eqnarray}
Here we have used \eqref{4.2.41} with \eqref{4.2.42a}--\eqref{4.2.42j} for the vector operator and the well known heat kernel coefficient for scalars given by DeWitt~\citep{DeWittdynamical} (or the more general expressions given by Gilkey~\cite{Gilkey75}). The vector coefficients have been simplified using \eqref{5.5} and \eqref{5.6}. (Note that $\gamma=q/(1-q)=\xi-1$ here.) This result agrees with that given by \citep{endo1984gauge} and by \citep{BarvinskyVilkovisky}. Only the coefficient of $\Box R$ depends on the gauge fixing condition, and it can be ignored because it only contributes a total derivative to the integrand in \eqref{5.7}; there is no contribution from the $\Box R$ term in \eqref{5.7} to the trace anomaly as will be discussed below.(This must be the case if the anomaly is universal because we could choose to work with a Riemannian manifold without boundary for which we can safely ignore integrals of total derivatives. Contributions from the $\Box R$ term to boundary contributions in the trace anomaly is a separate issue that we do not consider here.)

In order to remove the pole terms in the one loop effective action \eqref{5.7} it is necessary to include quadratic curvature counterterms to the usual Einstein-Hilbert gravitational action. We will take
\begin{equation}
S_G=-\int d^Nxg^{1/2}\Big\lbrack \lambda+\kappa\,R+\alpha_1\,R^{\mu\nu\lambda\sigma}R_{\mu\nu\lambda\sigma} + \alpha_2\,R^{\mu\nu}R_{\mu\nu} + \alpha_3\,R^{2}\Big\rbrack\;.\label{5.8}
\end{equation}
Here all of the coupling constants are regarded as bare quantities that can be expanded in terms of renormalized ones plus counterterms. We will concentrate on dimensional regularization here, so that all counterterms are expanded as a series of poles as $N\rightarrow4$. If we wish to discuss spacetimes of dimension higher than 4 then higher order curvature invariants must be added to \eqref{5.8}. For the Maxwell theory there will be no renormalization of $\lambda$ and $\kappa$. We adopt the viewpoint that the expectation value of the stress-energy-momentum tensor should be defined from the semi-classical Einstein equations. These read
\begin{equation}
G_{\mu\nu}+\Lambda\,g_{\mu\nu}=8\pi G\langle T_{\mu\nu}\rangle\,,\label{5.9}
\end{equation}
where
\begin{eqnarray}
\kappa&=&\frac{1}{16\pi G}\,,\label{5.10a}\\
\lambda&=&-\frac{\Lambda}{8\pi G}\,.\label{5.10b}
\end{eqnarray}
Because of the necessity for the quadratic terms in $S_G$ defined in \eqref{5.8} there is a contribution to $T_{\mu\nu}$ not only from the matter field part of the action, but also from the quadratic curvature terms in \eqref{5.8}. It follows that
\begin{equation}
\langle T_{\mu\nu}\rangle=\langle T^{Max}{}_{\mu\nu}\rangle+2g^{-1/2}\frac{\delta S_{\rm quad}}{\delta g^{\mu\nu}}\,,\label{5.11}
\end{equation}
where 
\begin{equation}
T^{Max}{}_{\mu\nu}=-2g^{-1/2}\frac{\delta S_{Max}}{\delta g^{\mu\nu}}\,,\label{5.12}
\end{equation}
with $S_{Max}$ the Maxwell part of the action made up of the sum of \eqref{5.1}--\eqref{5.3}, and the second term in \eqref{5.11} is the contribution from the quadratic curvature terms in \eqref{5.8}. Because of the Ward identity~\cite{dewitt1960radiation,adler1977regularization,BarvinskyVilkovisky} it can be shown~\cite{adler1977regularization} that the parts of the expectation value of the stress tensor coming from the gauge fixing condition \eqref{5.2} and the ghost fields \eqref{5.3} cancel exactly. This leaves only the contribution from the Maxwell field \eqref{5.1} which gives the standard result familiar from classical general relativity of
\begin{equation}
T^{Max}{}_{\mu\nu}=F_{\mu}{}^{\lambda}F_{\nu \lambda}-\frac{1}{4}g_{\mu\nu}F_{\lambda\sigma}F^{\lambda\sigma}\,.\label{5.13}
\end{equation}
If we concentrate on just the trace, it is clear that formally without any consideration of regularization that in four dimensions the trace of \eqref{5.13} vanishes. However this is a bit too glib because it is not clear that after regularization we have $\langle T^{Max\;\mu}{}_{\mu}\rangle=0$. The reason is that with $N=4+\epsilon$ we have
\begin{equation}
\langle T^{Max\;\mu}{}_{\mu}\rangle=-\frac{\epsilon}{4}\langle F_{\mu\nu}F^{\mu\nu}\rangle\;.\label{5.14}
\end{equation}
In order that we end up with zero for this result it is necessary, but not obvious, that $\langle F_{\mu\nu}F^{\mu\nu}\rangle$ not have a pole as $\epsilon\rightarrow0$. We will use the local momentum space expansion in the next subsection to show that $\langle F_{\mu\nu}F^{\mu\nu}\rangle$ is finite and hence that $\langle T^{Max\;\mu}{}_{\mu}\rangle=0$ as $\epsilon\rightarrow0$.

\subsection{Proof that $\langle T^{Max\;\mu}{}_{\mu}\rangle=0$ as $\epsilon\rightarrow0$}\label{sec5.1}

Writing $\langle F_{\mu\nu}F^{\mu\nu}\rangle$ in terms of the gauge field results in
\begin{equation}
\langle F_{\mu\nu}F^{\mu\nu}\rangle=2(g^{\mu\lambda}g^{\nu\sigma}-g^{\mu\sigma}g^{\nu\lambda})\,T_{\mu\nu\lambda\sigma}\,,\label{5.1.1}
\end{equation}
where we have defined
\begin{equation}
T_{\mu\nu\lambda\sigma}=\langle \nabla_\mu A_\nu \nabla_\lambda A_\sigma\rangle\,.\label{5.1.2}
\end{equation}
In order to evaluate this expression we define the right hand side using the coincidence limit of a point separated expression, familiar from point splitting regularization. Specifically we define
\begin{eqnarray}
T_{\mu\nu\lambda\sigma}(x')&=&\frac{1}{2} \Big\lbrack\langle \nabla_\mu A_\nu(x)\nabla^\prime_{\lambda}A_\sigma(x')\rangle + \langle \nabla_\lambda A_\sigma(x)\nabla^\prime_{\mu}A_\nu(x')\rangle\Big\rbrack\label{5.1.3a}\\
&=&\frac{1}{2} \Big\lbrack \nabla_\mu\nabla^\prime_\lambda G_{\nu\sigma}(x,x') + \nabla_\lambda\nabla^\prime_\mu G_{\sigma\nu}(x,x')\Big\rbrack\,.\label{5.1.3b}
\end{eqnarray} 
The symmetrization here ensures that the relation $T_{\mu\nu\lambda\sigma}=T_{\lambda\sigma\mu\nu}$ evident from \eqref{5.1.2} holds after regularization. The square brackets in this subsection are used to denote that the coincidence limit $x\rightarrow x'$ is taken, in conformity with standard point splitting regularization notation~\cite{Synge,fulling1989aspects}. We choose to relate all expressions to the origin of normal coordinates $x'$ to facilitate calculations. Because the Christoffel connection vanishes at $x'$ we have
\begin{equation}
\big\lbrack \nabla_\mu \nabla^\prime_\lambda G_{\nu \sigma}(x,x')\big\rbrack = \big\lbrack \partial_\mu \partial^\prime_\lambda G_{\nu \sigma}(x,x')\big\rbrack\,.\label{5.1.4}
\end{equation}

Because of the presence of the factor of $\epsilon$ in \eqref{5.14} we only need to evaluate the pole part of $T_{\mu\nu\lambda\sigma}$ and we can do this by using the local momentum space expansion of the Green function found in Sec.~\ref{sec4}. It is easy to see from \eqref{5.1.4} that the pole part required for $N\rightarrow4$ is given by
\begin{equation}
{\rm PP}\big\lbrack \nabla_\mu \nabla^\prime_\lambda G_{\nu \sigma}(x,x')\big\rbrack = \int\frac{d^Np}{(2\pi)^N} \Big\lbrace p_\mu p_\lambda\,G_{4\,\nu\sigma} + ip_\mu\nabla^\prime_\lambda\,G_{3\,\nu\sigma}\Big\rbrace\,.\label{5.1.5}
\end{equation}
Here $G_3$ is given by \eqref{4.3} and $G_4$ is given by \eqref{4.4}. It is worth remarking that this result can also be obtained by an application of Synge's theorem~\cite{Synge,christensen1976vacuum,christensen1978regularization,fulling1989aspects}. The calculation of \eqref{5.1.5} although straightforward enough is somewhat lengthy. The pole part makes use of the standard results from dimensional regularization,
\begin{eqnarray*}
S^2&\rightarrow&-\frac{1}{8\pi^2\epsilon}\\
p_\mu p_\nu S^3&\rightarrow&-\frac{1}{32\pi^2\epsilon}\delta_{\mu\nu}\\
p_\mu p_\nu p_\lambda p_\sigma S^4&\rightarrow&-\frac{1}{192\pi^2\epsilon}\delta_{\mu\nu\lambda\sigma}\\
p_\mu p_\nu p_\lambda p_\sigma p_\alpha p_\beta S^5&\rightarrow&-\frac{1}{1536\pi^2\epsilon}\delta_{\mu\nu\lambda\sigma\alpha\beta}\\
p_\mu p_\nu p_\lambda p_\sigma p_\alpha p_\beta p_\gamma p_\delta S^6&\rightarrow&-\frac{1}{15360\pi^2\epsilon}\delta_{\mu\nu\lambda\sigma\alpha\beta\gamma\delta}\,.
\end{eqnarray*}
The result for the pole part of $T_{\mu\nu\lambda\sigma}$ turns out to be
\begin{eqnarray}
{\rm PP}(T_{\mu\nu\alpha\beta})&=&-\frac{1}{8\pi^2\epsilon} \Big\lbrace \frac{(\xi +1)}{288}\,R^2 \delta_{\alpha \beta} \delta_{\mu \nu}   
+\frac{(\xi -1)}{288}\,R^2 (\delta_{\alpha \mu} \delta_{\beta \nu}   
+\delta_{\alpha \nu} \delta_{\beta \mu} )\nonumber\\
&&-\frac{(\xi +5)}{72}  RR_{\alpha \beta} \delta_{\mu \nu} -\frac{(\xi +1)}{72} R R_{\mu \nu} \delta_{\alpha \beta} \nonumber\\
&&  + \frac{(1-\xi)}{72}R(R_{\alpha \mu}  \delta_{\beta \nu} 
  + R_{\alpha \nu}  \delta_{\beta \mu}+R_{\beta \mu}  \delta_{\alpha \nu}+ R_{\beta \nu}\delta_{\alpha \mu} )\nonumber\\
&&+ \frac{(\xi +5)}{36}R_{\alpha \beta} R_{\mu \nu} 
  + \frac{(2 \xi +43)}{180}  R_{\alpha \lambda} R_{\beta \lambda} \delta_{\mu \nu} 
 \nonumber\\
&& 
+ \frac{(\xi -1)}{36}(R_{\alpha \mu} R_{\beta \nu} 
  + R_{\alpha \nu} R_{\beta \mu} )+ \frac{(\xi +1)}{90}R_{\mu \lambda} R_{\nu }{}^{\lambda} \delta_{\alpha \beta} 
  \nonumber\\
  &&+\frac{(\xi -1)}{90}( R_{\alpha \lambda} R_{\mu}{}^{ \lambda} \delta_{\beta \nu}
  +  R_{\alpha \lambda} R_{\nu}{}^{\lambda} \delta_{\beta \mu} 
+ R_{\beta \lambda} R_{\mu}{}^{ \lambda} \delta_{\alpha \nu} 
 + R_{\beta \lambda} R_{\nu}{}^{ \lambda} \delta_{\alpha \mu} )\nonumber\\
 &&
-\frac{(\xi +1) }{720} R_{\lambda \sigma} R^{\lambda \sigma} \delta_{\alpha \beta} \delta_{\mu \nu} 
  +\frac{(1-\xi) }{720}R_{\lambda \sigma} R^{\lambda \sigma} ( \delta_{\alpha \mu} \delta_{\beta \nu} + \delta_{\alpha \nu} \delta_{\beta \mu} )\nonumber\\
  &&
+ \frac{19 (\xi -1)}{1440}(R_{\alpha \lambda \beta \mu} R_{\nu}{}^{ \lambda} 
+ R_{\alpha \lambda \beta \nu} R_{\mu}{}^{ \lambda} + R_{\alpha \mu \beta \lambda} R_{\nu}{}^{ \lambda} 
 + R_{\alpha \nu \beta \lambda} R_{\mu}{}^{ \lambda} )\nonumber\\
 &&
+ \frac{(1-\xi) }{180}R^{\lambda \sigma} (R_{\alpha \lambda \beta \sigma}  \delta_{\mu \nu} 
 + R_{\alpha \lambda \mu \sigma}  \delta_{\beta \nu} 
 + R_{\alpha \lambda \nu \sigma} \delta_{\beta \mu} 
 + R_{\beta \lambda \mu \sigma} \delta_{\alpha \nu} 
 + R_{\beta \lambda \nu \sigma}  \delta_{\alpha \mu} 
  )\nonumber\\
  &&
+ \frac{1}{36}R(R_{\alpha \mu \beta \nu} + R_{\alpha \nu \beta \mu} )
-\frac{(\xi +1)}{180}  R_{\mu \lambda \nu \sigma} R^{\lambda \sigma} \delta_{\alpha \beta} 
  \nonumber\\
&&
  + \frac{(7 \xi +113)}{1440}(R_{\alpha \mu \nu \lambda} R_{\beta}{}^{ \lambda} 
  + R_{\alpha \nu \mu \lambda} R_{\beta}{}^{ \lambda} 
    + R_{\beta \mu \nu \lambda} R_{\alpha}{}^{ \lambda} 
  + R_{\beta \nu \mu \lambda} R_{\alpha}{}^{ \lambda} )\nonumber\\
   &&
+ \frac{(1-\xi) }{180}( R_{\alpha \lambda \beta \sigma} R_{\mu \lambda \nu \sigma} 
 + R_{\alpha \lambda \beta \sigma} R_{\mu}{}^{ \sigma}{}_{ \nu}{}^{ \lambda} 
  )
  \nonumber\\
  &&-\frac{(\xi -1) (\xi +10)}{540 \xi } (R_{\alpha \lambda \mu \sigma} R_{\beta}{}^{ \sigma}{}_{ \nu}{}^{ \lambda} + R_{\alpha \lambda \nu \sigma} R_{\beta}{}^{ \sigma}{}_{ \mu}{}^{ \lambda} 
 )\nonumber\\
  &&
+\frac{(22 \xi ^2+33 \xi -10)}{540 \xi }( R_{\alpha \lambda \mu \sigma} R_{\beta}{}^{  \lambda}{}_{  \nu}{}^{  \sigma} 
  + R_{\alpha \lambda \nu \sigma} R_{\beta}{}^{  \lambda}{}_{  \mu}{}^{  \sigma} 
  )\nonumber\\
  &&
-\frac{(2 \xi +13)}{360}  R_{\alpha \lambda \sigma \theta} R_{\beta}{}^{ \lambda \sigma \theta} \delta_{\mu \nu} 
-\frac{(\xi -1) (5 \xi +2)}{216 \xi }( R_{\alpha \mu \lambda \sigma} R_{\beta \nu}{}^{ \lambda \sigma} 
  + R_{\alpha \nu \lambda \sigma} R_{\beta \mu}{}^{ \lambda \sigma} 
  )\nonumber\\
  &&+\frac{(1-\xi )}{180}( R_{\alpha \lambda \sigma \theta} R_{\mu}{}^{ \lambda \sigma \theta} \delta_{\beta \nu} 
  + R_{\alpha \lambda \sigma \theta} R_{\nu}{}^{ \lambda \sigma \theta} \delta_{\beta \mu} 
 + R_{\beta \lambda \sigma \theta} R_{\mu}{}^{ \lambda \sigma \theta} \delta_{\alpha \nu} 
 + R_{\beta \lambda \sigma \theta} R_{\nu}{}^{ \lambda \sigma \theta} \delta_{\alpha \mu} )\nonumber\\
 &&
  +\frac{(\xi +1)}{720} R_{\lambda \sigma \theta \phi} R^{\lambda \sigma \theta \phi} \delta_{\alpha \beta} \delta_{\mu \nu}     
 +\frac{(\xi -1)}{720} R_{\lambda \sigma \theta \phi} R^{\lambda \sigma \theta \phi}( \delta_{\alpha \mu} \delta_{\beta \nu} 
   + \delta_{\alpha \nu} \delta_{\beta \mu} 
   )\nonumber\\
   &&
- \frac{(\xi +1) }{180}  R_{\mu \lambda \sigma \theta} R_{\nu}{}^{ \lambda \sigma \theta} \delta_{\alpha \beta} 
 + \frac{(\xi +1)}{120} \Box R  \delta_{\alpha \beta} \delta_{\mu \nu}    
 +\frac{(\xi -1)}{120} \Box R (\delta_{\alpha \mu} \delta_{\beta \nu} 
    + \delta_{\alpha \nu} \delta_{\beta \mu} 
  )\nonumber\\
   &&
 + \frac{(\xi -1)}{60}(R_{;\,\alpha\beta} \delta_{\mu \nu}  + R_{;\,\alpha\nu} \delta_{\beta \mu} 
   +R_{;\,\beta\mu} \delta_{\alpha \nu} )
    +\frac{(1-\xi) }{40}( R_{;\,\alpha\mu} \delta_{\beta \nu} + R_{;\,\beta\nu} \delta_{\alpha \mu} 
   )\nonumber\\
   &&
 + \frac{(\xi +1)}{60} R_{;\,\mu\nu} \delta_{\alpha \beta}    
 +\frac{(1-\xi) }{120}( R_{\mu \nu;\,\beta\alpha}    
   +R_{\mu \nu;\,\alpha\beta} )\nonumber\\
   &&
 +\frac{(1-\xi) }{480}( R_{\beta \nu;\,\mu\alpha}    
 + R_{\beta \mu;\,\nu\alpha}
  +R_{\alpha \nu;\,\mu\beta} 
     + R_{\alpha \mu;\,\nu\beta} )\nonumber\\
     &&
       +\frac{(1-\xi) }{120}( \Box R_{\alpha \mu} \delta_{\beta \nu}     
 + \Box R_{\alpha \nu} \delta_{\beta \mu} 
    + \Box R_{\beta \mu} \delta_{\alpha \nu} 
   + \Box R_{\beta \nu} \delta_{\alpha \mu} 
    + \Box R_{\mu \nu} \delta_{\alpha \beta} 
 )\nonumber\\
 && - \frac{(\xi +9)}{120} (\Box R_{\alpha \beta} \delta_{\mu \nu}
 + R_{\alpha \beta;\,\nu\mu}   
   +R_{\alpha \beta;\,\mu\nu}  )\nonumber\\
   &&
 +\frac{(21\xi +19)}{480}( R_{\beta \nu;\,\alpha\mu}  + R_{\alpha \mu;\,\beta\nu})  
 -\frac{(19\xi+21)}{480} (R_{\alpha \nu;\,\beta\mu}
  + R_{\beta \mu;\,\alpha\nu})  
 \nonumber\\
   &&
  +\frac{(1-\xi) }{120}( \Box R_{\alpha \mu \beta \nu} 
     + \Box R_{\alpha \nu \beta \mu})\Big\rbrace\,.\label{5.1.6}
    \end{eqnarray}
Using this result we can now evaluate the pole part of $\langle F_{\mu\nu}F^{\mu\nu}\rangle$ using \eqref{5.1.1} to be
\begin{eqnarray}
{\rm PP}(\langle F_{\mu\nu}F^{\mu\nu}\rangle)&=&2(g^{\mu\lambda}g^{\nu\sigma}-g^{\mu\sigma}g^{\nu\lambda})\,{\rm PP}(T_{\mu\nu\lambda\sigma})\nonumber\\
&=&0\,.\label{5.1.7}
\end{eqnarray}
This last line follows after a bit of calculation. This means that $\langle F_{\mu\nu}F^{\mu\nu}\rangle$ is finite as $\epsilon\rightarrow0$, and hence from \eqref{5.14} $\langle T^{Max\;\mu}{}_{\mu}\rangle=0$ as claimed. It is worth remarking that had the term in $G_3$ in \eqref{5.1.5} been neglected a non-zero pole term involving $\Box R$ would have resulted. 

\subsection{Trace anomaly}\label{sec5.2}

Having just established that $\langle T^{Max\;\mu}{}_{\mu}\rangle=0$, it follows from \eqref{5.11} that
\begin{equation}
\langle T^{\mu}{}_{\mu}\rangle=2g^{-1/2}g^{\mu\nu}\frac{\delta S_{\rm quad}}{\delta g^{\mu\nu}}\,.\label{5.2.1}
\end{equation}
Suppose that we write
\begin{equation}
S_{\rm quad}=\alpha_1\,I_1+\alpha_2\,I_2+\alpha_3\,I_3\;,\label{5.2.2}
\end{equation}
where
\begin{eqnarray}
I_1&=&\int d^Nx\,g^{1/2}\,R^{\mu\nu\lambda\sigma}R_{\mu\nu\lambda\sigma}\;,\label{5.2.3a}\\
I_2&=&\int d^Nx\,g^{1/2}\,R^{\mu\nu}R_{\mu\nu}\;,\label{5.2.3b}\\
I_3&=&\int d^Nx\,g^{1/2}\,R^{2}\;.\label{5.2.3c}
\end{eqnarray}
Here $\alpha_1,\alpha_2,\alpha_3$ are coupling constants that contain a finite part and the pole term necessary for renormalization as discussed at the start of Sec.~\ref{sec5}. Under an infinitesimal conformal transformation described by the factor $\delta\omega(x)$, we have
\begin{equation}
\delta g_{\mu\nu}(x)= 2\,\delta\omega(x)g_{\mu\nu}(x)\;.\label{5.3.6}
\end{equation}
The infinitesimal change in the inverse metric is
\begin{equation}
\delta g^{\mu\nu}(x)= -2\,\delta\omega(x)g^{\mu\nu}(x)\;.\label{5.3.7}
\end{equation}
The conformal transformation of $g$ is
\begin{equation}
\delta g(x)=2N\,\delta\omega(x)\,g(x)\;.\label{5.3.8}
\end{equation}
It is easy to establish the identity
\begin{equation}
2\,g^{\mu\nu}(x)\frac{\delta S_{\rm quad}}{\delta g^{\mu\nu}(x)}=-\,\frac{\delta S_{\rm quad}}{\delta \omega(x)}\;.\label{5.3.8d}
\end{equation}
This means we can concentrate on the conformal behaviour of $I_1,I_2,I_3$ in \eqref{5.2.3a}--\eqref{5.2.3c}. the following result is obtained 
\begin{eqnarray}
g^{-1/2}\frac{\delta S_{\rm quad}}{\delta\omega(x)}&=&(N-4)\Big\lbrack \alpha_1\,R^{\mu\nu\lambda\sigma}R_{\mu\nu\lambda\sigma} + \alpha_2\,R^{\mu\nu}R_{\mu\nu} + \alpha_3\,R^{2} \Big\rbrack\nonumber\\
&&-\lbrack4(\alpha_1-\alpha_3)+N(\alpha_2+4\alpha_3) \rbrack\,\Box R\;.\label{5.3.25}
\end{eqnarray}

If we look first at the classical theory, so that $\alpha_1,\alpha_2$ and $\alpha_3$ are finite expressions that describe the coupling to the quadratic curvature terms, then $S_{\rm quad}$ is only conformally invariant for $N=4$ if the coefficient of the $\Box R$ term vanishes. This requires the coupling constants to satisfy the constraint (choosing $N=4$)
\begin{equation}
\alpha_1+\alpha_2+3\alpha_3=0\;.\label{5.3.26}
\end{equation}
In the quantum theory a natural requirement is to demand that the renormalized coupling constants also satisfy this constraint, although ultimately this is only something that could be determined by observation. The associated counterterms for a $\lambda\phi^4$ theory satisfy this to at least two-loop order (see \cite{TomsPRDscalar} for example), so this is a reasonable, although not compulsory, requirement. It also means from the renormalization group that the constraint will hold at all energy scales.

There are now several things to notice with the result \eqref{5.3.25}. First of all if we look at the terms in the first line that involve the curvature squared terms, as we let $N\rightarrow4$ only the pole parts of $\alpha_1,\alpha_2$ and $\alpha_3$ will contribute. This part of the expression is directly related to the counterterms in the effective action that were necessary to renormalize it. If no counterterms are required (which is not the case in general) these terms would make no contribution in the $N\rightarrow4$ limit. For the coefficient of the $\Box R$ term, if we assume that \eqref{5.3.26} holds for the renormalized parameters, then the coefficient of the $\Box R$ term for $N=4$ is determined solely by the $\alpha_2$ and $\alpha_3$ counterterms. An important point is that even if we include a $\Box R$ counterterm in the renormalization of the effective action it does not contribute to the trace anomaly since its conformal change vanishes at $N=4$. To put it another way, the $\Box R$ term in the heat kernel coefficient that determines the $\Box R$ counterterm in the effective action at one-loop order does not necessarily correspond to the $\Box R$ term in the trace anomaly. 

The counterterms for $\alpha_1,\alpha_2,\alpha_3$ are easily found from \eqref{5.7}. A short calculation using these counterterms and \eqref{5.3.25} shows that
\begin{equation}
\langle T^{\mu}{}_{\mu}\rangle= \frac{1}{2880\pi^2}\Big(12\Box R -25R^2+88R_{\mu\nu}R^{\mu\nu}-13R_{\mu\nu\lambda\sigma}R^{\mu\nu\lambda\sigma}\Big)\;.\label{5.2.27}
\end{equation}
This is in complete agreement with Brown and Cassidy~\cite{brown1977stress} who established it in the Feynman gauge using dimensional regularization and a heat kernel method, and with Duff~\cite{duff1977observations}. Our result has been established in a way that shows independence of the gauge parameter and agrees with the result established by Endo~\cite{endo1984gauge} and implicit in \cite{BarvinskyVilkovisky}.

\section{Discussion}\label{sec6}

In Sec.~\ref{sec5} we showed that the conformal anomaly for the Maxwell field could be found using our expansions and gave agreement with previously known results. Our demonstration kept the gauge parameter general and it was shown that the result was independent of this parameter. We also showed that the result was not given by the traced heat kernel coefficient for the vector field as, in agreement with \cite{endo1984gauge}, the term that involved $\Box R$ did depend on the gauge parameter. The origin of the trace anomaly was the quadratic terms necessary to renormalize the effective action, and these turn out to be gauge parameter independent in four spacetime dimensions; our results for the necessary counterterms agree with previously known results \citep{endo1984gauge,BarvinskyVilkovisky}. Brown and Cassidy~\citep{brown1977stress} give a formal proof that the one loop effective action is independent of $\xi$; however it is clear from Sec.~\ref{sec5} that this assumes that total derivatives are discarded. The result that is found for the trace anomaly does agree with the trace of the heat kernel coefficients for the Feynman gauge ($\xi=1$). Thus it appears as if the formal identification of the trace anomaly with the heat kernel coefficient relies on the use of minimal operators. This requires further investigation and we hope to report on it elsewhere. It is also worth mentioning the analysis of Nielsen and van Nieuwenhuizen~\cite{nielsen1988gauge} who use heat kernel methods and a regularization of the proper time representation to show that the $\xi$-dependent contribution from the $\Box R$ part of the vector field heat kernel coefficient is cancelled by the ghost fields in their method. The net result of their calculation agrees with the standard result found using the Feynman gauge. 

Part of the apparent disagreement between different researchers resides in exactly how the expectation value of the stress-energy-momentum tensor is identified from formal expressions such as the effective action or other unregularized expressions. The point of view adopted above is that it should be identified as the source term in the effective Einstein field equations and this leads to a clear-cut definition. The physical content of any other approach should agree with our analysis, but the definition of what one calls the expectation value of the stress-energy tensor and how one identifies the geometric side of the effective field equations may differ.

\begin{acknowledgments}
I would like to thank M. J. Duff and I. G. Moss for motivating the application to the conformal anomaly in Sec.~\ref{sec5}.
\end{acknowledgments}

\appendix\section{Expansions for $(A^{\mu\nu})_{ab}, (B^\mu)_{ab}$ and $(C)_{ab}$}\label{ABC}

By taking $x=x'$ in \eqref{3.11} it can be seen that
\begin{equation}
(A^{\mu\nu}_{0})_{ab}=-\delta^{\mu\nu}\delta_{ab}+\frac{q}{2}\left(e_{a}{}^{\mu}e_{b}{}^{\nu} + e_{a}{}^{\nu}e_{b}{}^{\mu}\right),\label{4.1.21a}
\end{equation}
Here $e_{a}{}^{\mu}$ appearing in \eqref{4.1.21a} is understood to be evaluated at the origin of Riemann normal coordinate. We will omit the spacetime argument $x'$ except when it is necessary to distinguish it from $x$. 

We now need to evaluate the next terms in the Riemann normal coordinates expansions of \eqref{3.11}. We use \eqref{3.5} and \eqref{3.8} to find 
\begin{eqnarray}
(A^{\mu\nu}{}_{\alpha\beta})_{ab}&=&  - \frac{1}{3}\, {R}^{\mu}{}_{ \alpha}{}^{ \nu}{}_{ \beta} {\delta}_{a b} - \frac{q}{12}\, {R}^{\nu}{}_{ \alpha \beta \lambda}\left( {e}_{a}{}^{ \mu} {e}_{b}{}^{ \lambda}+{e}_{a}{}^{ \lambda} {e}_{b}{}^{ \mu}\right)\nonumber\\
&&  - \frac{q}{12}\, {R}^{\mu}{}_{ \alpha \beta \lambda} \left({e}_{a}{}^{ \nu} {e}_{b}{}^{ \lambda} + {e}_{a}{}^{ \lambda} {e}_{b}{}^{ \nu} \right)  .\label{4.1.21}
\end{eqnarray}
Again $e_{a}{}^{\mu}$ appearing in \eqref{4.1.21a} is understood to be evaluated at the origin of Riemann normal coordinates. If we like the notation can be simplified by replacing the repeated spacetime indices with those for the orthonormal frame.
Strictly speaking the result in \eqref{4.1.21} should really be symmetrized in $\alpha$ and $\beta$; however as the result will always be contracted with a symmetric expression we will not explicitly write out the symmetrized expression.

For the next two terms in the expansion of $(A^{\mu\nu})_{ab}$ we find
\begin{eqnarray}
(A^{\mu \nu}{}_{\alpha\beta\gamma})_{ a b}&=&  - \frac{1}{6}\, {{R}^{\mu}{}_{ \alpha}{}^{ \nu}{}_{ \beta\,;\gamma}}\,  {\delta}_{a b}- \frac{q}{24}\, {{R}^{\mu}{}_{ \alpha \beta \sigma\,;\gamma }}\left(  {e}_{a}{}^{ \nu} {e}_{b}{}^{ \sigma}+ {e}_{a}{}^{ \sigma} {e}_{b}{}^{ \nu}  \right) \nonumber\\
&& - \frac{q}{24}\, {{R}^{\nu}{}_{ \alpha \beta \sigma\,;\gamma }}\left(  {e}_{a}{}^{ \mu} {e}_{b}{}^{ \sigma}+ {e}_{a}{}^{ \sigma} {e}_{b}{}^{ \mu}  \right),\label{4.1.22}
\end{eqnarray}
and
\begin{eqnarray}
(A^{\mu \nu}{}_{\alpha\beta\gamma\delta})_{ a b}&=&
 - \frac{1}{20}\, {{R}^{\mu}{}_{ \alpha}{}^{ \nu}{}_{ \beta\,;\gamma\delta}}\,  {\delta}_{a b} - \frac{1}{15}\, {R}^{\mu}{}_{ \alpha \beta \lambda} {R}^{\nu}{}_{ \gamma \delta}{}^{ \lambda}\, {\delta}_{a b}\nonumber\\
  &&- \frac{q}{80}\, {{R}^{\nu}{}_{ \alpha \beta \lambda\,;\gamma\delta}}\big( {e}_{a}{}^{ \mu} {e}_{b}{}^{ \lambda}+  {e}_{a}{}^{ \lambda} {e}_{b}{}^{ \mu}  \big) \nonumber\\
 &&- \frac{q}{80}\, {{R}^{\mu}{}_{ \alpha \beta \lambda\,;\gamma\delta}}\big( {e}_{a}{}^{ \nu} {e}_{b}{}^{ \lambda}+  {e}_{a}{}^{ \lambda} {e}_{b}{}^{ \nu}  \big) \nonumber\\
 &&
 - \frac{7q}{720}\, {R}^{\nu}{}_{ \alpha \beta \lambda} {R}_{\gamma}{}^{ \lambda}{}_{ \delta \sigma} \big( {e}_{a}{}^{ \mu} {e}_{b}{}^{ \sigma} +{e}_{a}{}^{ \sigma} {e}_{b}{}^{ \mu} \big)  \nonumber\\
 && - \frac{7q}{720}\, {R}^{\mu}{}_{ \alpha \beta \lambda} {R}_{\gamma}{}^{ \lambda}{}_{ \delta \sigma} \big( {e}_{a}{}^{ \nu} {e}_{b}{}^{ \sigma} +{e}_{a}{}^{ \sigma} {e}_{b}{}^{ \nu} \big)  \nonumber\\
 &&+ \frac{q}{72}\, {R}^{\mu}{}_{ \alpha \beta \lambda} {R}^{\nu}{}_{ \gamma \delta \sigma}\big( {e}_{a}{}^{ \lambda} {e}_{b}{}^{ \sigma}
 + {e}_{a}{}^{ \sigma} {e}_{b}{}^{ \lambda}\big),\label{4.1.23}
\end{eqnarray}
with a similar note about symmetrization in the indices $\alpha,\beta,\gamma,\delta$.

The first three terms in the Riemann normal coordinate expansion for $(B^\mu)_{ab}$ are (with a similar warning about symmetrization)
\begin{eqnarray}
(B^{\mu}{}_{\alpha})_{ab}&=&\frac{2}{3}\, {R}^{\mu}{}_{ \alpha} {\delta}_{a b} + {R}^{\mu}{}_{ \alpha \lambda \beta} {e}_{a}{}^{ \lambda} {e}_{b}{}^{ \beta}\nonumber\\
&&+ \frac{q}{6}\, {R}^{\mu}{}_{ \lambda \alpha \beta} \big({e}_{a}{}^{ \beta} {e}_{b}{}^{ \lambda}-2\,{e}_{a}{}^{ \lambda} {e}_{b}{}^{ \beta}  \big)  - \frac{q}{2}\, {R}_{\alpha \lambda} {e}_{a}{}^{ \mu} {e}_{b}{}^{ \lambda},\label{4.1.24}
\end{eqnarray}
\begin{eqnarray}
(B^{\mu}{}_{\alpha\beta})_{ab}&=&\frac{1}{2} {R_{\mu \beta\,;\alpha}} \delta_{a b} - \frac{1}{12}{R_{\alpha \beta\,;\mu}} \delta_{a b}+ \frac{2}{3} {R_{\mu \beta \nu \lambda\,;\alpha}} e_{a}{}^{ \nu} e_{b}{}^{ \lambda} \nonumber\\
&&+ \frac{q}{6}\,{R^{\mu}{}_{ \nu \beta \lambda\,;\alpha}}\big( e_{a}{}^{ \lambda} e_{b}{}^{ \nu} -  e_{a}{}^{ \nu} e_{b}{}^{ \lambda}\big)\nonumber\\
&& - \frac{q}{12}\, {R^{\mu}{}_{ \alpha \beta \lambda\,;\nu}}\big( e_{a}{}^{ \nu} e_{b}{}^{ \lambda} + e_{a}{}^{ \lambda} e_{b}{}^{ \nu}\big)\nonumber\\
&& - \frac{q}{12}\, {R_{\alpha \nu \beta \lambda}{}^{;\mu}} e_{a}{}^{ \nu} e_{b}{}^{ \lambda}  
- \frac{q}{3} {R_{\beta \nu\,;\alpha}} e_{a}{}^{ \mu} e_{b}{}^{ \nu} ,\label{4.1.25}
\end{eqnarray}
\begin{eqnarray}
(B^{\mu}{}_{\alpha\beta\gamma})_{ab}&=&
\frac{1}{5}\, {{R}^{\mu}{}_{ \alpha\,;\beta\gamma}}\,   {\delta}_{a b} - \frac{3}{40}\, {{R}_{\alpha \beta}{}^{;\mu}{}_{\gamma}}\,   {\delta}_{a b} - \frac{23}{180}\, {R}_{\alpha}{}^{ \lambda} {R}^{\mu}{}_{ \beta \gamma \lambda} {\delta}_{a b}\nonumber\\
&& + \frac{4}{45}\, {R}^{\mu \lambda}{}_{ \alpha \sigma} {R}_{\beta \lambda \gamma}{}^{ \sigma} {\delta}_{a b} + \frac{1}{40}\, {R}_{\beta \gamma\,;\alpha}{}^{\mu}\, {\delta}_{a b} + \frac{1}{4}\, 
{R}^{\mu}{}_{ \gamma \lambda \sigma\,;\alpha\beta}\,  {e}_{a}{}^{ \lambda} {e}_{b}{}^{ \sigma}\nonumber\\
&& + \frac{1}{4}\, {R}^{\mu}{}_{ \alpha \beta \lambda} {R}_{\gamma}{}^{ \lambda}{}_{ \sigma \tau}\, {e}_{a}{}^{ \sigma} {e}_{b}{}^{ \tau} - \frac{q}{20}\, {R}^{\mu}{}_{ \lambda \gamma \sigma\,;\alpha\beta}\,  {e}_{a}{}^{ \lambda} {e}_{b}{}^{ \sigma} \nonumber\\
&& + \frac{3q}{40}\, {R}^{\mu}{}_{ \lambda \gamma \sigma\,;\alpha\beta}\,  {e}_{a}{}^{ \sigma} {e}_{b}{}^{ \lambda}  - \frac{q}{40}\, {R}_{\mu \beta \gamma \sigma\,;\lambda\alpha}\big(  {e}_{a}{}^{ \lambda} {e}_{b}{}^{ \sigma} q + {e}_{a}{}^{ \sigma} {e}_{b}{}^{ \lambda}\big) \nonumber\\
&& - \frac{q}{40}\, {R}^{\mu}{}_{ \beta \gamma \sigma\,;\alpha\lambda}\big(  {e}_{a}{}^{ \lambda} {e}_{b}{}^{ \sigma} +  {e}_{a}{}^{ \sigma} {e}_{b}{}^{ \lambda}\big)+ \frac{q}{24}\, {R}_{\alpha \lambda \beta \sigma} {R}_{\gamma \lambda \sigma \tau}\, {e}_{a}{}^{ \mu} {e}_{b}{}^{ \tau}\nonumber\\
&& - \frac{q}{40}\big( {R}_{\beta \lambda \gamma \sigma\,;\mu\alpha} + {R}_{\beta \lambda \gamma \sigma\,;\alpha\mu}\big)\,  {e}_{a}{}^{ \lambda} {e}_{b}{}^{ \sigma} \nonumber\\
&& - \frac{q}{45}\, {R}^{\mu}{}_{ \alpha \lambda \sigma} {R}_{\beta}{}^{ \lambda}{}_{ \gamma \tau} \big({e}_{a}{}^{ \sigma} {e}_{b}{}^{ \tau}  + {e}_{a}{}^{ \tau} {e}_{b}{}^{ \sigma} \big)\nonumber\\
&&+ \frac{q}{180}\, {R}^{\mu}{}_{ \alpha \beta \lambda} {R}_{\gamma \sigma}{}^{ \lambda}{}_{ \tau}\big( {e}_{a}{}^{ \sigma} {e}_{b}{}^{ \tau}  +16\, {e}_{a}{}^{ \tau} {e}_{b}{}^{ \sigma} \big)\nonumber\\
&&  - \frac{q}{30}\, {R}^{\mu}{}_{ \lambda \alpha \sigma} {R}_{\beta}{}^{ \lambda}{}_{ \gamma \tau} \big({e}_{a}{}^{ \sigma} {e}_{b}{}^{ \tau}  + {e}_{a}{}^{ \tau} {e}_{b}{}^{ \sigma} \big)\nonumber\\
&& + \frac{q}{360}\, {R}^{\mu}{}_{ \lambda \alpha \sigma} {R}_{\beta \sigma \gamma \tau}\big(11\, {e}_{a}{}^{ \tau} {e}_{b}{}^{ \lambda} - 4\, {e}_{a}{}^{ \lambda} {e}_{b}{}^{ \tau} \big)\nonumber\\
&&  - \frac{q}{8}\, {R}_{\gamma \lambda\,;\alpha\beta}\,  {e}_{a}{}^{ \mu} {e}_{b}{}^{ \lambda} 
+ \frac{q}{12}\, {R}_{\alpha \lambda} {R}^{\mu}{}_{ \beta \gamma \sigma}\, {e}_{a}{}^{ \sigma} {e}_{b}{}^{ \lambda}   \;. \label{4.1.26}
\end{eqnarray}

For $(C)_{ab}$ we need to expand $Q_{ab}$ as well as the geometric terms in \eqref{3.13}. The results are
\begin{equation}
(C_0)_{ab}={Q}_{a b} - \frac{q}{2}\, R_{\alpha \beta} e_{a}{}^{ \alpha} e_{b}{}^{ \beta}\;,\label{4.1.35a}
\end{equation}
\begin{equation}
(C_{\alpha})_{ab}=\nabla_\alpha{Q}_{a b} + \frac{1}{3}\, {{R}_{\alpha \beta\,;\lambda}}\big(  {e}_{a}{}^{ \lambda} {e}_{b}{}^{ \beta} - {e}_{a}{}^{ \beta} {e}_{b}{}^{ \lambda}\big) - \frac{q}{3}\big( {{R}_{\alpha \beta\,;\lambda}} +{{R}_{\lambda \beta\,;\alpha}} \big)\,  {e}_{a}{}^{ \lambda} {e}_{b}{}^{ \beta}\;,\label{4.1.35b}
\end{equation}
\begin{eqnarray}
(C_{\alpha\beta})_{ab}&=&\frac{1}{2}\nabla_{\alpha}\nabla_\beta Q_{a b}
 + \frac{1}{4}\,R_{\beta \nu\,;\mu\alpha}\big(e_{a}{}^{ \mu} e_{b}{}^{ \nu} - e_{a}{}^{ \nu} e_{b}{}^{ \mu}\big)
 + \frac{1}{6}\, R_{\alpha}{}^{ \mu} R_{\beta \mu \nu \lambda} e_{a}{}^{ \nu} e_{b}{}^{ \lambda} \nonumber\\
&&
+ \frac{1}{4}\, R_{\alpha}{}^{ \mu\nu}{}_{\lambda} R_{\beta \mu \nu \sigma} e_{a}{}^{ \lambda} e_{b}{}^{ \sigma}
 - \frac{q}{8}\big(R_{\beta \nu\,;\alpha\mu}+R_{\beta \nu\,;\mu\alpha}+R_{\mu \nu\,;\alpha\beta} \big) e_{a}{}^{ \mu} e_{b}{}^{ \nu} \nonumber\\
&&
 - \frac{q}{8}\, R_{\alpha}{}^{ \mu \nu}{}_{ \lambda} R_{\beta \nu \mu \sigma} e_{a}{}^{ \lambda} e_{b}{}^{ \sigma}  
+ \frac{q}{24}\, R_{\alpha}{}^{ \mu \nu}{}_{ \lambda} R_{\beta \mu \nu \sigma} e_{a}{}^{ \lambda} e_{b}{}^{ \sigma} 
 - \frac{q}{24} R_{\alpha}{}^{ \mu}{}_{ \beta}{}^{ \nu} R_{\mu \lambda \nu \sigma} e_{a}{}^{ \lambda} e_{b}{}^{ \sigma}\nonumber\\
&&
 - \frac{q}{12}\, R^{\mu}{}_{ \nu} R_{\alpha \mu \beta \lambda} e_{a}{}^{ \lambda} e_{b}{}^{ \nu} 
 - \frac{q}{24}\, R_{\alpha\nu}{}^{ \mu \lambda} R_{\beta \sigma \mu \lambda} e_{a}{}^{ \nu} e_{b}{}^{ \sigma} \;.\label{4.1.35c}
\end{eqnarray}
The last term should really be symmetrized in $\alpha$ and $\beta$ as discussed above for the other coefficients.

\section{Expression for the auxiliary Green function}\label{auxG4}

For $G_3$ we use \eqref{3.1.24}--\eqref{3.1.27}, with the relevant expressions for the expansion coefficients given in \eqref{4.1.22},\eqref{4.1.25}and \eqref{4.1.35b} to find
\begin{eqnarray}
e^{a}{}_{\alpha}(x')e^{b}{}_{\beta}(x')(G_3)_{ab}&=&2i\, {Q}_{\alpha \beta}{}^{; \mu}\, S{}^{3}\, {p}_{\mu} -iq\, {Q}_{\alpha \mu}{}^{; \mu}\,S{}^{2} T\,  {p}_{\beta}\nonumber\\
&& -iq\, {Q}_{\alpha}{}^{ \mu}{}_{; \beta} \,S{}^{2} T\, {p}_{\mu} -iq{}^{2}\,{Q}^{\mu \nu}{}_{; \nu}\, S{}^{2} T{}^{2}\,  {p}_{\alpha} {p}_{\beta} {p}_{\mu} \nonumber\\
&&+ 2iq\,{Q}_{\alpha}{}^{ \mu\;; \nu} \big\lbrack  S  
 +(1 -  q) T \big\rbrack\,S^2T\, {p}_{\beta} {p}_{\mu} {p}_{\nu}\nonumber\\
 && + 2iq\, {Q}^{\mu}{}_{ \beta}{}^{; \nu}\, S{}^{3} T\, {p}_{\alpha} {p}_{\mu} {p}_{\nu} \nonumber\\
 && -iq{}^{2}\, {Q}^{\mu \nu}{}_{; \beta}\, S{}^{2} T{}^{2}\, {p}_{\alpha} {p}_{\mu} {p}_{\nu} \nonumber\\
 &&+ 2iq^2\,{Q}^{\mu \nu\;; \lambda} \big\lbrack  S  
 +  (1 -  q)T  \big\rbrack\,S^2T^2\,{p}_{\alpha} {p}_{\beta} {p}_{\mu} {p}_{\nu} {p}_{\lambda}\nonumber\\
 &&+ \frac{iq}{2}\, {R_{;\alpha}}\,  S{}^{2} T\, {p}_{\beta}  + \frac{iq}{3}\,{R_{;\beta}}\big\lbrack  S  + (1-q) T\big\rbrack\,ST\,p_{\alpha}\nonumber\\
 &&- i{R^{;\mu}}\Big\lbrace  S{}^{3} {\delta}_{\alpha \beta} {p}_{\mu}\nonumber\\
 &&\quad +\frac{q}{2}\,\big\lbrack2\, S{}^{2}   + (  2 -{3q})S T   +  2(1-q)^2T{}^{2} \big\rbrack\,ST\,{p}_{\alpha} {p}_{\beta} {p}_{\mu}\Big\rbrace\nonumber\\
 &&+ \frac{i}{6}\,{{R}_{\beta}{}^{ \mu}{}_{;\alpha}}\big\lbrack 4\,  S{}^{2} +  q\,S T   +q(1-q)  T{}^{2}  \big\rbrack\,S\,p_{\mu}\nonumber\\
 && + \frac{i}{6}\,{{R}_{\alpha}{}^{ \mu}{}_{;\beta}}\big\lbrack - 4\,  S{}^{2} + 7 q\,S T   +q(1-q)  T{}^{2}  \big\rbrack\,S\,p_{\mu}\nonumber\\
 &&-\frac{iq}{3}\, {{R}^{\mu \nu}{}_{;\beta}}\big\lbrack 2\, S{}^{2}    + ( 2 -5  q)S T   +2(1-q)^2  T{}^{2} \big\rbrack\,ST\, {p}_{\alpha} {p}_{\mu} {p}_{\nu}\nonumber\\
 &&-\frac{iq}{2}\, {{R}_{\alpha \beta}{}^{;\mu}}\big\lbrack  S  +  (1 -  q)T \big\rbrack \,ST\,p_\mu\nonumber\\
 &&-2iq\, {{R}_{\alpha}{}^{ \nu\;;\mu}}\big\lbrack  S  +  (1 -  q)T \big\rbrack\,S^2T\,{p}_{\beta} {p}_{\mu} {p}_{\nu}\nonumber\\
 &&- 2iq\, {{R}_{\beta}{}^{ \nu\;;\mu}}\,  S{}^{3} T\, {p}_{\alpha} {p}_{\mu} {p}_{\nu}  
 + 2i\,{{R}^{\mu\nu\;; \lambda}}\Big\lbrace S{}^{4} {\delta}_{\alpha \beta} {p}_{\mu} {p}_{\nu} {p}_{\lambda} \nonumber\\
 &&\hspace{-72pt} +q\big\lbrack  S{}^{3}  + ( 1 - 2 q)S{}^{2} T   + (1-q)(1-2q)S T{}^{2}   + (1-q)^3 T{}^{3}  \big\rbrack\,ST\,{p}_{\alpha} {p}_{\beta} {p}_{\mu} {p}_{\nu} {p}_{\lambda}\Big\rbrace\nonumber\\
 &&\hspace{-72pt} +\frac{2iq}{3}\, {{R}_{\alpha}{}^{ \mu}{}_{ \beta}{}^{ \nu\;;\lambda}}\big\lbrack S{}^{2}    + (1-q)S T   + (1-q)^2T{}^{2} \big\rbrack\,ST\,{p}_{\mu} {p}_{\nu} {p}_{\lambda}\;.
\end{eqnarray}

The results for $S$ and $T$ are given in \eqref{4.1.1.2} and \eqref{4.1.1.3} respectively.

Lastly we turn to the much lengthier expression for $G_4$. We use \eqref{djt1.24}--\eqref{djt1.30} along with the result for $G_0$ in \eqref{4.1.1.1} and the relevant expressions for $A,B,C$ given in Appendix~\ref{ABC}. The net result is
\begin{eqnarray}
e^{a}{}_{\mu}(x')e^{b}{}_{\nu}(x')(G_{4})_{ab}&=&\frac{1}{9}\,R{}^{2} \Big\lbrace S{}^{3} {\delta}_{\mu \nu} + q\, \big\lbrack S{}^{2}    + (1-q)S T\nonumber\\
&&\qquad  + (1-q)^2 T{}^{2}  \big\rbrack\,ST\,{p}_{\mu} {p}_{\nu}\Big\rbrace \nonumber\\
&&\hspace{-24pt}- \frac{2}{3}\, R\,{R}^{\alpha \beta} \Big\lbrace  S{}^{4} {\delta}_{\mu \nu} {p}_{\alpha} {p}_{\beta} 
+ q\,\big\lbrack S{}^{3} +  (  1 -2 q)S{}^{2} T \nonumber\\
&&\quad
+ (1-q)(1-2q)S T{}^{2}  + (1-q)^3 T{}^{3}  \big\rbrack\,ST\,{p}_{\alpha} {p}_{\beta} {p}_{\mu} {p}_{\nu}\Big\rbrace\nonumber\\
&&+\frac{q}{3} R\,{R}_{\mu}{}^{ \alpha} \big\lbrack  2\,S  +   (1 -  q)T \big\rbrack\,S^2T\,{p}_{\alpha} {p}_{\nu}\nonumber\\
&&+\frac{q}{3} R\,{R}_{\nu}{}^{ \alpha} \big\lbrack  2\,S  +   (1 -  q)T \big\rbrack\,S^2T\,{p}_{\alpha} {p}_{\mu}\nonumber\\
&& +\frac{q}{18}\,  R\,{R}_{\mu \nu}  \big\lbrack S + (1-q) T \big\rbrack \,ST\nonumber\\
&& +\frac{1}{45}\, {R}_{\alpha \beta} {R}^{\alpha \beta} \Big\lbrace 2\, S{}^{3} {\delta}_{\mu \nu} + q\, \big\lbrack 2\,S{}^{2}    +(2 - 17 q) S T\nonumber\\
&&\qquad   +2(1-q)^2 T{}^{2} \big\rbrack\,ST\,{p}_{\mu} {p}_{\nu}\Big\rbrace \nonumber\\
&&-\frac{1}{5}\,{R}_{\alpha}{}^{ \beta} {R}^{\alpha \lambda} \Big\lbrace 8\, S{}^{4} {\delta}_{\mu \nu} {p}_{\beta} {p}_{\lambda} +q\,\big\lbrack 8\, S{}^{3}    +( 8 -23 q) S{}^{2} T \nonumber\\
&&
 + 2(1-q)(4-9q)S T{}^{2}  + 8(1-q)^3 T{}^{3}  \big\rbrack\,ST\,{p}_{\beta} {p}_{\lambda} {p}_{\mu} {p}_{\nu}\Big\rbrace\nonumber\\
 &&\hspace{-72pt}+\frac{1}{3}\, {R}^{\alpha \beta} {R}^{\lambda \sigma} \Big\lbrace 4\, S{}^{5} {\delta}_{\mu \nu} {p}_{\alpha} {p}_{\beta} {p}_{\lambda} {p}_{\sigma} + q\,\big\lbrack 4\, S{}^{4}    +2(2-5q) S{}^{3} T \nonumber\\
 &&+ (2-3q)(2-5q)S{}^{2} T{}^{2}  +2(1-q)^2(2-5q) S T{}^{3}\nonumber\\
 &&   + 4(1-q)^4 T{}^{4}  \big\rbrack\,ST\,{p}_{\alpha} {p}_{\beta} {p}_{\lambda}  {p}_{\sigma}{p}_{\mu} {p}_{\nu}\Big\rbrace \nonumber\\
 &&\hspace{-72pt}+\frac{q}{9}\, {R}_{\alpha}{}^{ \beta} {R}_{\mu}{}^{ \alpha} \big\lbrack 7\, S{}^{2}   +(7- 4 q) S T  -2(1-q)^2T{}^{2} \big\rbrack\,ST\, {p}_{\beta} {p}_{\nu}\nonumber\\
 &&\hspace{-72pt}-\frac{q}{3}\,{R}^{\alpha \beta} {R}_{\mu}{}^{ \lambda} \big\lbrack 6\, S{}^{2}    +( 4- 7 q) S T   + 2(1-q)^2T{}^{2}  \big\rbrack\,S^2T \,{p}_{\lambda} {p}_{\alpha} {p}_{\beta} {p}_{\nu}\nonumber\\
 &&\hspace{-72pt}-\frac{q}{3}\, {R}_{\alpha \beta} {R}_{\nu}{}^{ \lambda} \big\lbrack 6\, S{}^{2}   + (4 -7q)S T  + 2(1-q)^2 T{}^{2}  \big\rbrack\,S^2T\, {p}_{\lambda} {p}_{\alpha} {p}_{\beta} {p}_{\mu}\nonumber\\
 &&\hspace{-72pt}-\frac{q}{18}\, {R}^{\alpha \beta} {R}_{\mu \nu} \big\lbrack 2\, S{}^{2}   + (2 - 5 q)S T   + 2(1-q)^2 T{}^{2} \big\rbrack \,ST\,{p}_{\alpha} {p}_{\beta}\nonumber\\
 &&\hspace{-72pt}+\frac{q}{9}\, {R}^{\alpha \beta} {R}_{\nu \alpha} \big\lbrack 13\, S{}^{2}   +2(2-5q) S T  + 4(1-q)^2T{}^{2}  \big\rbrack \,ST\,{p}_{\beta} {p}_{\mu}\nonumber\\
  &&\hspace{-72pt}-\frac{q}{180}\, {R}_{\mu}{}^{ \alpha} {R}_{\nu \alpha} \lbrack 41 S + 11(1-q)T\rbrack\,ST+ q\,{R}_{\mu}{}^{ \alpha} {R}_{\nu}{}^{ \beta} S{}^{3} T {p}_{\alpha} {p}_{\beta} \nonumber\\
  &&\hspace{-72pt} +\frac{2}{15} {R}^{\alpha \lambda \beta \sigma} {R}_{\alpha \beta} \Big\lbrace6\, S{}^{4} {\delta}_{\mu \nu} {p}_{\lambda} {p}_{\sigma} +q\,\big\lbrack 6\, S{}^{3}   + (6-11q)S{}^{2} T \nonumber\\
  &&+ (1-q)(6-11q)S T{}^{2}   + 6(1-q)^3T{}^{3} \big\rbrack\,ST\, {p}_{\lambda} {p}_{\mu} {p}_{\nu} {p}_{\sigma}\Big\rbrace \nonumber\\
  &&\hspace{-72pt}-\frac{q}{9}\, {R}_{\mu}{}^{ \alpha \beta \lambda} {R}_{\alpha \beta} \big\lbrack  2\, S{}^{2}   - (7 - 4 q)S T  +2(1-q)^2  T{}^{2}  \big\rbrack\,ST\, {p}_{\lambda} {p}_{\nu}\nonumber\\
  &&\hspace{-72pt}-\frac{2q}{9}\, R\,{R}_{\mu}{}^{ \alpha }{}_{\nu}{}^{ \beta}  \big\lbrack  S{}^{2}   + (1-q)S T   +  (1-q)^2T{}^{2}  \big\rbrack\,ST\,{p}_{\alpha} {p}_{\beta}\nonumber\\
  &&\hspace{-72pt} +\frac{q}{10} {R}_{\mu \alpha \nu \beta}{R}^{\alpha \beta} \big\lbrack 6 S + (1-q)T\big\rbrack\,ST\nonumber\\
  &&\hspace{-72pt}- \frac{q}{90}\,{R}_{\mu}{}^{ \alpha}{}_{ \nu}{}^{ \lambda} {R}_{\alpha}{}^{ \beta} \big\lbrack  20\, S{}^{3}  +( 84 + 15 q)S{}^{2} T   + 2(  2 - 17 q)S T{}^{2}\nonumber\\
  &&\quad  + 38(1-q)^2 T{}^{3} \big\rbrack\,S\,{p}_{\beta} {p}_{\lambda}\nonumber\\
  &&\hspace{-72pt}- \frac{q}{45}\, {R}_{\mu}{}^{ \beta \alpha \lambda} {R}_{\nu \alpha} \Big\lbrace 37\, S{}^{2}    + 22(1-q)S T   + 7(1-q)^2 T{}^{2} 
  \Big\rbrace\,ST\,{p}_{\beta} {p}_{\lambda}\nonumber\\
  &&\hspace{-72pt}+\frac{2q}{15}\, {R}_{\mu}{}^{ \lambda \alpha \sigma} {R}_{\alpha}{}^{ \beta} \Big\lbrace S{}^{3}   + (1 - 11 q)S{}^{2} T   +(1 - q)(1 - 11 q) S T{}^{2}\nonumber\\
  &&\quad  +10(1-q)^3  T{}^{3}  \Big\rbrace\,ST \,{p}_{\nu} {p}_{\beta} {p}_{\lambda}  {p}_{\sigma}\nonumber\\
  &&\hspace{-72pt}+ \frac{q}{90}\, {R}_{\mu}{}^{ \lambda}{}_{ \nu \alpha} {R}^{\alpha \beta} \Big\lbrace10\, S{}^{3}   -( 34 - 15 q) S{}^{2} T   -34(1-q) S T{}^{2}\nonumber\\
  &&\quad   -16(1-q)^2  T{}^{3}  \Big\rbrace\,S\,{p}_{\beta} {p}_{\lambda} \nonumber\\
  &&\hspace{-72pt}+\frac{2q}{3}\, {R}_{\mu}{}^{ \lambda}{}_{ \nu}{}^{ \sigma} {R}^{\alpha \beta} \Big\lbrace S{}^{3}   + (1 - 2 q)S{}^{2} T   +(1 - q)(1 - 2 q) S T{}^{2}\nonumber\\
  &&\quad   +(1 - q)^3 T{}^{3} \Big\rbrace\,ST \,{p}_{\alpha} {p}_{\beta} {p}_{\lambda} {p}_{\sigma}\nonumber\\
  &&\hspace{-72pt}+\frac{q}{18} {R}_{\mu \nu \alpha}{}^{ \lambda} {R}^{\alpha \beta} \Big\lbrace 4\, S{}^{3}   + 3q\, S{}^{2} T  + 6q\, S T{}^{2}   -4(1-q)^2  T{}^{3}  \Big\rbrace\,S \,{p}_{\beta} {p}_{\lambda}\nonumber\\
  &&\hspace{-72pt}+\frac{2q}{9}\, {R}_{\nu \alpha \beta}{}^{ \lambda} {R}^{\alpha \beta} \Big\lbrace2\, S{}^{2}   +(2-5q) S T  +2(1-q)^2  T{}^{2} \Big\rbrace\,ST\, {p}_{\lambda} {p}_{\mu}\nonumber\\
  &&\hspace{-72pt}+\frac{q}{45}\, {R}_{\nu}{}^{ \beta \alpha \lambda} {R}_{\mu \alpha} \Big\lbrace 8\, S{}^{2}    -22(1-q) S T   -7(1-q)^2 T{}^{2} \Big\rbrace\,ST \,{p}_{\beta} {p}_{\lambda} \nonumber\\
  &&\hspace{-72pt}+\frac{q}{15}\, {R}_{\nu}{}^{ \lambda \alpha \sigma} {R}_{\alpha}{}^{ \beta} \big\lbrack 20\, S{}^{3}   + (2 + 5q)S{}^{2} T   +(1-q)(2 + 5q) S T{}^{2}  \nonumber\\
  &&\quad + 2(1- 3q{}^{2} +  q{}^{3})T{}^{3}  \big\rbrack\,ST\,{p}_{\beta} {p}_{\lambda}  {p}_{\sigma}{p}_{\mu}\nonumber\\
  &&\hspace{-72pt} +\frac{1}{15}\, {R}_{\alpha \beta \lambda \sigma} {R}^{\alpha \beta \lambda \sigma} \Big\lbrace  S{}^{3} {\delta}_{\mu \nu} + q\,\big\lbrack S{}^{2}  + (1-q)S T\nonumber\\
  &&\quad  + (1-q)^2T{}^{2}  \big\rbrack\,ST\,{p}_{\mu} {p}_{\nu}\Big\rbrace \nonumber\\
  &&\hspace{-72pt}-\frac{8}{15}\, {R}_{\alpha \beta \lambda}{}^{ \sigma} {R}_{\alpha \beta \lambda}{}^{ \tau} \Big\lbrace   S{}^{4}\, {\delta}_{\mu \nu}\, {p}_{\sigma} {p}_{\tau} +q\,\big\lbrack S{}^{3}   +(1-q) S{}^{2} T  + (1-q)^2S T{}^{2}\nonumber\\
  &&\quad   +  (1-q)^3T{}^{3} \big\rbrack\,ST\,{p}_{\mu} {p}_{\nu} {p}_{\sigma} {p}_{\tau} \Big\rbrace\nonumber\\
  &&\hspace{-72pt}+\frac{16}{15}\, {R}_{\alpha}{}^{ \beta \lambda \sigma} {R}_{\alpha}{}^{ \tau}{}_{ \lambda}{}^{ \rho} \Big\lbrace  S{}^{5}\, {\delta}_{\mu \nu}\, {p}_{\beta} {p}_{\rho} {p}_{\sigma} {p}_{\tau} + q\,\big\lbrack  S{}^{4}  + (1-q)S{}^{3} T   + (1-q)^2S{}^{2} T{}^{2}\nonumber\\
  &&\quad   + (1-q)^3S T{}^{3}   +(1-q)^4T{}^{4}  \big\rbrack\,ST\,  {p}_{\mu} {p}_{\nu}{p}_{\beta} {p}_{\rho} {p}_{\sigma} {p}_{\tau}\Big\rbrace\nonumber\\
  &&\hspace{-72pt}   + \frac{q}{5}\,{R}_{\alpha}{}^{ \beta \lambda \sigma} {R}_{\mu}{}^{ \alpha}{}_{ \nu \lambda} \big\lbrack  S{}^{2}   +(1-q) S T +2(1-q)^2  T{}^{2} \big\rbrack\,ST\, {p}_{\beta} {p}_{\sigma}\nonumber\\
  &&\hspace{-72pt}+\frac{q^2}{18}\, {R}^{\alpha \beta \lambda \sigma} {R}_{\nu \alpha \beta \lambda} \big\lbrack S  + (1-q) T \big\rbrack\,ST^2\, {p}_{\mu} {p}_{\sigma}\nonumber\\
  &&\hspace{-72pt}+\frac{q^2}{36}\, {R}^{\alpha \beta \lambda \sigma} {R}_{\nu \alpha \lambda \sigma} \big\lbrack  S   +(1-q) T \big\rbrack\,ST^2\, {p}_{\mu} {p}_{\beta}\nonumber\\
&&\hspace{-72pt} -\frac{1}{120}\, {R}_{\mu \alpha \beta \lambda} {R}_{\nu}{}^{ \alpha \beta \lambda} \Big\lbrace 60\, S{}^{3}-7q\big\lbrack  S  +(1-q)  T \big\rbrack\,ST\Big\rbrace\nonumber\\
&&\hspace{-72pt}+\frac{1}{45}\, {R}_{\mu \alpha \beta}{}^{ \lambda} {R}_{\nu}{}^{ \alpha \beta \sigma} \big\lbrack 30(1+q) S{}^{3}   +q(4+5q) S{}^{2} T\nonumber\\
&&   + q(1-q)(4+5q)S T{}^{2}  -7q(1-q)^2 T{}^{3}  \big\rbrack\,S\,{p}_{\lambda} {p}_{\sigma}\nonumber\\
&&\hspace{-72pt}  -\frac{q}{45}\, {R}_{\mu \alpha \beta \lambda} {R}_{\nu}{}^{ \beta \alpha \sigma} \big\lbrack 30\, S{}^{3}   + 5( 4 + q) S{}^{2} T   + 5( 1- q)( 4 + q) S T{}^{2}\nonumber\\
&&   +18(1-q)^2  T{}^{3}  \big\rbrack\,S \,{p}_{\lambda} {p}_{\sigma}\nonumber\\
&&\hspace{-72pt} -\frac{q}{90}\, {R}_{\mu}{}^{ \alpha \beta \lambda} {R}_{\nu \beta \lambda}{}^{ \sigma} \big\lbrack ( 27+5q)S{}^{2}    +(1-q)( 27+5q) S T \nonumber\\
&& + 16(1-q)^2 T{}^{2}  \big\rbrack\,ST\,{p}_{\alpha} {p}_{\sigma} \nonumber\\
&&\hspace{-72pt} +\frac{q}{90}\, {R}_{\mu \alpha \beta}{}^{ \lambda} {R}_{\nu}{}^{ \sigma \alpha \beta} \big\lbrack  (3 + 10 q)S{}^{2}  + (1- q)(3 + 10 q)S T \nonumber\\
&& -16(1-q)^2  T{}^{2} \big\rbrack\,ST\,{p}_{\lambda} {p}_{\sigma}\nonumber\\
&&\hspace{-72pt} -\frac{q}{180}\, {R}_{\mu}{}^{ \alpha \beta \lambda} {R}_{\nu}{}^{ \sigma}{}_{ \beta \lambda}\big\lbrack 60\, S{}^{3}   +( 84+5q) S{}^{2} T   +( 1-q)( 84+5q) S T{}^{2}\nonumber\\
&&   + 56(1-q)^2 T{}^{3} \big\rbrack\,S\, {p}_{\alpha} {p}_{\sigma}\nonumber\\
&&\hspace{-72pt}+\frac{14q}{15}\,  {R}_{\mu}{}^{ \alpha \rho \lambda} {R}_{\nu}{}^{ \sigma}{}_{ \rho}{}^{ \beta} \big\lbrack S{}^{3}    + (1 - q)S{}^{2} T   + (1-q)^2S T{}^{2}\nonumber\\
&&   + (1-q)(1+q^2) T{}^{3}  \big\rbrack\,ST\, {p}_{\alpha} {p}_{\beta} {p}_{\lambda} {p}_{\sigma} \nonumber\\
&&\hspace{-72pt}-\frac{q}{10}\, \Box{R}_{\mu}{}^{ \beta}{}_{ \nu}{}^{ \lambda}\big\lbrack  3\, S{}^{2}   +3(1-q) S T  +2(1-q)^2 T{}^{2}  \big\rbrack\,ST\, {p}_{\beta} {p}_{\lambda}\nonumber\\
&&\hspace{-72pt}+\frac{6q}{5}\, {R}_{\mu}{}^{ \lambda}{}_{ \nu}{}^{ \sigma\,;\alpha\beta}\big\lbrack  S{}^{3}   + (1-q)S{}^{2} T  +(1-q)^2 S T{}^{2}\nonumber\\
&&  +(1-q)^3  T{}^{3}  \big\rbrack\,ST\, {p}_{\alpha} {p}_{\beta} {p}_{\lambda} {p}_{\sigma}\nonumber\\
&&\hspace{-72pt} +\frac{2}{5}\, \Box{R}\,\Big\lbrace S{}^{3} {\delta}_{\mu \nu} +q\, \big\lbrack  S{}^{2}    +(1-q) S T   +(1-q)^2  T{}^{2} \big\rbrack\,ST\,{p}_{\mu} {p}_{\nu}\Big\rbrace \nonumber\\
&&\hspace{-72pt} -\frac{1}{5}\, R^{;\alpha\beta}\Big\lbrace 12\, S{}^{4} {\delta}_{\mu \nu} {p}_{\alpha} {p}_{\beta}+q\,\big\lbrack 12\, S{}^{3}    + ( 12 -17q)S{}^{2} T\nonumber\\
&&   + (1-q)(12-17q)S T{}^{2}  +12(1-q)^3 T{}^{3}  \big\rbrack\,ST\,{p}_{\alpha} {p}_{\beta} {p}_{\mu} {p}_{\nu}\Big\rbrace \nonumber\\
&&\hspace{-72pt} +q\, R_{;\mu}{}^{\alpha}\big\lbrack   S   + (1-q) T  \big\rbrack\,S^2T\,{p}_{\alpha} {p}_{\nu}  \nonumber\\
&&\hspace{-72pt}+q\, R_{;\nu}{}^{\alpha}\big\lbrack S^2+\frac{1}{2}(2-3q)ST+(1-q)^2T^2\big\rbrack\,ST\,p_{\mu}p_{\alpha}\nonumber\\
&&\hspace{-72pt}-\frac{11q}{20}\,R_{;\mu\nu}\,S^2T\nonumber\\
&&\hspace{-72pt} -\frac{1}{5}\, \Box{R}^{\beta \lambda}\Big\lbrace 4\, S{}^{4} {\delta}_{\mu \nu} {p}_{\beta} {p}_{\lambda} +q\,\big\lbrack4\, S{}^{3}    + (4-9q)S{}^{2} T\nonumber\\
&&   +(1-q)(4-9q) S T{}^{2}   +4(1-q)^3 T{}^{3}   
\big\rbrack\,ST\,{p}_{\beta} {p}_{\lambda} {p}_{\mu} {p}_{\nu}\Big\rbrace\nonumber\\
&&\hspace{-72pt} +\frac{4}{5}\, {R}^{\lambda \sigma\,;\alpha\beta}\Big\lbrace6\, S{}^{5} {\delta}_{\mu \nu}  +q\,\big\lbrack 6\, S{}^{4}    +  (6 - 11 q)S{}^{3} T\nonumber\\
&& +(1-q) (6 - 11 q) S{}^{2} T{}^{2}  + (1-q)^2(6-11q)S T{}^{3}\nonumber\\
&&   + 6(1-q)^4 T{}^{4} \big\rbrack\,ST\, {p}_{\mu} {p}_{\nu} \Big\rbrace\,{p}_{\alpha} {p}_{\beta} {p}_{\lambda} {p}_{\sigma}\nonumber\\
&&\hspace{-72pt} + q\,\Box{R}_{\mu}{}^{ \beta}\big\lbrack S+(1-q)T\rbrack \,S T\, {p}_{\beta} {p}_{\nu} \nonumber\\
&&\hspace{-72pt} +\frac{3q}{20} \Box{R}_{\mu \nu}\big\lbrack S +(1-q) T \big\rbrack\,ST\nonumber\\
&&\hspace{-72pt} + q\,\Box{R}_{\nu}{}^{ \beta}\,  S{}^{3} T\, {p}_{\beta} {p}_{\mu}  \nonumber\\
&&\hspace{-72pt} -4q\, {R}_{\mu}{}^{ \lambda\,;\alpha\beta}  \big\lbrack  S{}^{3}    +(1-q) S{}^{2} T  +(1-q)^2 S T{}^{2} \big\rbrack\,ST\,{p}_{\alpha} {p}_{\beta} {p}_{\lambda} {p}_{\nu} \nonumber\\
&&\hspace{-72pt} -\frac{q}{10}\, {R}_{\mu \nu}{}^{;\alpha\beta}  \big\lbrack 9\, S{}^{2}    +9(1-q) S T   +10(1-q)^2  T{}^{2}  \big\rbrack\,ST\,{p}_{\alpha} {p}_{\beta}\nonumber\\
&&\hspace{-72pt} - 4q\, {R}_{\nu}{}^{ \lambda\,;\alpha\beta}\,  S{}^{4} T\, {p}_{\alpha} {p}_{\beta} {p}_{\lambda} {p}_{\mu}  \nonumber\\
&&\hspace{-72pt} +\frac{3q}{5}\,{R}^{\beta \lambda}{}_{;\mu}{}^{\alpha}\big\lbrack S{}^{3}    +(1-q) S{}^{2} T   +(1-q)^2 S T{}^{2}\big\rbrack\,ST\, {p}_{\alpha} {p}_{\beta} {p}_{\lambda} {p}_{\nu} \nonumber\\
&&\hspace{-72pt} +\frac{2}{5}\,{R}_{\nu}{}^{ \beta}{}_{;\mu}{}^{\alpha}\big\lbrack 5\, S{}^{3}  +  q\,S{}^{2} T  +q(1-q) S T{}^{2}   +q(1-q)^2 T{}^{3}  \big\rbrack\,S\,{p}_{\alpha} {p}_{\beta} \nonumber\\
&&\hspace{-72pt} -\frac{q}{10}\, {R}^{\beta \lambda}{}_{;\nu}{}^{\alpha}\big\lbrack 
20\, S{}^{3}    + ( 14 -25 q)S{}^{2} T   
+(1-q)( 14 -25 q) S T{}^{2}  \nonumber\\
&&
+ 2 (1-q)^2 (7-6 q)T{}^{3}  \big\rbrack \,ST\,{p}_{\alpha} {p}_{\beta} {p}_{\lambda} {p}_{\mu} \nonumber\\
&&\hspace{-72pt} +\frac{1}{5}\, {R}_{\mu}{}^{ \beta}{}_{;\nu}{}^{\alpha}\big\lbrack - 10\, S{}^{3}  + 7q\, S{}^{2} T   +7q(1-q) S T{}^{2}   + 2q(1-q)^2T{}^{3}  \big\rbrack\,S\,{p}_{\alpha} {p}_{\beta} \nonumber\\
&&\hspace{-72pt} -\frac{3q}{5}\, {R}^{\beta \lambda \alpha}{}_{;\mu}\,  (  S{}^{2} +(1-q)S  T   +(1-q)^2  T{}^{2}  \big\rbrack\,S^2T\, {p}_{\alpha} {p}_{\beta} {p}_{\lambda} {p}_{\nu} \nonumber\\
&&\hspace{-72pt} -\frac{q}{10}\,{R}_{\nu}{}^{ \alpha\,;\beta}{}_{\mu}\big\lbrack  S   +(1-q)  T  \big\rbrack\,S^2T\,{p}_{\alpha} {p}_{\beta} \nonumber\\
&&\hspace{-72pt} + \frac{3q}{10}\,{R}^{\alpha \beta}{}_{;\mu\nu}\big\lbrack S + (1-q)T\big\rbrack\,S^2T\,{p}_{\alpha} {p}_{\beta} \nonumber\\
&&\hspace{-72pt} -\frac{q}{10}\, {R}^{\beta \lambda\,;\alpha}{}_{\nu}\big\lbrack ( 6 - 15 q)S{}^{2}    +(1-q)( 6 - 15 q)S T  \nonumber\\
&& + 2 (1-q)^2 (3-4 q)T{}^{2}  \big\rbrack\,ST^2\,{p}_{\alpha} {p}_{\beta} {p}_{\lambda} {p}_{\mu}\nonumber\\
&&\hspace{-72pt} +\frac{9q}{10}\, {R}_{\mu}{}^{\alpha\,; \beta}{}_{\nu}\big\lbrack S  + (1-q) T  \big\rbrack\,S^2T\,{p}_{\alpha} {p}_{\beta}\nonumber\\
&&\hspace{-72pt} - \frac{q}{5}\, {R}^{\alpha \beta}{}_{;\mu\nu}\big\lbrack  S   + (1-q) T \big\rbrack\,S^2T\,{p}_{\alpha} {p}_{\beta} \nonumber\\
&&\hspace{-72pt} -\frac{2q^2}{3}\, R\,{Q}^{\alpha \beta} \big\lbrack  S    + (1-q) T  \big\rbrack\,S^2T^2\,{p}_{\alpha} {p}_{\beta} {p}_{\mu} {p}_{\nu}\nonumber\\
&&\hspace{-72pt} 
 + \frac{q^2}{3}\, {R}_{\alpha \beta}Q^{\alpha\beta}\, S{}^{2} T{}^{2}\, {p}_{\mu} {p}_{\nu}  \nonumber\\
&&\hspace{-72pt}
 -q^2\, {R}^{\alpha}{}_{ \lambda}Q^{\lambda\beta}\big\lbrack  3\, S   + (1-q)T  \big\rbrack\,S^2T^2\,{p}_{\alpha} {p}_{\beta} {p}_{\mu} {p}_{\nu}\nonumber\\
&&\hspace{-72pt} 
 +\frac{2q^2}{3}\, {R}^{\lambda \sigma}Q^{\alpha\beta}\big\lbrack 3\, S{}^{2}    + (4-7q)S T   +3(1-q)^2  T{}^{2}  \big\rbrack\,S^2T^2\,{p}_{\alpha} {p}_{\beta} {p}_{\lambda} {p}_{\mu} {p}_{\nu} {p}_{\sigma}\nonumber\\
 &&\hspace{-72pt} - \frac{q{}^{2}}{3}\, {R}_{\mu \lambda}Q^{\lambda\beta} S{}^{2} T{}^{2}\, {p}_{\beta} {p}_{\nu}  -q{}^{2}\, {R}_{\mu}{}^{ \lambda}Q^{\alpha\beta}\, S{}^{3} T{}^{2}\, {p}_{\alpha} {p}_{\beta} {p}_{\lambda} {p}_{\nu}  \nonumber\\
&&\hspace{-72pt} 
 - \frac{q{}^{2}}{6}\, {R}_{\mu \nu}Q^{\alpha\beta}\, S{}^{2} T{}^{2}\,{p}_{\alpha} {p}_{\beta}  + \frac{q{}^{2}}{6}\, {R}_{\nu \lambda}Q^{\lambda\beta}\, S{}^{2} T{}^{2}\, {p}_{\beta} {p}_{\mu}  \nonumber\\
&&\hspace{-72pt} 
 -  q{}^{2}\,{R}_{\nu}{}^{ \lambda}Q^{\alpha\beta}\, S{}^{3} T{}^{2}\, {p}_{\alpha} {p}_{\beta} {p}_{\lambda} {p}_{\mu}\nonumber\\
 &&\hspace{-72pt} -\frac{q^2}{3} {R}^{\lambda \alpha \beta \sigma}Q_{\lambda\beta} \big\lbrack  S   +(1-q)  T  \big\rbrack\,S^2T^2\,{p}_{\alpha} {p}_{\mu} {p}_{\nu} {p}_{\sigma} \nonumber\\
&&\hspace{-72pt} 
 +\frac{q}{3}\, {R}_{\mu \alpha \beta}{}^{ \lambda}Q^{\alpha\beta} \big\lbrack 3\,S    - q\,T \big\rbrack\,S^2T\,{p}_{\lambda} {p}_{\nu}\nonumber\\
&&\hspace{-72pt} +\frac{q}{6}\, {R}_{\mu \alpha \nu}{}^{ \lambda}Q^{\alpha\beta} \big\lbrack (6 +  q)S  - 2q\, T \big\rbrack\,S^2T\, {p}_{\beta} {p}_{\lambda} \nonumber\\
&&\hspace{-72pt} 
 +\frac{4q^2}{3}\, {R}_{\mu}{}^{ \lambda}{}_{ \alpha}{}^{ \sigma}Q^{\alpha\beta} \big\lbrack  S  +(1 - q)T  \big\rbrack\,S{}^{2} T{}^{2}\,{p}_{\beta} {p}_{\lambda} {p}_{\nu} {p}_{\sigma} \nonumber\\
&&\hspace{-72pt} 
 - \frac{q{}^{2}}{6}\, {R}_{\mu}{}^{ \lambda}{}_{ \nu \alpha}Q^{\alpha\beta}\, S{}^{3} T\, {p}_{\beta} {p}_{\lambda}  \nonumber\\
&&\hspace{-72pt} 
 +\frac{2q{}^{2}}{3}\, {R}_{\mu}{}^{ \lambda}{}_{ \nu}{}^{ \sigma}Q^{\alpha\beta}\big\lbrack    S+(1-q) T  \big\rbrack\,S{}^{2} T{}^{2}\, {p}_{\alpha} {p}_{\beta} {p}_{\lambda} {p}_{\sigma} \nonumber\\
&&\hspace{-72pt} 
 - \frac{q{}^{2}}{6}\, {R}_{\mu \nu \alpha}{}^{ \lambda}Q^{\alpha\beta} \big\lbrack  S +2\, T  \big\rbrack\,S^2T\,{p}_{\beta} {p}_{\lambda} \nonumber\\
&&\hspace{-72pt} 
 + \frac{q{}^{2}}{6}\, {R}_{\nu \alpha \beta}{}^{ \lambda}Q^{\alpha\beta}\, S{}^{2} T{}^{2}\, {p}_{\lambda} {p}_{\mu}  \nonumber\\
&&\hspace{-72pt} 
  - \frac{2q{}^{2}}{3}\,  {R}_{\nu}{}^{ \lambda}{}_{ \alpha}{}^{ \sigma}Q^{\alpha\beta} \big\lbrack  S  + (1-q)T \big\rbrack\,S{}^{2} T{}^{2}\,{p}_{\beta} {p}_{\lambda} {p}_{\mu} {p}_{\sigma} \nonumber\\
&&\hspace{-72pt}  - \frac{q}{3}\,R\,{Q}_{\mu}{}^{ \alpha} \big\lbrack 2\, S +(1-q)  T  \big\rbrack\,S^2T\,{p}_{\alpha} {p}_{\nu} \nonumber\\
&&\hspace{-72pt}  
 - {R}_{\alpha}{}^{ \beta}{Q}_{\mu}{}^{ \alpha} S{}^{3} T {p}_{\beta} {p}_{\nu} q - \frac{q}{6}\, {R}_{\nu \alpha}{Q}_{\mu}{}^{ \alpha}\, S{}^{2} T   
 - q\,{R}_{\nu}{}^{ \beta}{Q}_{\mu}{}^{ \alpha}\, S{}^{3} T\, {p}_{\alpha} {p}_{\beta} \nonumber\\
&&\hspace{-72pt}+ \frac{q}{3}\,{R}^{\beta \lambda}{Q}_{\mu}{}^{ \alpha} \big\lbrack 3\, S{}^{2}    +(4 - 7 q) S T  +2(1-q)^2 T{}^{2}  \big\rbrack\,S^2T\,{p}_{\alpha} {p}_{\beta} {p}_{\lambda} {p}_{\nu}  \nonumber\\
&&\hspace{-72pt}  
  - \frac{q}{3}\, {R}_{\nu}{}^{ \beta}{}_{ \alpha}{}^{ \lambda}{Q}_{\mu}{}^{ \alpha} \big\lbrack S   -(1-q) T \big\rbrack\,S^2T\,{p}_{\beta} {p}_{\lambda} \nonumber\\
&&\hspace{-72pt}  - \frac{2}{3}\, R\,{Q}_{\mu \nu}\,S{}^{3} 
+ 2\, {R}^{\alpha \beta}{Q}_{\mu \nu}\, S{}^{4}\, {p}_{\alpha} {p}_{\beta}\nonumber\\
&&\hspace{-72pt}    - \frac{2q}{3}\, R\,{Q}_{\nu}{}^{ \alpha} \big\lbrack 2\, S   +(1-q)  T  \big\rbrack\,S{}^{2}T\,{p}_{\alpha} {p}_{\mu}\nonumber\\
&&\hspace{-72pt}  
 - 2q\, {R}_{\alpha}{}^{ \beta}{Q}_{\nu}{}^{ \alpha}\, S{}^{3} T\, {p}_{\beta} {p}_{\mu} \nonumber\\
 &&\hspace{-72pt} +\frac{q}{3}\, {R}^{\beta \lambda}{Q}_{\nu}{}^{ \alpha} \big\lbrack 3\, S{}^{2}   + (4-7q)S T  +2(1-q)^2  T{}^{2}  \big\rbrack\, S{}^{2} T\,{p}_{\alpha} {p}_{\beta} {p}_{\lambda} {p}_{\mu}\nonumber\\
 &&\hspace{-72pt} - \frac{q}{6}\, {R}_{\mu \alpha}{Q}_{\nu}{}^{ \alpha}\,\, S{}^{2} T  -q\, {R}_{\mu \beta}{Q}_{\nu}{}^{ \alpha}\, S{}^{3} T\, {p}_{\alpha} {p}_{\beta}  \nonumber\\
&&\hspace{-72pt}  
 + \frac{q}{3}\,{R}_{\mu \beta \alpha \lambda}{Q}_{\nu}{}^{ \alpha} \big\lbrack 2\, S   + (1-q) T  \big\rbrack\,S{}^{2} T\, {p}_{\beta} {p}_{\lambda}\nonumber\\
&&\hspace{-72pt}+q{}^{2}\, {Q}_{\alpha}{}^{ \beta} {Q}^{\alpha \lambda}\, S{}^{3} T{}^{2}\, {p}_{\beta} {p}_{\lambda} {p}_{\mu} {p}_{\nu}  \nonumber\\
&&\hspace{-72pt} 
 +  q{}^{3}\,{Q}^{\alpha \beta} {Q}^{\lambda \sigma}\, S{}^{3} T{}^{3}\, {p}_{\alpha} {p}_{\beta} {p}_{\lambda} {p}_{\mu} {p}_{\nu} {p}_{\sigma} \nonumber\\
&&\hspace{-72pt} 
 + q\,{Q}^{\alpha \beta} {Q}_{\mu \alpha}\, S{}^{3} T\, {p}_{\beta} {p}_{\nu}  +q{}^{2}\, {Q}6{\alpha \beta} {Q}_{\mu}{}^{ \lambda}\, S{}^{3} T{}^{2}\, {p}_{\lambda} {p}_{\alpha} {p}_{\beta} {p}_{\nu}  \nonumber\\
&&\hspace{-72pt} 
 + {Q}_{\mu \alpha} {Q}_{\nu}{}^{ \alpha}\, S{}^{3} +q\, {Q}_{\mu}{}^{ \alpha} {Q}_{\nu}{}^{ \beta} \,S{}^{3} T\, {p}_{\alpha} {p}_{\beta}  + q\,{Q}^{\alpha \beta} {Q}_{\nu \alpha}\, S{}^{3} T\, {p}_{\beta} {p}_{\mu} \nonumber\\
&&\hspace{-72pt} 
 +  q{}^{2}\,{Q}^{\alpha \beta} {Q}_{\nu}{}^{ \lambda}\, S{}^{3} T{}^{2}\, {p}_{\lambda} {p}_{\alpha} {p}_{\beta} {p}_{\mu} \nonumber\\
&&\hspace{-72pt}  -q{}^{2} {Q}^{\alpha \beta\,; \lambda}{}_{ \beta} \big\lbrack  S   +(1-q)  T  \big\rbrack\, S{}^{2} T{}^{2}\,{p}_{\alpha} {p}_{\lambda} {p}_{\mu} {p}_{\nu}\nonumber\\
&&\hspace{-72pt} -q{}^{2} {Q}^{\alpha \beta\,; \lambda}{}_{ \nu} \big\lbrack  S + (1-q) T  \big\rbrack\,S{}^{2} T{}^{2}\,{p}_{\alpha} {p}_{\beta} {p}_{\lambda} {p}_{\mu} \nonumber\\
&&\hspace{-72pt} 
 + 4 q{}^{2}\,{Q}^{\alpha \beta\,; \lambda \sigma} \big\lbrack  S{}^{2}    +(1-q) S T  + (1-q)^2 T{}^{2}\big\rbrack\,S{}^{2} T{}^{2}\,{p}_{\alpha} {p}_{\beta} {p}_{\lambda} {p}_{\mu} {p}_{\nu} {p}_{\sigma} \nonumber\\
&&\hspace{-72pt} 
 + \frac{q{}^{2}}{2}\, {Q}^{\alpha \beta}{}_{; \nu \beta}\, S{}^{2} T{}^{2}\, {p}_{\alpha} {p}_{\mu} \nonumber\\
&&\hspace{-72pt} 
-q{}^{2}\, {Q}^{\alpha \beta}{}_{; \nu}{}^{ \lambda} \big\lbrack  S  + (1-q) T \big\rbrack\,S{}^{2} T{}^{2}\,{p}_{\alpha} {p}_{\beta} {p}_{\lambda} {p}_{\mu}  \nonumber\\
&&\hspace{-72pt} 
 + 4q\, {Q}^{\alpha}{}_{ \nu}{}^{; \beta \lambda}\, S{}^{4} T\, {p}_{\alpha} {p}_{\beta} {p}_{\lambda} {p}_{\mu}  -q\,{Q}_{\mu \alpha}{}^{; \beta \alpha} \big\lbrack  S   + (1-q) T \big\rbrack\,S{}^{2} T\,{p}_{\beta} {p}_{\nu}\nonumber\\
&&\hspace{-72pt}  +4\,q {Q}_{\mu}{}^{ \alpha\,; \beta \lambda} \big\lbrack  S{}^{2}   +(1-q) S T  + (1-q)^2 T{}^{2}  \big\rbrack\,S^2T\,{p}_{\alpha} {p}_{\beta} {p}_{\lambda} {p}_{\nu}\nonumber\\
&&\hspace{-72pt} 
 -q\,{Q}_{\mu}{}^{ \alpha\,; \beta}{}_{ \nu} \big\lbrack S  + (1-q) T  \big\rbrack\,S{}^{2} T\,{p}_{\alpha} {p}_{\beta}\nonumber\\
&&\hspace{-72pt} 
 + \frac{q}{2}\, {Q}_{\mu}{}^{ \alpha}{}_{; \nu \alpha}\, S{}^{2} T  \nonumber\\
&&\hspace{-72pt} -q\, {Q}_{\mu}{}^{ \alpha}{}_{; \nu}{}^{ \beta} \big\lbrack  S   +(1-q)  T \big\rbrack\,S{}^{2} T\,{p}_{\alpha} {p}_{\beta} 
 + 4\, {Q}_{\mu \nu}{}^{; \alpha \beta}\, S{}^{4}\, {p}_{\alpha} {p}_{\beta} \nonumber\\
&&\hspace{-72pt}  
 - q{}^{2}\, \Box{Q}_{\alpha \beta} \big\lbrack  S  + (1-q) T  \big\rbrack\,S{}^{2} T{}^{2}\,{p}_{\alpha} {p}_{\beta} {p}_{\mu} {p}_{\nu} \nonumber\\
&&\hspace{-72pt}  
 -q\, \Box{Q}^{\alpha}{}_{ \nu}\, S{}^{3} T\, {p}_{\alpha} {p}_{\mu}  \nonumber\\
&&\hspace{-72pt}  -q\, \Box{Q}_{\mu}{}^{ \alpha} \big\lbrack  S   + (1-q)T \big\rbrack\,S{}^{2} T\,{p}_{\alpha} {p}_{\nu} - \Box{Q}_{\mu \nu}\, S{}^{3} \nonumber\\
&&\hspace{-72pt} 
 - q{}^{2}\, {Q}^{\alpha \beta}{}_{; \beta}{}^{ \lambda} \big\lbrack  S  + (1-q)T \big\rbrack\,S{}^{2} T{}^{2}\, {p}_{\alpha} {p}_{\lambda} {p}_{\mu} {p}_{\nu} \nonumber\\
&&\hspace{-72pt} 
 + \frac{q{}^{2}}{2}\, {Q}^{\alpha \beta}{}_{; \beta \nu}\, S{}^{2} T{}^{2}\, {p}_{\alpha} {p}_{\mu}  \nonumber\\
&&\hspace{-72pt} 
 -q\, {Q}_{\mu \alpha}{}^{; \alpha \beta} \big\lbrack S   +(1-q)T\big\rbrack\, S{}^{2} T\, {p}_{\beta} {p}_{\nu} \nonumber\\
&&\hspace{-72pt} 
 + \frac{q}{2}\, {Q}_{\mu}{}^{ \alpha}{}_{; \alpha \nu}\, S{}^{2} T \;.   \label{G4ab}
 \end{eqnarray}

\section{Results for the coefficients appearing in \eqref{4.2.37}}\label{Tresults}

The coefficients appearing in \eqref{4.2.37} are given for general spacetime dimensions by
\begin{eqnarray}
T_{21}&=&\lbrack 432 N+(-12 N^3+72 N^2-528 N-576) q\\
&&\qquad+(-5 N^4+10 N^3+20 N^2+104 N+288) q^2\rbrack/\lbrack 18 N (N^2-16) (N^2-4)  q^2\rbrack\nonumber\\
&&-(1-q)^{-{N}/{2}}\Big\lbrack 432 N- \left(12 N^3+144 N^2+528 N+576\right) q\nonumber\\
&&\qquad+\left(N^4+28 N^3+176 N^2+392 N+288\right) q^2\nonumber\\
&&\qquad+\left(-2 N^4-20 N^3-64 N^2-64 N\right) q^3\nonumber\\
&&\qquad+\left(N^4+4 N^3-4 N^2-16 N\right) q^4\Big\rbrack/\lbrack 18 N (N^2-16) (N^2-4)  q^2\rbrack\;,\label{4.2.38a}\\
T_{22}&=&\lbrack 2880 N+360 N(N-8) q-60 (N+4) \left(N^2-14 N+16\right) q^2\nonumber\\
&&\qquad+(-19 N^4+194 N^3+16 N^2-2576 N+960) q^3\rbrack/\lbrack 90N (N^2-16) (N^2-4)  q^3\rbrack\nonumber\\
&&+\quad2\,(1-q)^{-N/2}\lbrack -720N+90 N (3 N+8)q -30 (N-1) (N+4) (N+8) q^2\nonumber\\
&&\qquad+(N+2) (N+4) \left(N^2+43 N-30\right)q^3-N (N+2) (N+4) (2 N+11)q^4\nonumber\\
&&\qquad+(N-2) N (N+2) (N+4) q^5
/\lbrack 45N (N^2-16) (N^2-4)  q^3\rbrack\;,\label{4.2.38b}\\
T_{23}&=&\lbrack-1440N -720 (N-2) N q 
-60 \left(3 N^3-10 N^2-24 N-32\right) q^2\nonumber\\
&&\qquad-(N+2) \left(29 N^3-152 N^2-112 N+960\right) q^3\rbrack
/\lbrack 45N (N^2-16) (N^2-4)  q^3\rbrack\nonumber\\
&&\quad+(1-q)^{1-N/2}\lbrack 1440N -240 (N+2) (N+4) q^2 \nonumber\\
&&\qquad-(N-32) N (N+2) (N+4) q^3\nonumber\\
&&\qquad + (N^2-4) N  (4 + N) q^4\rbrack
/\lbrack 45N (N^2-16) (N^2-4)  q^3\rbrack\;,\label{4.2.38c}\\
T_{24}&=&-\,\frac{(15 N-64)}{180 (N-4)}-\frac{(1-q)^{2-N/2}}{45(N-4)}\;,\label{4.2.38d}\\
T_{25}&=&\lbrack - 1728+48 \left(N^2-3 N+44\right)q+(N^5-N^4-12 N^3\nonumber\\
&&\quad-8 N^2+104 N-672) q^2\rbrack
/\lbrack 72N (N^2-16) (N^2-4)  q^2\rbrack\nonumber\\
&&\quad+(1-q)^{-N/2}\lbrack 1728 -48 (N+4) (N+11) q \nonumber\\
&&\qquad+(N+2) (N+4) \left(N^2+10 N+84\right) q^2-2 N (N+2) (N+4)^2 q^3\nonumber\\
&&\qquad+(N-2) N (N+2) (N+4) q^4 \rbrack
/\lbrack 72N (N^2-16) (N^2-4)  q^2\rbrack\;,\label{4.2.38e}\\
T_{26}&=&\lbrack 2160N + 240 \left(2 N^2-15 N-20\right) q-(N+2) (N^4-3 N^3-48 N^2\nonumber\\
&&\qquad+580 N-1680) q^2
\rbrack
/\lbrack 180N (N^2-16) (N^2-4)  q^2\rbrack\nonumber\\
&&\quad+(1-q)^{-N/2}\lbrack - 2160N +600 (N+2) (N+4) q\nonumber\\
&&\qquad-(N+2) (N+4) \left(N^2+58 N+420\right) q^2\nonumber\\
&&\qquad+2 N (N+2) (N+4) (N+28) q^3\nonumber\\
&&\qquad-(N-2) N (N+2) (N+4) q^4\rbrack
/\lbrack 180N (N^2-16) (N^2-4)  q^2\rbrack\;,\label{4.2.38f}\\
T_{27}&=&\frac{(N-5)}{180 (N-4)}+ \frac{(1-q)^{2-\frac{N}{2}}}{180 (N-4)}\;,\label{4.2.38g}\\
T_{28}&=&\lbrack 160 (N-2) + 80 (N^2-3 N+8) q + 20 (N-4) \left(N^2+N+6\right) q^2\nonumber\\
&&\qquad+ (N-2) \left(3 N^3-2 N^2-36 N-80\right) q^3\rbrack
/\lbrack 5N (N^2-16) (N^2-4)  q^3\rbrack\nonumber\\
&&\quad+(1-q)^{-N/2}\lbrack -480 (N-2) + 240 (N-8) q + 60 (N+4) (N+6) q^2\nonumber\\
&&\qquad+(N+2) (N+4) \left(N^2-2 N-60\right) q^3\nonumber\\
&&\qquad+ N (N^2-4)(N+4) (q^5-2q^4)  \rbrack
/\lbrack 15N (N^2-16) (N^2-4)  q^3\rbrack\;,\label{4.2.38h}\\
T_{29}&=&\lbrack 960 (N+2)+ 240 \left(N^2-8 N-24\right) q -480 \left(N^2-2 N-12\right) q^2\nonumber\\
&& -\left(9 N^4+26 N^3-336 N^2-224 N+1920\right) q^3
\rbrack
/\lbrack 30N (N^2-16) (N^2-4)  q^3\rbrack\nonumber\\
&&\quad+(1-q)^{1-N/2}\lbrack -960 (N+2) +240 (N+4)^2 q -240 (N+2) (N+4) q^2\nonumber\\
&&\qquad-(N-12) N (N+2) (N+4) q^3\nonumber\\
&&\qquad + N(N^2-4)  (N+4) q^4
\rbrack
/\lbrack 30N (N^2-16) (N^2-4)  q^3\rbrack\;,\label{4.2.38i}\\
T_{210}&=&\lbrack-1920-480 (N-8) q+  960 (N-2) q^2\nonumber\\
&&\qquad+N \left(N^4-N^3-14 N^2+64 N-320\right) q^3
\rbrack
/\lbrack 30N (N^2-16) (N^2-4)  q^3\rbrack\nonumber\\
&&\quad+(1-q)^{-N/2}\lbrack 1920 -480 (N+8) q +480 (N+4) q^2\nonumber\\
&&\qquad+N (N+2) (N+4) (N+8) q^3\nonumber\\
&&\qquad-2 N (N+2) (N+3) (N+4) q^4\nonumber\\
&&\qquad+N(N^2-4) (N+4) q^5\rbrack
/\lbrack 30N (N^2-16) (N^2-4)  q^3\rbrack\;,\label{4.2.38j}\\
T_{211}&=& \lbrack 24 N +8 (N-4) (N+1) q+ (N-4) (N^2-4) q^2\rbrack\nonumber\\
&&\qquad
/\lbrack 2N (N^2-16) (N^2-4)  q^2\rbrack\nonumber\\
&&\quad-(1-q)^{-N/2}\lbrack 12 N- 2 (N+2) (N+4) q\nonumber\\
&&\qquad+ (N+2) (N+4) q^2
\rbrack/\lbrack N (N^2-16) (N^2-4)  q^2\rbrack\;,\label{4.2.38k}\\
T_{212}&=&\lbrack -24 N^2-12 (N-4) (N+2) N q\nonumber\\
&&\qquad+(N-4) (N-2) \left(N^3+3 N^2-4 N-8\right) q^2
\rbrack/\lbrack 2N (N^2-16) (N^2-4)  q^2\rbrack\nonumber\\
&&\quad+4\,(1-q)^{-N/2}\lbrack 3 N^2-3 (N+4) N q\nonumber\\
&&\qquad
+(N+2) (N+4) q^2\rbrack/\lbrack N (N^2-16) (N^2-4)  q^2\rbrack\;,\label{4.2.38l}\\
T_{213}&=&-\lbrack 48+12 (N-4) q+ \left(N^2-6 N+8\right) q^2
\rbrack/\lbrack 2N (N^2-16) (N^2-4)  q^2\rbrack\nonumber\\
&&\quad+(1-q)^{-N/2}\lbrack 48-12 (N+4) q\nonumber\\
&&\qquad+(N+2) (N+4) q^2
\rbrack/\lbrack 2N (N^2-16) (N^2-4)  q^2\rbrack\;,\label{4.2.38m}\\
T_{214}&=&-2\,\lbrack 36N + 2 \left(5 N^2-24 N-32\right) q\nonumber\\
&&\qquad + (N-5) (N-4) (N+2) q^2 
\rbrack/\lbrack 3N (N^2-16) (N^2-4)  q^2\rbrack\nonumber\\
&&\quad+(1-q)^{-N/2} \lbrack 72N -16 (N+2) (N+4) q\nonumber\\
&&\qquad +(N+2) (N+4) (N+10) q^2\nonumber\\
&&\qquad
-N (N+2) (N+4) q^3
\rbrack/\lbrack 3N (N^2-16) (N^2-4)  q^2\rbrack\;,\label{4.2.38n}\\
T_{215}&=&\lbrack 288 -4 \left(N^2-12 N+80\right) q\nonumber\\
&&\qquad+ (4-N) \left(N^2+20\right) q^2
\rbrack/\lbrack 6N (N^2-16) (N^2-4)  q^2\rbrack\nonumber\\
&&\quad-(1-q)^{-N/2}\lbrack 288 - 4 (N+4) (N+20) q\nonumber\\
&&\qquad+(N+2) (N+4) (N+10) q^2\nonumber\\
&&\qquad-N (N+2) (N+4) q^3
\rbrack/\lbrack 6N (N^2-16) (N^2-4)  q^2\rbrack\;,\label{4.2.38o}\\
T_{216}&=&2\,\lbrack 48N +24 (N-2) N q+ \left(6 N^3-20 N^2-48 N-64\right) q^2\nonumber\\
&&\qquad + (N-4) (N+2) \left(N^2-N-8\right) q^3
\rbrack/\lbrack 3N (N^2-16) (N^2-4)  q^3\rbrack\nonumber\\
&&\quad-2\,(1-q)^{1-N/2}\lbrack 48 N -8 (N+2) (N+4) q^2\nonumber\\
&&\qquad+ N (N+2) (N+4) q^3
\rbrack/\lbrack 3N (N^2-16) (N^2-4)  q^3\rbrack\;,\label{4.2.38p}\\
T_{217}&=&-2\,\lbrack 36N + 2 \left(5 N^2-24 N-32\right) q\nonumber\\
&&\qquad +(N-5) (N-4) (N+2) q^2
\rbrack/\lbrack 3N (N^2-16) (N^2-4)  q^2\rbrack\nonumber\\
&&\quad+(1-q)^{-N/2} \lbrack 72 N -16 (N+2) (N+4) q\nonumber\\
&&\qquad+(N+2) (N+4) (N+10) q^2\nonumber\\
&&\qquad -N (N+2) (N+4) q^3
\rbrack/\lbrack 3N (N^2-16) (N^2-4)  q^2\rbrack\;,\label{4.2.38q}\\
T_{218}&=&\lbrack -96 N +24 N (N+4) q\nonumber\\
&&\qquad+4 \left(5 N^3-14 N^2-56 N+32\right) q^2\nonumber\\
&&\qquad+ (N-4) \left(5 N^3+4 N^2-44 N-16\right) q^3
\rbrack/\lbrack 6N (N^2-16) (N^2-4)  q^3\rbrack\nonumber\\
&&\quad+(1-q)^{-N/2} \lbrack 48 N-12 N (3 N+4) q+2 (N+4) \left(N^2+16 N-8\right) q^2\nonumber\\
&&\qquad(N+2) (N+4) (3 N+4) q^3\nonumber\\
&&\qquad+N (N+2) (N+4) q^4
\rbrack/\lbrack 3N (N^2-16) (N^2-4)  q^3\rbrack\;,\label{4.2.38r}\\
T_{219}&=&-\,\lbrack 144 N-4 \left(N^3-8 N^2+32 N+32\right) q\nonumber\\
&&\qquad +(N-4) (N+2) \left(N^3-8 N-4\right) q^2
\rbrack/\lbrack 6N (N^2-16) (N^2-4)  q^2\rbrack\nonumber\\
&&\quad+(1-q)^{-N/2} \lbrack 
72 N -2 (N+2) (N+4)^2 q\nonumber\\
&&\qquad+(N+2) (N+4) (3 N+2) q^2\nonumber\\
&&\qquad
-N (N+2) (N+4) q^3
\rbrack/\lbrack 3N (N^2-16) (N^2-4)  q^2\rbrack\;,\label{4.2.38s}\\
T_{220}&=&\lbrack 24 N +8 (N-4) (N+1) q\nonumber\\
&&\qquad+
(N-4) (N-2) (N+2) q^2
\rbrack/\lbrack N (N^2-16) (N^2-4)  q^2\rbrack\nonumber\\
&&\quad-2\,(1-q)^{-N/2} \lbrack 12 N-2 (N+2) (N+4) q\nonumber\\
&&\qquad+ (N+2) (N+4) q^2
\rbrack/\lbrack N (N^2-16) (N^2-4)  q^2\rbrack\;,\label{4.2.38t}\\
T_{221}&=&-\,\lbrack 96 N+48 (N-2) N q\nonumber\\
&&\qquad+12 \left(N^3-3 N^2-6 N-8\right) q^2\nonumber\\
&&\qquad+(N-4) (N+2) \left(2 N^2-N-12\right) q^3
\rbrack/\lbrack 3N (N^2-16) (N^2-4)  q^3\rbrack\nonumber\\
&&\quad+(1-q)^{1-N/2}\lbrack 96 N -12 (N+2) (N+4) q^2\nonumber\\
&&\qquad+ N (N+2) (N+4) q^3
\rbrack/\lbrack 3N (N^2-16) (N^2-4)  q^3\rbrack\;,\label{4.2.38u}\\
T_{222}&=&\lbrack 192 +48 (N-8) q -96 (N-2) q^2\nonumber\\
&&\qquad  -N \left(N^2+6 N-40\right) q^3
\rbrack/\lbrack 6N (N^2-16) (N^2-4)  q^3\rbrack\nonumber\\
&&\quad-(1-q)^{1-N/2} \lbrack 192 -48 (N+4) q\nonumber\\
&&\qquad+  N (N+2) (N+4) q^3
\rbrack/\lbrack 6N (N^2-16) (N^2-4)  q^3\rbrack\;,\label{4.2.38v}\\
T_{223}&=&\lbrack -192 (N+2)-48 \left(N^2-8 N-24\right) q\nonumber\\
&&\qquad+ 96 \left(N^2-2 N-12\right) q^2\nonumber\\
&&\qquad-(N-4) \left(N^4+2 N^3-18 N^2-4 N+96\right) q^3
\rbrack/\lbrack 6N (N^2-16) (N^2-4)  q^3\rbrack\nonumber\\
&&\quad-(1-q)^{1-N/2}\lbrack -96 (N+2) +24 (N+4)^2 q\nonumber\\
&&\qquad-24 (N+2) (N+4) q^2\nonumber\\
&&\qquad+
 N (N+2) (N+4) q^3
\rbrack/\lbrack 3N (N^2-16) (N^2-4)  q^3\rbrack\;.\label{4.2.38w}
\end{eqnarray}

As before the case $N=2$ must be evaluated separately, and coincides with the limit as $N\rightarrow2$ of the above expressions.
\begin{eqnarray}
T_{21}&=&\frac{(2 q^3-20 q^2+39 q-18)}{72q(1-q)}+\frac{(1-q) (2 q-3) \log (1-q)}{12 q^2}\;,\label{4.2.39a}\\
T_{22}&=&-\,\frac{(8 q^4-93 q^3+85 q^2-150 q+120)}{360 (1-q) q^2}+\frac{(2 q^3-2 q^2+3 q-4) \log (1-q)}{12 q^3}\;,\label{4.2.39b}\\
T_{23}&=&-\,\frac{(q^3+50 q^2-15 q-30)}{90 q^2}+\frac{\left(q^3-4 q^2+2\right) \log (1-q)}{6 q^3}\;,\label{4.2.39c}\\
T_{24}&=&-\frac{1}{12}-\frac{q}{90}\;,\label{4.2.39d}\\
T_{25}&=&\frac{(36-56q+25q^2-2 q^3)}{288 (1-q) q} + \frac{(q-2) (2 q-3) \log (1-q)}{48 q^2}\;,\label{4.2.39e}\\
T_{26}&=&\frac{(2 q^3-148 q^2+251 q-90)}{720 (1-q) q} + \frac{(3-q) (2 q-1) \log (1-q)}{24 q^2}\;,\label{4.2.39f}\\
T_{27}&=&\frac{(q+2)}{360}\;,\label{4.2.39g}\\
T_{28}&=&\frac{(2 q^2+15 q-30)}{60 q} -\frac{(1-q) \log (1-q)}{2 q^2} \;,\label{4.2.39h}\\
T_{29}&=&-\,\frac{(120-210q+85q^2+3 q^3)}{180 q^2} + \frac{(q^3-12 q^2+18 q-8) \log (1-q)}{12 q^3} \;,\label{4.2.39i}\\
T_{210}&=&\frac{(3 q^3-19 q^2-60 q+60)}{180 q^2} +\frac{\left(q^3-6 q+4\right) \log (1-q)}{12 q^3}\;,\label{4.2.39j}\\
T_{211}&=&-\,\frac{(6-9q+2 q^2)}{48 (1-q) q} -\frac{(1-q) \log (1-q)}{8 q^2} \;,\label{4.2.39k}\\
T_{212}&=&\frac{(6-3q-q^2)}{24 (1-q) q} + \frac{(1-2 q) \log (1-q)}{4 q^2}\;,\label{4.2.39l}\\
T_{213}&=&\frac{(q^2-12 q+12)}{96 q(1-q)}+\frac{(2-q) \log (1-q)}{16 q^2}\;,\label{4.2.39m}\\
T_{214}&=&\frac{(6 q^2-13 q+6)}{24 q(1-q)} + \frac{(q^2-5 q+3) \log (1-q)}{12 q^2}\;,\label{4.2.39n}\\
T_{215}&=&-\,\frac{(5 q^2-16 q+12)}{48 (1-q) q} -\frac{(q-3) (q-2) \log (1-q)}{24 q^2}\;,\label{4.2.39o}\\ 
T_{216}&=&-\frac{(6+3 q-10 q^2)}{18 q^2}  -\frac{\left(q^3-4 q^2+2\right) \log (1-q)}{6 q^3}\;,\label{4.2.39p}\\
T_{217}&=&-\frac{(6+3 q-10 q^2)}{18 q^2}  -\frac{(2-4 q^2+q^3) \log (1-q)}{6 q^3}\;,\label{4.2.39q}\\
T_{218}&=&\frac{(12-24 q+31 q^2+1225 q^3)}{72 (1-q) q^2} + \frac{(2-3 q+4 q^2+2q^3) \log (1-q)}{12 q^3}\;,\label{4.2.39r}\\
T_{219}&=&\frac{(6-13q+6 q^2)}{24 q(1-q)} + \frac{(3-3q+q^2) \log (1-q)}{12 q^2}\;,\label{4.2.39s}\\
T_{220}&=&-\,\frac{(6-9q+2 q^2)}{24 (1-q) q} -\frac{(1-q) \log (1-q)}{4 q^2} \;,\label{4.2.39t}\\
T_{221}&=&\frac{(6+3q-7 q^2)}{18 q^2} + \frac{(q^3-6 q^2+4) \log (1-q)}{12 q^3}\;,\label{4.2.39u}\\
T_{222}&=&\frac{(5 q^2+12 q-12)}{72 q^2}-\frac{(4-6q+q^3) \log (1-q)}{24 q^3}\;,\label{4.2.39v}\\
T_{223}&=&\frac{(24-42q+11 q^2)}{36 q^2} + \frac{\left(8-18q+12q^2- q^3\right) \log (1-q)}{12 q^3}\;,\label{4.2.39w}
\end{eqnarray}

\end{document}